\documentclass[11pt,amsmath,floats,floatfix,nofootinbib,secnumarabic, superscriptaddress, tightenlines]{revtex4}
\usepackage{graphicx} \usepackage{bm} \usepackage{epsfig}
\usepackage{pstricks}
\usepackage{setspace}
\usepackage{graphicx}
\usepackage{bm}
\usepackage{hyperref}
\usepackage{epsfig}
\usepackage{color}
\usepackage{colordvi}
\newcommand{\beq}{\begin{equation}}
\newcommand{\eeq}{\end{equation}}
\newcommand{\bea}{\begin{eqnarray}}
\newcommand{\eea}{\end{eqnarray}}
\newcommand{\bit}{\begin{itemize}}
\newcommand{\eit}{\end{itemize}}

\newcommand{\nn}{\nonumber}

\newcommand{\be}{\begin{equation}}
\newcommand{\ee}{\end{equation}}
\newcommand{\lp}{\left(}
\newcommand{\rp}{\right)}
\newcommand{\lb}{\left[}
\newcommand{\rb}{\right]}
\newcommand{\ber}{\begin{eqnarray}}
\newcommand{\eer}{\end{eqnarray}}
\newcommand{\beal}{\begin{align}}
\newcommand{\eal}{\end{align}}
\newcommand{\bes}{\begin{split}}
\newcommand{\es}{\end{split}}

\newcommand{\calo}{{\cal O}}
\newcommand{\calL}{{\cal L}}
\newcommand{\bphi}{{\bar\phi}}

\begin{document}

\title{Dissipative effects in the Effective Field Theory of Inflation}

\begin{abstract}
We generalize the effective field theory of single clock inflation to include dissipative effects. Working in unitary gauge we couple a set of composite operators, $\calo_{\mu\nu\ldots}$, in the effective action which is constrained solely by invariance under time-dependent spatial diffeomorphisms. We restrict ourselves to situations where the degrees of freedom responsible for dissipation do not contribute to the density perturbations at late time. The dynamics of the perturbations is then modified by the appearance of `friction' and noise terms, and assuming certain locality properties for the Green's functions of these composite operators, we show that there is a regime characterized by a large friction term $\gamma \gg H$ in which the $\zeta$-correlators are dominated by the noise and the power spectrum can be significantly enhanced. We also compute the three point function $\langle \zeta\zeta\zeta\rangle$ for a wide class of models and discuss under which circumstances large friction leads to an increased level of non-Gaussianities. In particular, under our assumptions, we show that strong dissipation together with the required non-linear realization of the symmetries implies $|f_{\rm NL}| \sim \frac{\gamma}{c_s^2H} \gg 1$. As a paradigmatic example we work out a variation of the `trapped inflation' scenario with local response functions and perform the matching with our effective theory.  A detection of the generic type of signatures that result from incorporating dissipative effects during inflation, as we describe here, would teach us about the dynamics of the early universe and also extend the parameter space of inflationary models.
\end{abstract}

\author{Diana L\'opez Nacir}
\affiliation{Departamento de F\'isica, Facultad de Ciencias Exactas y Naturales, UBA and IFIBA, CONICET. Ciudad Universitaria, Pabell\'on 1, 1428, Buenos Aires, Argentina.\vskip 0.25cm}
\author{Rafael A. Porto}
\affiliation{School of Natural Sciences, Institute for Advanced Study, Einstein Drive, Princeton, NJ 08540, USA\vskip 0.25cm}
\affiliation{Department of Physics \& ISCAP, Columbia University, New York, NY 10027, USA \vskip 0.25cm}
\author{Leonardo Senatore} 
\affiliation{Stanford Institute for Theoretical Physics, Stanford University, Stanford, CA 94305\vskip 0.25cm} 
\affiliation{KIPAC, Stanford University \& SLAC, Stanford, CA 94305\vskip 0.3cm} 
\author{Matias Zaldarriaga}
\affiliation{School of Natural Sciences, Institute for Advanced Study, Einstein Drive, Princeton, NJ 08540, USA\vskip 0.25cm}

\maketitle
\newpage
\tableofcontents

\section{Introduction \& main results}\label{warm}

The Effective Field Theory (EFT) paradigm is one of the cornerstones of theoretical physics, from the standard model to condensed matter systems \cite{iraeft,joeeft}. EFT ideas have recently gathered thrust also in the realm of gravitational physics. For example, EFT techniques have been introduced in \cite{nrgr1,nrgr1p,radgo} to solve for the dynamics of coalescing binary systems to great accuracy \cite{nrgr2,nrgr2p,nrgr2p2,andi,foffa,eftrev};  and an EFT setup has been proposed for the study of cosmological perturbations in \cite{eftfluid}.

The EFT of inflation for the case of single field (one clock) models was developed in \cite{eft1, Cheung:2007sv, Senatore:2009gt, Senatore:2009cf, Senatore:2010jy, Creminelli:2006xe, Bartolo:2010bj, Bartolo:2010di,dan2,strongc,dan2new}. The starting point is an action in unitary gauge (where all the fluctuating degrees of freedom are encoded in the metric) which is required solely to be invariant under time dependent spatial diffeomorphisms. The advantage of this approach is that it enables us to parameterize all possible signatures of inflation in terms of a set of coefficients for (`higher-dimensional') operators in a Lagrangian built with the low energy (large distance) degrees of freedom, and constrained only by the symmetries of the theory. Within the EFT it is possible to describe the fluctuations around an approximate de Sitter background without any assumption about the fundamental degree of freedom that is driving inflation\footnote{One may wonder about the underlying theory that produces the background. However, once the inflationary paradigm is accepted, it is ultimately the theory of fluctuations that is directly tested by observations.} (which may as well be strongly coupled).
To that end it is useful to restore time diffeomorphisms (broken by the existence of a preferred time slicing) by means of the St\"uckelberg field $\pi$, which is the Goldstone boson that realizes time reparameterizations non-linearly. 

There are two important consequences of introducing the $\pi$ field which turn out to be extremely helpful. First of all, one notices that at sufficiently large energies ($E \gg \sqrt{\epsilon} H$) the Goldstone boson captures all the information about the physical scalar mode ($\zeta\simeq-H\pi$), namely the `longitudinal mode' in unitary gauge.\footnote{This is similar to what occurs in gauge theories, for example in longitudinal $WW$ (or $Z$) scattering, whose amplitude can be obtained in terms of processes involving the associated Goldstone bosons (`eaten' by the $W$'s and $Z$'s) at high enough energies, $E \gg m_W$ (`equivalence theorem') \cite{peskin, ben}.} And secondly, there is a limit (decoupling limit, i.e. $\epsilon \to 0$) in which we may ignore all the effects induced by gravitational interactions.\footnote{This is also equivalent to sending the gauge coupling $g$ to zero while keeping the symmetry breaking scale $v$ fixed, or in our case taking $M_p\to \infty$ and keeping $M_p^2|\dot H|$ finite.} These two observations allow us to concentrate on a theory of Goldstone bosons, whose interactions are dictated by symmetry, which greatly simplifies the computations.

An EFT for multi-field inflation was recently introduced in \cite{multieft}, where new (light) degrees of freedom were included. Multi-field inflation can reproduce the signals from single field models, but can also give rise to new ones which (provided certain shapes are detected) may allow us not only to separate between the two, but also distinguish amongst different realization in multi-field scenarios. In \cite{multieft} the authors concentrated on the case in which additional degrees of freedom (ADOF) affect directly the overall curvature perturbation (or isocurvature perturbations). This can happen for example if these extra fields modify the reheating time and consequently the duration of inflation, or if they affect the composition of the plasma in the reheating phase. Because of this, the fields considered in \cite{multieft} were light scalars so that they acquired scale invariant perturbations.

In this paper we also consider situations in which ADOF other than the Goldstone boson are excited. However, contrary to the models in \cite{multieft}, we concentrate on cases where this extra sector does not directly affect the duration of inflation, or the composition of the plasma, but it alters the dynamics of inflation by directly coupling to the clock around or before the time the modes we observe cross the horizon. This includes, for instance, the `trapped inflation' scenario \cite{trapped} where the production of particles modifies the evolution of the inflaton $\phi$, while producing negligible direct contributions to (late time) density fluctuations due to dilution.

Since in general these new particles will contribute to the stress energy tensor of the background, the {\it true} Goldstone boson includes not only the physical clock, which we assume uniquely controls the physics of inflation, but also a component that depends on the fluctuations of these ADOF. To isolate the relevant component whose perturbations control the observed density fluctuations, here we will not work with this Goldstone boson (namely the field whose quadratic Lagrangian is uniquely fixed by the background), and thus reserve $\pi$ for the fluctuations of the clock that determines the end of inflation (for example, the inflaton $\phi$ in the model of \cite{trapped}). This choice will slightly modify the construction of the EFT, and in particular the choice of `unitary gauge' and overall normalization of our $\pi$. However, it maintains the relationship between $\pi$ and $\zeta$, i.e. $\zeta \simeq -H\pi$ (at linear order), which we find more convenient.\\

Given that the ADOF do not explicitly contribute to observable quantities, this suggests we may integrate them out and obtain an effective action in terms only of $\pi$. However, this procedure is not straightforward, mainly for two reasons. First of all, even though we do not observe the fluctuations of the ADOF as `external states', they are produced during inflation and in general are not in the vacuum; and secondly, these fluctuations may not be gapped, or in other words, we are allowing for very soft (essentially gapless) collective excitations. This means that the effective action cannot be described with only one degree of freedom, and in turn this will be linked to dissipation whose effective description is one of the goals of the present paper.  Notice that this does not mean that the ADOF are necessarily `light'. For example,  as shown in \cite{trapped}, heavy particles (compared to the Hubble scale) can be ultimately produced by the time dependence in the Hamiltonian induced by the physical clock $\phi$.\footnote{In the case of trapped inflation particles are produced when the adiabaticity condition is violated \cite{beauty,trapped}.} However, their influence in the dynamics of the perturbations of the clock, which includes dissipation and noise, remains active even at low(er) frequencies, i.e. $\omega \simeq H$.\\

One can study dissipative effects using the `in-in' closed-path-time formalism \cite{calzetta}, however, here we will resort to a different setup. Similarly to the EFT for dissipation introduced in \cite{dis1,dis2} (to deal with gravitational wave absorption in binary black hole systems), we will incorporate dissipative effects in the EFT of inflation by coupling the metric in unitary gauge to a set of (scalar, vector and tensor) composite operators, $\cal O_{\mu\nu\ldots}$, constrained solely by the symmetries of the EFT. All the information about the dissipative sector is thus encoded in a set of correlation functions which can be matched against observation, e.g. the power spectrum, non-Gaussianities, etc. (A similar formalism  \cite{hydro} can be used to describe dissipative effects in the EFT for hydrodynamics developed in \cite{efthydro}.)
This approach is clearly very ample, and for that reason it is also difficult to treat in complete generality. However, it is possible to identify a physical regime in parameter space where many simplifications occur. We will consider physical situations where the time scale for dissipation and fluctuation induced by the ADOF is much smaller than a Hubble time, with negligible memory effects. Under this condition fluctuation and dissipative effects become local, which allows us to study many possible realizations in complete generality. We spell out our assumptions in detail throughout the paper.\\

Two separate type of contributions enter in the dynamics of $\pi$. There is the noise induced by the ADOF, and there is also the back reaction due to the mutual interaction, i.e. `friction', namely a $\gamma\dot\pi$ term. (This is due to the physical fact that during inflation energy is being damped into the ADOF.) 
Here we concentrate in the strong dissipative regime, where $\gamma$ is taken to be much larger than $H$. In such circumstances, we show the memory on the initial conditions washes out and the power spectrum and non-Gaussianities are dominated by the noise induced by the ADOF. Moreover, the former can be significantly enhanced with respect to the quantum fluctuations in the Bunch-Davies state. The reason is twofold: first of all the size of the fluctuations for the normalized $\pi$ field ($\pi_c \equiv \sqrt{N_c}\pi$) is larger than Hubble, this is because these are not produced by the vacuum (also they freeze out at a larger value of $k/a$, i.e. $c_s k/a_\star = c_s k_\star \simeq \sqrt{\gamma H} \gg H$, for $\gamma \gg H$); and secondly, the normalization scale $N_c$ can be smaller than what is required by Einstein equations in the absence of any other contribution to the stress energy tensor, i.e. $N_c \leq 2 M_p^2|\dot H|/c_s^2$.\\

In the EFT of inflation~\cite{eft1} the non-linear realization of time diffeomorphisms allows us to relate different observables (such as the two and three point $\zeta$-correlation functions). For instance, considering the case of scalar couplings and taking the noise to be Gaussian, we show that terms of the form $S_{\calo g^{00}} \equiv \int \sqrt{-g}\calo g^{00}$ (in unitary gauge) can increase the level of non-Gaussianities by a factor of  $\gamma/H$ with respect to the result for single field inflation without ADOF \cite{eft1}, yielding $|f_{\rm NL}| \simeq \gamma/(Hc_s^2)$. In the spirit of the EFT for multifield inflation of \cite{multieft} we discuss a class of models which fit into this category.
On the other hand, operators of the form $S_{f\calo}\equiv \int \sqrt{-g} f(t)\calo$ produce two types of non-linearities. Either sourced by direct non-linear couplings to $\calo$, or induced by contributions beyond linear response theory. We show there is a class of models where the computation of dissipation and non-Gaussianities are linked. Assuming as we do the existence of a preferred clock that drives inflation, we show that the linear and non-linear response are indeed related in such a way to produce non-Gaussianities of order $|f_{\rm NL}| \simeq \gamma/(c_s^2H)$.

The basic idea is perhaps better illustrated in the standard inflationary scenario of a slowly rolling scalar field  $\phi$. As we know from our classical mechanics intuition, to induce dissipation we need factors of $\dot\phi$ in the equation of motion (EOM). On the other hand, (non-linearly realized) general covariance requires $\partial_t\phi \to n^\mu \partial_\mu \phi$ where $n^\mu$ is the normal vector orthogonal to the equal time surfaces given by $n^\mu = g^{\mu\nu}\partial_\nu \phi/\sqrt{-(\partial\phi)^2}$. At the level of the perturbations of the clock, namely $\phi \to \bphi + \delta\phi$, this induces a dissipative term $\gamma\dot{\delta\phi}$ but also non-linear interactions, and in particular a term $\gamma (\partial_i\delta\phi)^2/\dot\bphi$ (properly normalized). The latter gives non-Gaussianities of order
\beq \frac{\gamma}{\dot\bphi}\frac{(\partial_i \delta\phi)^2}{c_s^2\partial_i^2\delta\phi} \sim f_{\rm NL}\zeta \quad \to \quad |f_{\rm NL}| \sim \frac{\gamma}{c_s^2 H},\eeq
where we used $\zeta \simeq -H\delta\phi/\dot\bphi$, and allowed for $c_s \leq 1$.  We will make this argument more precise and at the same time generic for all models of single clock inflation. We conclude that in either case, $S_{\calo g^{00}}$ or $S_{f(t)\calo}$, the non-linear interactions are significantly enhanced in the strong dissipative regime with $\gamma \gg H$, as one would have naively expected.

Finally, vector couplings such as $\int \sqrt{-g} \calo_\mu g^{\mu0}$, may induce a large friction term but without the addition of non-linear couplings between $\pi$ and $\calo_\mu$. However, once again depending on the model, the non-linear response will produce large non-linearities as above.\\

This paper is organized as follows: In the remaining of sec. \ref{warm} we discuss the basic ideas and results, putting emphasis on the overall picture rather than the technicalities of the calculations. Then in secs. \ref{eftsup} - \ref{nongaus} we explicitly construct the EFT to include dissipation in inflation and provide detailed support for our claims. In sec. \ref{matching} we perform the matching for a key example: (a local version of) trapped inflation. The idea of including dissipative effects during inflation is also a key element of the warm inflation paradigm \cite{berera1,berera2,warm}, which we will briefly comment upon towards the end. We relegate other examples and more technical points to appendices. Everywhere we set $c=\hbar=k_B=1$ and adopt the mostly plus sign convention.

\subsection{Preliminaries}
Let us imagine adding a friction term to the (one-dimensional) harmonic oscillator. For convenience of notation let us denote the displacement from equilibrium as $\pi(t)$. The EOM reads
\beq
\label{eom1}
 \ddot \pi + \gamma\dot \pi + \omega_0^2 \pi = J,
\eeq
where for future purposes we added a stochastic force with $\langle J\rangle=0$.
As it is well known, this seemingly innocuous equation does not derive from a Lagrangian of the form $\calL(\pi,\dot \pi)$.\footnote{
One can, nonetheless, construct models where a dissipative equation results from an effective description where the ADOF responsible for dissipation are `integrated out' in a `in-in' formalism with twice as many degrees of freedom \cite{calzetta,weiss,Zhang}. This is not the route we follow in this paper (see next sub-section).}  The reason is simple, energy is not conserved. In fact,
\beq
\label{eneloss}
\frac{dE}{dt} =-\gamma \dot\pi^2.
\eeq

The expression in Eq. (\ref{eom1}) is local in time, however, in general the effective EOM for $\pi$ takes the so called `Langevin' form, which is non-local, i.e.
\beq
\ddot\pi +\omega_0^2 \pi +  \int dt' \tilde\gamma(t-t')\pi(t') = J(t).
\eeq

Clearly,  our treatment would greatly simplify if we were allowed to perform a local approximation. This in turn amounts to making an assumption about the ADOF leading to $\tilde\gamma(t)$ in the above equation. In particular, to get the EOM in the form of Eq.~(\ref{eom1}) we need
\beq
\label{imgam}
{\rm Im} \tilde\gamma(\omega) \simeq  \gamma \omega,
\eeq
with $\gamma$ a constant. The relationship in Eq.~(\ref{imgam}) is sometimes referred in the literature as {\it Ohmic} behavior \cite{weiss}. We illustrate some examples in appendix \ref{app0}.\\

In practice we do not expect Eq. (\ref{imgam}) to hold up to arbitrarily high frequencies, and a more realistic (though phenomenological) approach is provided by Drude's model, with Eq. (\ref{imgam}) replaced by \cite{weiss}
\beq
\label{imdrude}
{\rm Im} \tilde\gamma_D(\omega) = \gamma \omega \left(1+ \omega^2/\omega_D^2\right)^{-1},
\eeq
where $\omega_D$ serves as a cutoff. In this  case there is a memory time on the scale $\tau_D \sim \omega_D^{-1}$, and an exponential damping \beq \gamma_D(t) \sim \Theta(t)\gamma_D
\frac{e^{-t/\tau_D}}{\tau_D}.\label{drude}\eeq

Instead of relying on assumptions about the physics of the ADOF, sometimes it is possible to connect the properties of the noise and Green's functions by means of some well known theorems; for example if we assume the noise satisfies the following conditions:
\bea
\langle J(t) \rangle &=& 0 \\
\langle J(t)J(t') \rangle &\simeq& \nu_J \delta(t-t') \label{whiten},
\eea
 with $\nu_J$ a constant. This is the case, for instance, if the ``environment" (i.e. the dissipative ADOF) is placed at a (sufficiently large) temperature $T$. Then, using the Fluctuation-Dissipation (FD) theorem \cite{calzetta, weiss} one can show 
 \beq
 \label{fdth}
 {\rm Im} \tilde\gamma(\omega) \simeq  \frac{\nu_J}{T} \omega
\eeq
(assuming equilibration occurs sufficiently fast after the perturbations are turned on), or in other words $\nu_J \simeq \gamma T.$ In this scenario the EOM becomes local and memory effects are washed away.\\

To keep the treatment as simple as possible, we will study situations where the local approximation applies. Later on we will discuss some specific examples. However, at the level of the EFT we refrain from adopting any model for the underlying dynamics of the ADOF. We introduce the basic idea of our approach next.

\subsection{The story of $\cal O$}\label{storyo}

 We consider now the generic situation where we have a theory for $\pi$ that describe small (long wavelength) perturbations of a dynamical system with Lagrangian ${\cal L}_\pi$. Following \cite{dis1,dis2} to include dissipation we couple $\pi$ to a composite operator $\cal O$ such that
\beq
\label{calopi}
S_{\rm int} =- \int d^4x \calo(x)\pi(x).
\eeq
For the cases where there is a shift symmetry, $\pi \to \pi +c$, the interaction takes the form
\beq
\label{tcalop}
\tilde S_{\rm int} =\int d^4 x \tilde\calo(x)\dot\pi(x),
\eeq
which can be described as in Eq. (\ref{calopi}) by replacing $\calo \to \dot{\tilde\calo}$.\\

The addition of $S_{\rm int}$ allows us to study the response of our system to the interaction with a dissipative sector (represented by $\calo$)
 in complete generality. The virtue of this approach lies in that we do not need to assume any specific representation for the dynamics of $\calo$ (which could in principle represent a strongly coupled sector), and we just need to make sure interactions such as Eq. (\ref{calopi}) respect the symmetries of the long distance physics described by  $\pi$  \cite{dis1,dis2}.\\
 
 We define now an operator $\delta\calo\equiv \calo-\bar\calo$, with $\bar\calo$ the background expectation value of $\calo$. We then split $\delta\calo$ into two pieces, schematically, 
\beq
\delta\calo = \delta\calo_S(x) + \delta\calo_R(x),
\eeq
with $\delta\calo_S(x)$ representing the stochastic part of $\calo(x)$ in the absence of $\pi$,
whereas $\delta\calo_R(x)$  corresponds to the change of the expectation value of $\calo$ that results as a response through the interaction to a $\pi$ fluctuation. We start by computing $\delta\calo_R(x)$ within linear response theory, with $\pi$ playing the role of  the external `force' that disturbs the dynamics of the degrees of freedom associated with the operator $\calo$. (For an introductory account of response theory see for
instance \cite{Fetter}.) Namely,
\beq
\label{dret0}
\delta\calo_R(x) =  -\int d^4y G_{\rm ret}^\calo(x,y)\pi(y),
\eeq
where
\beq
\label{dret}
G_{\rm ret}^\calo(x,y) = i\langle[\delta\calo(x),\delta\calo(y)]\rangle \theta(t_x-t_y).
\eeq

If we denote as $D_\pi \pi=0$ the linearized EOM that derives from $\cal L_\pi$ in the absence of ADOF, then the addition of Eq. (\ref{calopi}) leads to
\beq
D_\pi \pi =- \calo  + O(\pi^2).
\eeq

After we solve for $\calo$ we get
\beq
\label{dpig}
D_\pi \pi -  \int d^4y G_{\rm ret}^\calo(x-y)\pi(y) = -\delta\calo_S + O(\pi^2),
\eeq
or in Fourier space
\beq
\left(D_\pi({\bf q},\omega)- G_{\rm ret}^\calo({\bf q},\omega)\right) \pi_q(\omega) =-\delta\calo_S({\bf q},\omega) + \ldots.
\eeq

In the above expression $D_\pi({\bf q},\omega)$ is an analytic function, but presumably this may not always be the case for $G_{\rm ret}^\calo({\bf q},\omega)$. However, we will assume we can use a local approximation for the dynamics induced by the Green's function for a vast range of frequencies up to corrections of order $({\bf q}/M_\calo, \omega/\Gamma_\calo)$, where $M_\calo,\Gamma_\calo$ are the typical scales at which the non-locality starts to be non-negligible. We will make this more precise when we study the inflationary case. At any rate, it is clear that in order to recover an equation as in (\ref{eom1}) we need\footnote{As one would expect dissipation is associated with the imaginary part of the Green's functions in Fourier space, a.k.a. the optical theorem, see appendix \ref{appA}.}
\beq
\label{expG}
{\rm Im} G_{\rm ret}^\calo (\omega,{\bf q}) \simeq {\rm Im} G_{\rm ret}^\calo (\omega,{\bf 0}) \simeq  \gamma \omega,
\eeq
for $\omega>0$, with $J \equiv -\delta\calo_S$.  While an expansion in derivatives is justified by the fact that we can work at $\omega\lesssim \Gamma_{\cal O}$, the order at which the Taylor expansion starts is an assumption about the UV physics. (Notice that this is the lowest analytic order since the imaginary part of the Green's function has to be odd in $\omega$.)

Contrary to the imaginary part, the real part of the Green's function is even in $\omega$, and therefore need not vanish as $\omega \to 0$.\footnote{Moreover, the real part of the Green's function $G^\calo_{\rm ret}(\omega,{\bf q})$ may also contribute to the speed of sound via a term quadratic in ${\bf q}$.} This would lead to a mass for $\pi$ (and consequently for $\zeta$), hence to an evolution for $\zeta$ outside the horizon. This is not forbidden by any principle, but if this was the case it would lead to an effect on curvature perturbations at late times from the ADOF, which ought to be negligible by our main assumption. Therefore, in this paper we consider situations where the ADOF are sensitive only to derivatives of $\pi$ rather than the value of $\pi$ itself. This requires an effective shift symmetry at the level of the response. (This is the case for example in the model of \cite{trapped} where the production of ADOF is connected with a breaking of adiabaticity \cite{beauty}, due to the time dependent background, and fluctuations in the ADOF are related to derivatives of $\pi$.) If for instance inflation is driven by a scalar field $\phi$, in principle the response of the $\calo$ operators might be very non-linearly related to $\phi$, and the Green's function for $\pi$ may as well depend non-linearly on the coupling constants in the Lagrangian that includes both $\phi$ and $\calo$. In fact, the shift symmetry may not be even present at this stage. For this reason we will  employ the term `emergent shift symmetry' to refer to cases when the real part of the Green's function vanishes as $\omega\rightarrow 0$. We give an example of this phenomenon in sec.~\ref{localtrap} where we discuss a simplified version of the trapped inflation model~\cite{trapped}.\\

In principle if we had a specific UV model of the ADOF in mind, we could compute the exact Green's function and compare with the one derived from the EFT. This is typical in EFTs and it is often referred as {\it matching} \cite{iraeft}.  We will provide a realization of this matching procedure when we study some specific models later on in sec.~\ref{matching}. (See also appendices \ref{appB} and \ref{mag00}.) Nevertheless, even though we do not have an explicit description for $\calo$ (which might be rather involved), the key point of the EFT approach is that we can still study the dynamics of $\pi$ in terms of the Green's function of the type of Eq. (\ref{dret}) under the approximation of Eq.~(\ref{expG}), where $\gamma$ is kept as a free parameter.\\

Notice that the condition for a $\gamma\dot\pi$ local dissipative dynamics implies
\beq
\label{tildG}
{\rm Im} \tilde G_{\rm ret}^\calo (\omega) \sim 1/\omega,
\eeq
for the operator $\tilde\calo$ in Eq. (\ref{tcalop}). This behavior is not allowed near $\omega\simeq 0$ by some basic analytic properties of the Green's functions. One can nonetheless imagine situations where Eq. (\ref{tildG}) holds at intermediate frequencies, i.e. $\mu_\calo \ll \omega \ll \Gamma_{\calo}$ (with $\mu_\calo, \Gamma_\calo$ related to the response functions of the $\calo$'s), while at very low frequencies the Green's function is analytic. Since we are assuming the relevant energy scales are smaller than the typical ones in the ADOF sector, that we take to be of order $\Gamma_\calo$, in order for such behavior to occur it requires the Green's function to have a mass scale anomalously low compared to $\Gamma_\calo$. This is (most probably) a sign of tuning in the effective theory.\footnote{We show in appendix \ref{mag00} how, upon tuning, one may obtain the scaling of Eq. (\ref{tildG}), in particular 
\beq
\omega^2{\rm Im}{\tilde G}_{\rm ret}^{\calo}(\omega) \simeq \gamma \frac{\omega^3}{\omega^2+\mu_\calo^2} + O(\omega/\Gamma_\calo) \simeq  \gamma\omega + O(\mu_\calo/\omega, \omega/\Gamma_\calo)
\eeq
for $\mu_\calo \ll \omega \ll \Gamma_{\calo}$, such as it is required for a $\gamma\dot\pi$ term stemming from the coupling $\tilde\calo\dot\pi$.}
In any case, it is reasonable to assume this tuning would affect only one parameter (see appendix \ref{mag00} for more details).\\

An example the reader may be familiar with is the so called Abraham-Lorentz-Dirac (ADL) force, which arises from the (velocity dependent) interaction $A \dot\pi$  after we integrate out the electromagnetic field (here $A$ plays the role of $\calo$). The EOM turns out to be local in time \cite{radre} provided we choose boundary conditions where all radiation is outgoing (that is we do not include `mirrors').\footnote{Non-local dynamics appears in the so called `memory effect' in gravitational wave radiation off coalescent binary inspirals \cite{memory}. The latter is entirely due to the non-linear interactions of the gravitational field which are not present in electromagnetism.} However, in this case the dissipative term depends on the third time derivative of the position, i.e. $\dddot \pi/\Lambda_e$ with $\Lambda_e \simeq m_e/e^2$ a cutoff scale related to the unitarity bound of the theory. (This follows from an expression similar to Eq. (\ref{expG}) applied to the vector potential.) Even though in this paper we mostly concentrate on dissipative effects represented by Eq. (\ref{eom1}) we will also comment on higher derivative couplings in sec. \ref{localo2}.\\

As we mentioned earlier another setup where a local approximation appears naturally is to consider a white noise for the $\delta\calo_S$'s, as it would be the case in thermal equilibrium at large temperatures. In such scenario
\beq
\label{dbarh}
\langle \delta\calo_S({\bf{k}},t)\delta\calo_S({\bf{q}},t') \rangle = (2\pi)^3 \nu_{\calo} \delta(t-t')\delta^{(3)}({\bf{k}}+{\bf{q}}),
\eeq
and using the FD theorem we get (see Eq. (\ref{fdth}))
\beq
\label{fdth2}
\frac{{\rm Im}G_{\rm ret}^\calo(\omega)}{\omega} \equiv \gamma = \frac{\nu_{\calo}}{T},
\eeq
 as required.


\subsection{The two-point function}\label{twopoint}

One of the most important observables we are interested in this paper is the two-point function of the $\pi$ field at horizon crossing, which is related to the two-point function for the curvature perturbation $\zeta$ ($\zeta \simeq -H\pi$ \cite{eft1}), that is conserved outside the horizon. In the standard scenario of an expanding universe the linearized EOM for the $\pi$ field is equivalent to Eq. (\ref{eom1}) with $\gamma \to 3H$ and $\omega_0 = c_s k_{\rm ph} \equiv \frac{c_s k}{a(t)}$,
that is (notice that now we have a time dependent $\omega_0$)
\beq
\label{eom1p}
\ddot \pi_k +3H\dot\pi_k + c_s^2 \frac{k^2}{a^2} \pi_k =0.
\eeq

The reason the mode freezes out is due to the fact that the term proportional to $\omega_0$ goes to zero as $t\to +\infty$, for
 a fixed (co-moving) $k$, and a constant value for $\pi$ solves the equation. The time at which this happens is determined
by the condition $k_\star \equiv k/a(t^\star) \sim H(t^\star)/c_s$, or $\omega_\star \sim H_\star$. (The $\star$ denotes a quantity at freeze out.) If we impose the Bunch-Davis state as initial condition, the well known result is \cite{eft1}:
\beq
\label{powersh}
\langle \zeta_k \zeta_q\rangle_{\rm BD} = (2\pi)^3 \frac{H_\star^2}{4c_s^\star\epsilon_\star M^2_pk^3} \delta^{(3)}{({\bf q + k})},
\eeq
with $\epsilon \equiv -\dot H/H^2$. Since these are the quantum zero-point energy fluctuations, this expression follows straightforwardly from
\beq
\label{s2}
{\cal S}_2= \epsilon M_p^2\frac{H^2}{c_s^2}\int d^4 x \dot\pi^2 \sim \epsilon_\star M_p^2 c^\star_s\frac{\zeta^2}{\omega_\star^2} \sim 1 \quad\to\quad \zeta \sim \frac{\omega_\star}{\sqrt{2{c_s^\star}\epsilon_\star}M_p}.
\eeq

Naively one would expect that the addition of a friction term to Eq. (\ref{eom1p}), of the form $\gamma \dot \pi_k$, will modify the crossing condition to $k/a(t^*) \sim \gamma(t^*)/c_s$, leading to $\omega_\star \sim \gamma$. Hence, from Eq. (\ref{s2}), it appears as if it would produce a larger two-point function. However, this is incorrect, and the contribution from the homogenous equation turns out to be negligible. We will show this in detail later on (see sec. \ref{homog}), but the basic idea is rather simple as we argue next.\footnote{Notice that the term in $\gamma\dot\pi$ does not have the same role as the standard $3H\dot\pi$ one in an expanding universe, since for the latter the frequency of all modes redshift at the same rate $H$, while that is not case with $\gamma$.}

\subsubsection{Homogeneous solution}\label{hompart}

Let us take Eq. (\ref{eom1p}) but assume there is an extra dissipative term $\gamma\dot\pi_k$, with $\gamma \gg H$. To gain some intuition we will solve the equation adiabatically starting with constant values of $\omega_0 = c_s k_{\rm ph}$. (For reasons that will be clear in sec. \ref{homog}, we also parameterize time running from $[-|t_0|,0]$.) This is a good approximation as long as $\dot\omega_0/\omega_0^2 \ll 1$, which holds provided $c_s k_{\rm ph} \ge H$. (In fact, as we will see, $c_s k_\star \simeq \sqrt{\gamma H} \gg H$.) It is easy to see there are two independent solution, namely
\beq
f_\mp (t) = A_{\mp} \exp\left[\frac{-\gamma t}{2}\left(1\mp \sqrt{1-4\frac{\omega_0^2}{\gamma^2}}\right)\right].
\eeq

We will consider two regimes. First, at some early time ($|t_0| \gg 1/\gamma$) we assume we are in the solution with $\omega_0 = c_s k_{\rm ph} \gg \gamma$, so that we match it with the usual oscillatory behavior normalized to the Bunch-Davies vacuum
\beq
f^\pm_{\rm BD} = \frac{1}{\sqrt{2\omega_0}} e^{\pm i\omega_0t_0}.
\eeq
This is justified by realizing that by going sufficiently back in time, the mode begins to oscillate fast enough to decouple from the ADOF. We expect this to happen for $\omega\gtrsim \Gamma_\calo$. This fixes the overall coefficient to
\beq
\label{anormp}
A_\pm \simeq  \frac{e^{\frac{-\gamma|t_0|}{2}}}{\sqrt{2\omega_0}} + O(\gamma/\omega_0).\\
\eeq
As time progresses, we enter our second regime, where $\omega^2_0$ decreased to the point the mode freezes out, $\omega_0 \to \omega_0^\star = c_s k_\star$. One can then show that after matching both regimes the solution that dominates scales like\footnote{The other solution decays faster with time.}
\beq
\label{fminus}
f_- \sim \frac{e^{-\frac{\gamma |t_0|}{2}}}{\sqrt{\omega^\star_0}} e^{\frac{-(\omega^\star_0)^2 t}{\gamma}}.
\eeq
As a consequence the homogenous solution acquires a damping factor $e^{-\frac{\gamma |t_0|}{2}} \ll 1$. A detailed analysis shows that indeed it acquires this type of exponential suppression (see section \ref{homog}).\\

From here we conclude that the contribution to the two-point function from the homogenous part becomes exponentially small as $t \to 0$, which opens the door for the source noise to dominate. 

\subsubsection{Noise}\label{noise}

In what follows we present a basic physical argument for the computation of $\langle \pi \pi \rangle$ due to the noise $\delta\calo_S$. (For ease of notation here we return to the more traditional range for time, $t \in [0,\infty]$.) The detailed analysis will be presented in section \ref{power}.\\

In the EOM of (\ref{eom1}) we now have to deal with an extra term, namely
\beq
\ddot \pi_k +\gamma\dot\pi_k + \omega_0^2 \pi_k =- N^{-1}_c \delta\calo_S,
\eeq
where $N_c$ is a normalization factor. Since we take $\gamma \gg H$, we work in the limit where
\beq
\omega_0 \ll \gamma .
\eeq
Under this condition we can in principle find the exact Green's function, however it is easier to look at the simplified version that holds in the overdamped limit
 \beq
\left(\frac{d}{dt} +\frac{\omega_0^2}{\gamma} \right) G^k_\gamma(t-t') = \frac{1}{\gamma} \delta(t-t'),
\eeq
where we drop the factor $\frac{d^2 G^k_\gamma}{dt^2} \ll \gamma \frac{d G^k_\gamma}{dt}.$ The solution reads 
\beq
G^k_\gamma(t-t') = \frac{1}{\gamma} e^{- \frac{\omega_0^2}{\gamma}(t-t')}\theta(t-t').
\eeq
(The $k$ dependence is implicit in $\omega_0$.) We see that the response induced by the Green's function is approximately constant for sources concentrated on late times, while it becomes exponentially damped for very early sources. This allows us to define an `equilibration time' as the scale controlling the exponential suppression:
\beq
\tau^{-1}_{\rm eq} \sim \frac{\omega_0^2}{\gamma} .
\eeq

The solution for $\pi$ now reads
\beq
\label{piknoise}
\pi_k(t) =- N^{-1}_c \int_0^\infty dt' G^k_\gamma(t-t')\delta\calo_S({\bf{k}},t').
\eeq
(This is all what is left at late times $t\gg\tau_{eq}$ since the homogeneous solution dies away.)
Assuming a white noise spectrum,
\beq\label{eq:localnoise}
\langle \delta\calo_S({\bf{k}},t')\delta\calo_S({\bf{q}},t)\rangle \simeq (2\pi)^3 \nu_{\calo} \delta(t-t') \delta^{(3)}({\bf q}+{\bf k}),
\eeq
the two-point function turns into
\bea
\langle \pi_k(t)\pi_q(t) \rangle &=& N_c^{-2} \int_0^\infty \int_0^\infty dt'' dt' G^k_\gamma(t-t')G^q_\gamma(t-t")\langle \delta\calo_S({\bf{k}},t')\delta\calo_S({\bf{q}},t'')\rangle \\
&\simeq& (2\pi)^3\delta^{(3)}({\bf k}+{\bf q}) \nu_{\calo} N_c^{-2} \int_0^\infty dt' \left(G^k_\gamma(t-t')\right)^2.\nn
\eea
Performing the integral we obtain
\beq
\label{pikq}
\langle \pi_k(t) \pi_q(t) \rangle \simeq \frac{\nu_{\calo} (2\pi)^3}{N_c^2\omega_0^2\gamma} \left(1-  e^{- \frac{2\omega_0^2}{\gamma}t}\right)\delta^{(3)}({\bf k}+{\bf q}) \quad\to\quad \frac{\nu_{\calo}(2\pi)^3}{N_c^2\gamma\omega_0^2}\delta^{(3)}({\bf k}+{\bf q}),
\eeq
which tends to a constant as $t\to +\infty$ as we expected.

In thermal equilibrium, when the FD theorem applies, we can relate the amplitude of the noise $\nu_{\cal O}$ to the damping scale of the Green's function and the temperature $T$. If that was the case we would then have
\beq
\label{nubar}
\nu_{\calo} = N_c \gamma  T \quad\Rightarrow\quad \langle \pi^2 \rangle \sim \frac{ T}{N_c\omega_0^2},
\eeq
or equivalently
\beq
N_c \omega_0^2 \langle \pi^2 \rangle \sim  T.
\eeq
This expression is suggestive because it reminds us of the equipartition of energy in thermal equilibrium. If we interpret $N_c\omega_0^2$ as a spring constant $k_s$ and $\pi$ as an harmonic oscillator, then $k_s\langle\pi^2\rangle \sim  T$. Indeed the factor of $N_c$ is the canonical normalization for the field $\pi$, and in a sense it represents the `mass' of the harmonic field $\pi$ (which does not have a mass in the strict sense).\\

The above expressions allow us to understand the properties of the Green's functions in the expanding universe, which is the case of interest here. The equilibration time represents the time it takes for the interactions to cancel out the effect of an initial fluctuation. This effect is due to dissipation. Indeed, in the absence of dissipation the effect of the initial conditions never disappear. The expression in Eq. (\ref{pikq}), which was obtained in a Minkowski background, is also valid in the limit in which we can neglect the time scale of variation of $\omega_0$, given by $H^{-1}$, with respect to the time scale of the Green's function, i.e. $\tau_{eq}^{-1}$. When this condition is violated we cannot trust the solution any longer. However, we can still look at the EOM for $\pi$ and realize that since it contains only derivatives, if the noise is sufficiently concentrated at short distances, the correlation function becomes a constant.
Using $\omega_0=c_s k/a$ this happens when
\beq
\frac{\omega_0^2}{\gamma}\sim H\quad\Rightarrow\quad \omega_0\sim\sqrt{\gamma H}\quad\Rightarrow \quad k_\star \sim \sqrt{\frac{\gamma H}{c_s^2}}.
\eeq

Notice that at freezing the physical momentum is much larger than $H$ for $\gamma\gg H$. Then from Eq. (\ref{pikq}) and matching the solution deep inside the horizon and at horizon crossing we obtain (after re-inserting the factors of $a$, the scale factor)
\beq
\label{tpnuo}
\langle \pi_k \pi_q \rangle (t_\star) \sim \frac{ \nu_{\calo}(2\pi)^3}{c_s^2N^2_c \gamma (k/a_\star)^2}\frac{1}{a_\star^3}\delta^{(3)}({\bf k}+{\bf q}) \sim \frac{\sqrt{H_\star/\gamma}  \nu_{\calo}}{N^2_c (c^\star_sk)^3} (2\pi)^3\delta^{(3)}({\bf k}+{\bf q}),
\eeq
where we used $(\omega^\star_0)^2= (c_sk/a_\star)^2$ and $1/a_\star = \sqrt{\gamma H_\star}/(c^\star_s k)$. For instance if we use the relation in Eq. (\ref{nubar}) we get 
\beq
\label{tpwarm}
\langle \pi_k \pi_q \rangle (t_\star)  \simeq (2\pi)^3\frac{\sqrt{\gamma H_\star}  T}{N_c (c^\star_sk)^3} \delta^{(3)}({\bf k}+{\bf q}).
\eeq 

These are indeed the results we find in the full computation (see Eqs. (\ref{pow1gamma},~{\ref{tempz})). Notice that depending on the value of $\nu_\calo$, $\gamma$ and $N_c$ (and/or $T$), the two point function can be significantly enhanced with respect to the standard result.

\subsection{Non-linear effects}\label{nonlineff}

To finish our summary let us briefly study possible non-linear interactions along the same line of reasoning. We will analyze these effects
 in great detail in the forthcoming sections. However, it is instructive to study a few simple cases which turn out to be paradigmatic examples.

\subsubsection{Shift symmetry}\label{shsym}

Let us start by considering interactions that respect a shift symmetry for $\pi$. Then the first operator we may introduce is
of the form $\tilde\calo\dot\pi$ (so that $\calo=\dot{\tilde\calo}$). The structure of the Lagrangian induced by the non-linear realization of  time-diffeomorphisms implies that this term comes attached with:\footnote{We will show this term arises from a $-\calo g^{00}$ coupling, and we particularize to $(\partial_i\pi)^2$ since it dominates over other terms, such as $\dot\pi^2$, for $k_\star \sim \sqrt{\gamma H} /c_s\gg H$.}
\beq
\label{dipi}
-\frac{1}{2}\tilde\calo (\partial_i \pi)^2.
\eeq

For simplicity we remove the tildes from now on. Under the assumption that the linear piece induces dissipation (see Eq. (\ref{tildG})), it is straightforward to
show the EOM becomes\footnote{As we discuss momentarily, the non-linear coupling
proportional to $\gamma$ can also arise from an emergent shift symmetry in the non-linear response for $\delta\calo$.}
\beq
\label{secondt}
\ddot\pi_k + \gamma \left(\dot\pi_k - \frac{1}{2}[\partial_i \pi\partial_i \pi]_k\right) + \omega_0^2(k)\pi_k = -N^{-1}_c\left(\delta\dot{\calo}^S_k -[\partial_i(\tilde\calo\partial_i\pi)]_k\right),
\eeq where $[\,\,]_k$ stands for the convolution. We will not attempt a detailed account at this stage, but instead we provide some heuristic arguments to isolate the basic bits of the full computation (see sec. \ref{Nolocalo2} otherwise). There are at least two effects due to the non-linearities to take into account (ignoring the homogenous solution, which as we discussed decays away), namely
\bea
\pi_k(t) &=& N^{-1}_c \int_0^\infty dt' G^k_\gamma(t-t')\left\{-\delta\dot{\calo}^S_k (t') - \frac{\gamma k^2}{N_c} \left[\int_0^\infty dt'' G^k_\gamma(t'-t'')\delta\dot{\calo}^S_k(t'')\right]^2\right. \nn \\
&& \left. -\frac{k^2}{N_c} \int_0^\infty d\tilde t G^k_\gamma(t'-\tilde t)\delta\calo^S_k (\tilde t) \delta\calo^S_k(t') \right\}. \label{eompig}
\eea
The second term in the first line comes from the quadratic term $(\partial_i\pi)^2$ in Eq. (\ref{secondt}) after substituting
the forced solution for $\pi$, i.e. $\pi\sim -\int G\, \delta{\cal O}_S$. The piece in the second line comes instead from the last term on the right hand side (RHS) of  Eq. (\ref{secondt}). Let us compute the contribution from the first non-linear term. Assuming the noise is Gaussian, e.g.
\beq \langle \delta\dot{\calo}^S_1\delta\dot{\calo}^S_2\delta\dot{\calo}^S_3\delta\dot{\calo}^S_4\rangle \sim \langle \delta\dot{\calo}^S_1\delta\dot{\calo}^S_2\rangle\langle\delta\dot{\calo}^S_3\delta\dot{\calo}^S_4\rangle+\ldots,\eeq and using the local properties of the two-point functions together with Eq.~(\ref{eq:localnoise}), we obtain (for $k_1\sim k_2\sim k_3\sim k$)\footnote{To simplify the notation, here and elsewhere in this section we omit the momentum conserving delta functions.}
\beq
\langle \pi_k \pi_k \pi_k \rangle_{(\gamma)} = -\frac{\gamma\nu_{\calo}^2 k^2}{N_c^4}\int dt'dt'' dt'''\left(G^k_\gamma(t-t''')\right)^2 G^k_\gamma(t-t')  \left(G^k_\gamma(t'-t'')\right)^2 + \ldots.
\eeq
If we now multiply by $-H^3$ to transform to $\zeta$ ($\zeta \simeq -H\pi$) and divide by $\langle \zeta\zeta\rangle^2$, using
 \beq
 \label{gkscal} \int dt G^k_\gamma  \sim 1/\omega_0^2 \sim 1/(c_sk)^2,
 \eeq
this simplified analysis indicates a value for the non-Gaussianities of order
\beq
\label{fnl1}
|f_{\rm NL}| \sim \frac{\gamma}{c_s^2H}.
\eeq

Unfortunately we cannot use the local approximation for the last term in Eq. (\ref{secondt}) since we assumed it applies for its time derivative, however, let us try to estimate its value by comparing with the one we just computed. If we take the ratio between the two at the level of the EOM we get (schematically)
\beq
\label{ratiogk}
\frac{ \gamma k^2 \int dt \dot G^k_\gamma \delta\calo_k\pi_k}{k^2\pi_k\delta\calo_k} \sim 1,
\eeq
which supports the value of $f_{\rm NL}$ in Eq. (\ref{fnl1}) also for this operator.\footnote{For the estimate in (\ref{ratiogk}) we used the linear part of Eq. (\ref{eompig}), integrated by parts the time derivative, and used $\int dt \dot G^k_\gamma \sim 1/\gamma.$}\\

Notice that we can also estimate the size of the non-Gaussianities by taking the ratio
\beq
\left.\frac{\calo (\partial_i \pi)^2}{\calo\dot\pi}\right|_{k_\star\sim \sqrt{\gamma H/c_s^2},~\omega_\star \sim H} \sim \frac{k_\star^2 \zeta^2}{H^2\zeta}\sim \frac{\gamma}{c_s^2H} \zeta \to |f_{\rm NL}| \sim \frac{\gamma}{c_s^2H},
\eeq
which is consistent with the more detailed result of Eq. (\ref{fnlNL}).\\

From here we conclude that a large value for $\gamma$ is linked to large non-Gaussianities, provided the operator responsible for dissipation also induces terms such as in Eq. (\ref{dipi}), or more generally a $\gamma (\partial_i \pi)^2$ piece in the EOM as in Eq. (\ref{secondt}). As we shall see throughout the paper, this is indeed the case for a vast class of models.

\subsubsection{Approximate shift symmetry}\label{appsh}

Let us assume the shift symmetry $\pi \to \pi+c$ is softly broken by a parameter $\epsilon \ll 1$, as it happens due to the slow-roll approximation. Without this invariance we can have a coupling of the form \beq \label{dotfdecalo}-\dot f(t)\delta\calo \pi, \eeq responsible for a (local) dissipative term, plus a source noise of the form $-\dot f \delta\calo_S$. (As we shall see these terms arise from a $-f(t+\pi)\calo$ coupling in the effective action.) Notice at linear level this operator is of the same type we studied previously, except for the overall factor of $\dot f$, which we assume is (approximately) constant to preserve the shift symmetry. In this scenario, and assuming the noise is Gaussian, contributions to the three point function will be induced by $\delta\calo \ddot f(t) \pi^2$ at linear order in the response. Then the level of non-Gaussianity can be estimated to be (see sec. \ref{ftcalo} for more details)
\beq
\label{estfto}
\frac{\ddot f(t)\calo \pi^2}{\dot f(t)\calo\pi} \sim -\frac{\ddot f(t)}{\dot f(t) H} \zeta \to f_{\rm NL} \sim -\frac{\ddot f(t)}{\dot f (t)H} \sim O(\epsilon),
\eeq
that is in practice very small, provided $\epsilon \ll 1$.

\subsubsection{Non-linear response}\label{nlrep}

Let us continue with the coupling $f(t+\pi)\calo$ but include now the response beyond linear theory, in which case we do not necessarily have the constraint of Eq. (\ref{estfto}). Hence we have to include contributions to $\delta^{(2)}\calo_R$ at second order in $\pi$ which arise from the intrinsic three-point function of the $\delta\calo$'s, i.e. $\langle [\delta\calo(z),[\delta\calo(y),\delta\calo(x)]]\rangle$.
In general, making use of the local approximation, we have (schematically)
\beq
\dot f \delta^{(2)}\calo^R_k(\omega) \sim N_c g_\calo(k,\omega)\pi_k^2,
\eeq
with $g_\calo(k,\omega)$ depending on the specific dynamics of the model. From here we get non-Gaussianities of order
\beq
\label{gk2}
 \frac{g_\calo(k,\omega) \pi_k^2}{c_s^2k^2\pi_k}\sim f_{\rm NL}\zeta \to f_{\rm NL} \sim \frac{g_\calo({k_\star},\omega_\star)}{(c_s k_\star)^2 H},
\eeq
which is not suppressed by factors of $\ddot f/(H\dot f)$. Unfortunately, it is not possible in general to relate the level of non-Gaussianities and the dissipative
 coefficient $\gamma$, unless the different terms in $\delta^{(n)} \calo_R$ are somehow related. However, there are specific situations where this happens, in which case we expect a connection between dissipation and non-linear interactions. 
 
 For instance let us consider the case in which inflation is driven by a scalar field $\phi$ and the dynamics of the interaction is such that in the background
\beq
\label{calodotphi}
 \bar\calo = F(\dot\bphi).
\eeq
Intuitively Eq. (\ref{calodotphi}) follows from some basic requirement of a velocity dependence to induce dissipation. 

Now we perturb $\phi \to \bphi +\delta\phi$. Given that $\calo$ is a scalar, then (provided $\delta\phi$ is a {\it smooth} perturbation)
\beq
\label{calophi2}
 \langle\calo\rangle \simeq F\left[\sqrt{\left(-\partial\phi\right)^2}\right]\quad \to\quad \dot f\delta\calo_R \simeq N_c \gamma \left(\dot \pi + \frac{\alpha}{2}\dot\pi^2 -\frac{1}{2} (\partial_i\pi)^2+\ldots\right),
\eeq
where $\pi = \delta\phi/\dot\bphi$, and the factor of $\dot f$ appears in order to properly normalize the coupling to $\pi$. In this expression we also assumed the linear piece is responsible for dissipation. (The coefficient $\alpha$ is an order one number which vanishes for the special case $F(x)=|x|$.)\footnote{The argument applies to a generic function $F(\dot\bphi)$, in which case $\gamma \equiv \gamma(\dot\bphi)$. See sec. \ref{nonlrep} for more details.}
One might wonder whether there is any way to get $\dot\bphi$'s other than through $\sqrt{-\left(\partial\phi\right)^2}$. Certainly the background breaks time diffeomorphisms, and therefore we have a natural timelike vector $n_\mu \sim \partial_\mu t$ at our disposition. However, if the response of the $\calo$'s is predominately determined by the field $\phi$, we have $n_\mu \sim \partial_\mu\phi$, then 
\beq \partial_t \phi \to n^\mu \partial_\mu \phi = \sqrt{-(\partial\phi)^2}.\eeq
The extra terms in Eq. (\ref{calophi2}) thus appear from the fact that the equal time surfaces set by the inflaton also fluctuate. This case is now similar to the one we discussed in sec. \ref{shsym}, and we end up with a $\gamma (\partial_i \pi)^2$ term in the EOM. Hence we get $g_\calo \simeq \gamma k^2$, and plugging it back into Eq. (\ref{gk2}) we obtain
\beq
|f_{\rm NL}| \sim \frac{\gamma}{c_s^2 H},
\eeq
as in Eq. (\ref{fnl1}). Therefore, in this example large dissipation is also connected with an enhancement of non-linear effects.\\

There is a subtle point in the above argument. As we mentioned, in general large non-Gaussianities do not necessarily follow from a dissipative term. We obtain large effects in cases where the $\calo$ operators are sensitive only to fluctuations of the clock that controls the end of inflation, namely the field $\phi$ in the above example. We refer to this as having a {\it preferred clock}. We discuss this in more detail in sec.~\ref{nonlrep}.

Note also that these estimations apply under the assumption of locality (in which case we have a well defined derivative expansion). Non-local effects can potentially increase even more the level of non-Gaussianities, such as it happens in the model analyzed in \cite{trapped}. However in this regime the EFT treatment becomes more difficult. We do not explore this scenario in this paper.

\subsubsection{Non-Gaussian noise}

Going over the possible sources of non-linearties, we are finally led to consider the case in which correlation functions of the noise are themselves not Gaussian. If, for simplicity, we assume that the three-point function is local, i.e.
\beq
\label{ngnoise}
\langle\delta\calo_S(\tilde t)\delta\calo_S(t')\delta\calo_S(t'')\rangle \sim \nu_{\calo^3} \delta(\tilde t-t')\delta(t'-t''),
\eeq
we get from the $-\dot f(t) \delta\calo \pi$ interaction
\beq
\langle \pi_k\pi_k\pi_k \rangle_{(\gamma)} \sim -\frac{\dot f^3 \nu_{\calo^3}}{N_c^3} \int dt' (G^k_\gamma(t-t'))^3.
\eeq
Then using Eq. (\ref{gkscal}) we can estimate
\beq
\label{fnlnoise}
f_{\rm NL} \sim \frac{\gamma\nu_{\calo^3}N_c}{\dot f(t) \nu_{\calo}^2},
\eeq
which depends on various parameters, although clearly it can also be large. See sec. \ref{ngaussnoise} for more details.

\vskip 0.5cm

Adding the expansion of the universe changes things a little bit, in particular we will have to deal with exponential dilution. However, once we assume the dissipative effects are taking place at a faster pace than the Hubble expansion our results in flat space are a good guidance to understand the basic features of the full computation. As we shall see, most of our previous analysis remains essentially unchanged (provided the dissipative mechanism acts periodically over the inflationary epoch). 

The new ingredient is the construction of an EFT formalism from which we will obtain the type of terms we discussed, and more importantly the non trivial connections between the linear and non-linear effects. We start with the EFT setup next.

\section{Effective field theory setup}\label{eftsup}

As shown in \cite{eft1}, for single clock inflation the action in the unitary gauge is given by
\begin{eqnarray}
\label{act1}
S&=&\frac{M_p^2}{2}\int d^4 x\sqrt{-g} R+\frac{1}{2}\int d^4 x \sqrt{-g}(\overline{p}-\overline{\rho}-(\overline{p}+\overline{\rho}) g^{00})\nonumber\\
&+&\frac{1}{2} \int d^4 x \sqrt{-g}\,M_2^4(t) (1+ g^{00})^2-\frac{1}{2}\int d^4 x \sqrt{-g}\,\overline{M}_1^3(t)\delta K_{\mu}^{\mu}(1+ g^{00})\nonumber\\
&-&\frac{1}{2}\int d^4 x \sqrt{-g}\,\overline{M}_2^2(t) (\delta K_{\mu}^{\mu})^2-\frac{1}{2}\int d^4 x \sqrt{-g}\,\overline{M}_3^2(t) \delta K^{\mu}_{\nu}\delta K^{\nu}_{\mu}\nonumber\\
&+&\frac{1}{6}\int d^4 x \sqrt{-g}\,{M}_3^4(t) (1+ g^{00})^3-\frac{1}{2}\int d^4x \sqrt{-g} \overline{M}^2_4 \hat{g}^{\mu}_{\nu}\partial_{\mu} \delta g^{00} \hat{g}^{\nu\rho}\partial_{\rho} \delta g^{00}\nonumber\\
&-&\frac{1}{2}\int d^4 x \sqrt{-g}\,\overline{M}_4^2(t)(\delta K_{\mu}^{\mu})^2(1+ g^{00})\ldots\label{action}
\end{eqnarray} where the spatially flat FRW background metric is given by
\beq d\overline{s}^2=\overline{g}_{\mu\nu}dx^{\mu}dx^{\nu}=-dt^2+a^2(t)\delta_{ij}dx^idx^j,
\eeq
and the unit vector perpendicular to surfaces of constant time  ${t}$,
\beq \label{nmu} n_{\mu}=\frac{-\partial_{\mu}{t}}{\sqrt{-g^{\nu\rho}\partial_{\nu}{t}\partial_{\rho}{t}}},
\eeq
takes the form $ n_{\mu}=-\delta^{0}_{\mu} (-g^{00})^{-1/2}$. The  extrinsic curvature of the surfaces is
 \beq
K^{\mu}_{\nu}=\hat{g}^{\mu\rho}\nabla_{\rho}n_{\nu}\eeq
where  $\hat{g}_{\mu\rho}=g_{\mu\rho}+n_{\mu}n_{\rho}$ is the induced spatial metric. Thus $\delta K^{\mu}_{\nu}=K^{\mu}_{\nu}- H \hat{g}^{\mu}_{\nu}$ is the variation of the extrinsic curvature of constant time surfaces with respect to the unperturbed FRW. The ellipses in (\ref{act1}) account for any additional term that respect (time dependent) spatial diffeomorphisms. Defining
\beq T_{\mu\nu}=-\frac{2}{\sqrt{-g}}\frac{\delta S}{\delta{g^{\mu\nu}}},
\eeq
Einstein equations imply (a bar over any quantity denotes its unperturbed value)
\begin{subequations}
\begin{align}
\label{tadsuba}
\overline{\rho}&=3M_p^2 H^2,\\
\label{tadsubb}
\overline{p}&=-M_p^2 (2\dot{H}+3 H^2).
\end{align}
\end{subequations}where $H=\dot{a}/a$, $M_{p}^2=(8\pi G_N)^{-1}$.\\

To introduce the $\pi$ field  in the EFT we follow St\"uckelberg's trick,
\beq
t \to \tilde{t}=t-\pi,~ x^i\to \tilde{x}^i=x^i,\label{pionstx}\eeq
so that $g^{00}$ can be written as
\begin{equation}
\label{pionstx2}
g^{00}(x)=\tilde{g}^{00}(\tilde{x}) (1+\dot{\pi})^2+2 \partial_i\pi \tilde{g}^{0i}(\tilde{x})(1+\dot{\pi})+\tilde{g}^{ij}(\tilde{x}) \partial_i\pi\partial_j\pi,
\end{equation} and $\delta K^{i}_{j}$ is
\begin{eqnarray}
\delta K^{i}_{j}(x)&=&-\frac{\partial_i\partial_j\pi}{a^2}-\frac{1}{2}[\partial_j \delta\tilde{g}^{0i}+\partial_i \delta\tilde{g}^{0j}]-\frac{\partial_t(a^4\delta\tilde{g}^{ij})}{2 a^2}\nn \\ &+& a^2 H\delta\tilde{g}^{ij}-\left(\frac{H}{2}\delta\tilde{g}^{00}-\dot{H}\pi\right)\delta_{ij},
\end{eqnarray}
to linear order in the perturbations. From now on tildes will be omitted.  In addition, we can choose coordinates so that  the metric in the perturbations is given by
\be ds^2=-N^2dt^2+h_{ij} (dx^i+N^i dt) (dx^j+N^j dt).\ee
In this paper we will ignore tensor perturbations.\\

In general the metric perturbations $\delta N$ and $N_i$ are determined by the momentum and Hamiltonian constraints. The action for $\pi$ is obtained after introducing their solution back into the action. For single field inflation, it was shown in \cite{eft1} that in certain regimes the metric fluctuations can be ignored, and indeed these are suppressed either by slow roll parameters, or by ratios of $H^2/M_p^2$. The same occurs when we include ADOF. We discuss the details of this decoupling limit in appendix \ref{appdec}. 

The quadratic contribution to the action for $\pi$ can thus be written as
\begin{eqnarray}\label{cuad}
S_{\pi}=\frac{1}{2}\int d^4x a^3\left\{(\overline{p}+\overline{\rho}+4M_2^4)\dot{\pi}^2-(\overline{p}+\overline{\rho}+H\overline{M}_1^3)\frac{(\partial_i\pi)^2}{a^2}-(\overline{M}^2_2+\overline{M}^2_3)\frac{(\nabla^2\pi)^2}{a^4}\right\}.
\end{eqnarray}
(To arrive at this result we have performed integrations by parts.\footnote{The differences with \cite{eft1} are due to the fact that here we are using  a term $\delta g^{00}\delta K^{\mu}_{\mu}$ instead of $\delta N\delta E^{\mu}_{\mu}$, where $K^{\mu}_{\mu}=\sqrt{-g^{00}}E^{\mu}_{\nu}$. The relations between the coefficients are $\overline{M}_1^3=d_1 M^3/2$, $M^4_2=M^4-3/4 d_1 H M^3$, $\overline{M}_2+\overline{M}_3=M^2(d_2+d_3)$.}) The above expression can be re-arranged as follows (ignoring $\overline{M}_2$ and $\overline{M}_3$)
\begin{eqnarray}\label{l2pi}
S_{\pi}=\int d^4x a^3\frac{N_c}{2}\left\{\dot{\pi}^2-c_s^2\frac{(\partial_i\pi)^2}{a^2}\right\},
\end{eqnarray}where \beq \label{ncs} c_s^2=\frac{(\overline{p}+\overline{\rho}+H\overline{M}_1^3)}{(\overline{p}+\overline{\rho}+4M_2^4)},\,\, N_c=(\overline{p}+\overline{\rho}+H\overline{M}_1^3)/c_s^2.\eeq

For more details on the EFT formalism see \cite{Creminelli:2006xe, eft1, multieft}. As we show next, the introduction of ADOF changes the relations in (\ref{ncs}), since in general they can have a non-vanishing background stress energy tensor.\footnote{As we will explain later, this in not in contradiction with the results of \cite{eft1}, where it was shown that the tadopole coefficients are uniquely fixed by $H$ and $\dot H$. This is due to a different choice for the field $\pi$.}

\section{Adding new degrees of freedom}\label{newdof}

In order to include dissipative effects in our system we will follow the procedure of sec. \ref{storyo} and introduce a set of (composite) operators that behave as an effective environment. Since we are dealing with gravity, we should take into account that the stress tensor corresponding to these new degrees of freedom, $T^{\mu\nu}_\calo$, may contribute significantly to the background. This is the case, for example, in the trapped inflation model of \cite{trapped} where particles are created while the inflaton slow rolls (see sec. \ref{matching}).

The fact that there is more than one field whose stress energy density takes an expectation value slightly complicates the construction of the  EFT. The basic idea of the EFT of inflation is rooted in the necessity of having an end point for the accelerated expansion, and that there is a physical clock that defines a special time-slicing where the clock is taken to be uniform. This is the so-called unitary gauge. Time translations are spontaneously broken during inflation by the presence of this preferred clock, which means that there is a Goldstone boson that non-linearly realizes the symmetry. In the case where we add the ADOF, there is an ambiguity in the definition of the clock field, as the additional fields may have non-vanishing background expectation value that also break time-translation invariance. Nevertheless, there are two natural definitions of the field that interpolates for the Goldstone boson, both equally good. The first one follows the approach of \cite{multieft}, in which  one introduces the Goldstone boson of time-translations, $\tilde\pi$, such that the action takes the form \cite{eft1,multieft}
\beq
\label{orunit}
\int \sqrt{-\tilde g} \left(-M_p^2(3H^2(t+\tilde\pi)+\dot H(t+\tilde\pi)) +M_p^2\dot H(t+\tilde\pi) \tilde g^{00}(\tilde\pi)\right)+\ldots \eeq 
(The ellipses include other (non-Goldstone) combinations that depend on the ADOF.) In this approach the only tadoples (namely terms that are linear in $\delta g^{\mu\nu}$) are the ones associated with the operators $\sqrt{-g}g^{00}$ and $\sqrt{-g}$, whose coefficients are uniquely fixed by the geometry as shown in \cite{eft1,multieft}.

A second alternative, which is the one we take in this paper, is to define a unitary gauge in which the physical clock that controls the end of inflation is taken to be uniform. Then by performing a time diffeomorphism we introduce a different St\"uckelberg field, that we will denote as $\pi$. The main difference between the two gauges relies on the fact that now the coefficients for the tadpole operators, $\sqrt{-g} g^{00}$ and $\sqrt{-g}$, are not determined by the geometry and will in general depend on contributions from the ADOF in the background (see Eqs. (\ref{tadpolec})--(\ref{c2}) below). This is the case because we also need to include tadpole operators induced by the ADOF. The two different $\pi$'s are related by a mixing that involves the ADOF fluctuations, schematically: $\tilde\pi\sim \pi+\delta\calo$. The field $\tilde \pi$ has a simpler Lagrangian, because the coefficients of the two tadpole terms are fixed, as shown in~\cite{eft1}. However, it is not convenient for us because $\tilde\pi$ is not sufficient to determine the end of inflation. If we were working with $\tilde \pi$, then the curvature perturbation would be related to the latter by a relationship of the form $\zeta\sim H\tilde\pi+\delta \calo$. Instead, by taking the second choice, we have a slightly more complicated Lagrangian, yet the link between $\zeta$ and $\pi$ is simply $\zeta\simeq -H\pi$ (at linear order), with no dependence on the $\calo$ operators. This is the case because, as we emphasized in the introduction, we work under the assumption that the ADOF do not contribute to density fluctuations at late times. 

This is the main difference between the cases we are studying here and the analysis of~\cite{multieft}, where additional light fields were included to the EFT of inflation of \cite{eft1}. It is easy to convince oneself that the two gauge choices are equivalent. We take the second.\\

Let us continue with the construction of the effective Lagrangian. Following \cite{eft1} our starting point is an effective action in a unitary gauge in which we write 
\bea
\label{tadpolec}
\int \sqrt{-g} (\Lambda(t) - c(t) g^{00}) + S_{\calo},
\eea	
where $c(t),\Lambda(t)$ are certain tadpole coefficients soon to be fixed by enforcing Einstein equations. Note we added $S_{\calo} = \int d^4x\sqrt{-g}\calL_\calo$ to account for the dynamics of $\calo$ independent of $\pi$. (We will incorporate the couplings between $\pi$ and $\calo$ in the next section.) Also, by construction, $S_\calo$ is a scalar under diffeomorphisms. Then with
\beq
\label{tmno}
\overline{T}^{\calo}_{\mu\nu}=\mathrm{diag}(\overline{\rho}_\calo,a^2 \overline{p}_\calo,a^2 \overline{p}_\calo,a^2 \overline{p}_\calo),
\eeq
it is straightforward to show
\bea
\label{lambt}
\Lambda(t) + \frac{1}{2} (\bar p_\calo-\bar \rho_{\calo}) &=& - M_p^2(3H^2+\dot H) \\
c(t) + \frac{1}{2} (\bar p_\calo+\bar \rho_{\calo}) &=& -M_p^2\dot H \label{c2}.
\eea

As explained in \cite{eft1} we introduce $\pi$ following the St\"uckelberg trick (see Eqs. (\ref{pionstx}) and (\ref{pionstx2})), which for the action in Eq. (\ref{tadpolec}) means that only $c(t)$ contributes to the normalization of the quadratic Lagrangian in $\pi$, so that\footnote{Terms linear in $\pi$ cancel out once the background EOM are satisfied.}
\beq\label{normnc}
c_s^2N_c \equiv 2c(t) = -2M_p^2\dot H -(\bar \rho_\calo +\bar p_\calo).
\eeq
(Note this coefficient would be given by $-2M_p^2\dot H$ in the absence of ADOF, in which case both definitions of $\pi$ would agree.)\\

As an example, let us consider once again inflation described by a scalar field with the following action 
\beq
S_{\rm tot} = \int \sqrt{g} \left(-\frac{1}{2}(\partial\phi)^2-V(\phi) + \calL_\calo\right).
\eeq
Since we assume the ADOF do not contribute significantly to density fluctuations at late time, our unitary gauge is the one where $\delta\phi=0$. Then we obtain
\beq
S_{\rm tot} = \int \sqrt{g} \left(-\frac{1}{2}\dot\bphi^2g^{00}-V(\bphi) + \calL_\calo\right),
\eeq
where $\bphi(t)$ is the background value. On the other hand, Friedmann equations (including the ADOF) tell us
$-\dot\bphi^2/2 = -N_c/2$ and $V(\bphi) = \Lambda(t)$, with $(\Lambda, N_c)$ defined in Eqs. (\ref{lambt}) and (\ref{normnc}) (for $c_s=1$).
Hence the action takes the form of the expression in Eq. (\ref{tadpolec}) with the aforementioned coefficients. Moreover, we also get the usual  Lagrangian for $\pi$ normalized by $N_c$, after identifying $\pi = \delta\phi/\dot\bphi$. We notice in passing that assuming the stress energy tensor that follows from $\calL_\calo$ obeys the null energy condition, i.e. $\bar\rho_\calo+\bar p_\calo \geq 0$, then \beq \label{foot18} c_s^2N_c \leq  -2M_p^2\dot H.\eeq

Adding higher dimensional operators will shift the normalization of $\pi$, like in Eq. (\ref{ncs}). In particular we will generate a non-zero correction to the speed of sound, so that $c_s \leq 1$. Therefore, before including interactions with the ADOF, at quadratic order the action is given by Eq.~(\ref{l2pi}), with $(N_c, c_s)$ some matching coefficients, defined as in Eq. (\ref{ncs}).\\

We will not adopt any particular model for the ADOF, rather we will attempt to produce correlations between different observables, such as the power spectrum and non-Gaussianities, under some mild assumptions about the $n$-point functions of the type $\langle \calo\ldots \calo\rangle$. But first let us start by constraining the type of operators that we may add to the effective action in the unitary gauge.

\section{The interaction terms in unitary gauge}\label{InterADOF}

We move now to the description of the type of operators that we can add to our Lagrangian in the unitary gauge that will induce couplings between the ADOF and the fluctuations of the clock. In general, the operators will have some tensorial transformation properties under space-time diffeomorphisms, and so they will be classified according to their rank.  As it was shown in the analysis of \cite{eft1, multieft}, one can write down operators containing only free upper 0 indices. In our case, however, there is a subtlety we need to address since the $\calo$'s are composite operators that may also contain the metric. Since the metric can be used to contract tensors made out of several different fields, we define tensor operators $\calo_{\alpha\beta\ldots}$ always with indices down, and so that $\delta {\calo}_{\alpha\beta\ldots}/\delta g^{\mu\nu}=0$.

Let us give an example. Let us consider two operators, ${\calo}_1=\psi^2$ and ${\calo}_2=g^{\mu\nu}\partial_\mu\psi\partial_\nu\psi$, with $\psi$ a scalar field. These are both scalar operators, however, according to our prescription we should write:
${\calo}_2=g^{\mu\nu}\tilde{\calo}_2{}_{\mu\nu}$, with $\tilde{\calo}_2{}_{\mu\nu}=\partial_\mu\psi\partial_\nu\psi$. In this way the ambiguity with respect to metric factors is removed. Operators are then classified as a Taylor expansion in fluctuations and derivatives as usual. We now proceed to illustrate the leading ones.

\subsection{Scalars}\label{scaladof}

In an expansion in metric fluctuations and derivatives, the most relevant operator is given by
\begin{equation}\label{inter1n}
S^{\calo}_1=-\int d^4 x \sqrt{-g}\,   f_1(t)\calo_1,
\end{equation}
where $\calo_1$ is a scalar under full space-time diffeomorphisms. The next type of operators can be organized as follows
\begin{align}
S^{\calo}_2&=-\int d^4 x \sqrt{-g}\, \left\{f_2(t)\delta g^{00} \calo_2+ f_3(t)(\delta g^{00})^2 \calo_3+ f_4(t)(\delta g^{00})^3 \calo_4+\ldots\right\}, \label{inter2}
\end{align}
where $\calo_a$, $a=1,2\ldots$, are also scalars and the ellipses include pieces involving higher power of the fluctuations as well as higher derivative terms  such as $\partial^0 \delta g^{00}$ or $\delta K$. In appendix~\ref{dodk} we discuss briefly operators of the form \beq
\label{extcalotr}
\int d^4x \sqrt{-g}~s(t) \frac{1}{M_K} \hat\calo \delta K^\nu_\nu .
\eeq

As we discussed, the operators $\calo_{1,2}$ may have (time dependent) background values, e.g. $\calo_{1,2} = \bar\calo_{1,2}(t) +\delta O_{1,2}$, which could also contribute to the background Einstein equations. These lead to corrections to $\bar T_{\mu\nu}$ from the interaction between the ADOF and the one responsible for inflation. For example, let us take once again the example of a slowly rolling scalar inflaton and add the coupling
\beq
\label{chiop}
\frac{1}{2} \int d^4x \sqrt{-g}~\frac{\chi^2}{\Lambda^2_\chi} g^{\alpha\beta} \partial_\alpha \phi \partial_\beta \phi \to -\frac{1}{2}\int d^4x \sqrt{-g}\left(\frac{\chi^2}{\Lambda^2_\chi}\dot\bphi^2\right)  g^{00},
\eeq
to a scalar field $\chi$ (say we have a shift symmetry $\phi\to \phi+c$ to prevent other couplings).
Then the new contribution to ${\bar T}_{\mu\nu}$ is given by
\beq
{\bar T}^{\chi\phi}_{\mu\nu} =\left( (\bar \rho_{\chi\phi}+\bar p_{\chi\phi})\delta_\mu^0\delta_\nu^0 + \bar p_{\chi\phi} \bar g_{\mu\nu}\right)
\eeq
with $\bar \rho_{\chi\phi} =\bar p_{\chi\phi} = \frac{1}{2}\dot\bphi^2 \bar\chi^2/\Lambda^2_\chi$. Einstein equations require
 \beq
 -c(t) = \left[M_p^2\dot H + \frac{1}{2}(\bar \rho_\calo+\bar p_\calo) + \frac{1}{2}(\bar \rho_{\chi\phi}+\bar p_{\chi\phi})\right],
 \eeq
and similarly for $\Lambda(t)$. Notice that the term in Eq. (\ref{chiop}) now contributes to the quadratic action for $\pi$, and we get
\beq \left[-c(t)-\frac{1}{2}(\bar\rho_{\chi\phi}+\bar p_{\chi\phi})\right] g^{00}\eeq
thus the canonical normalization coefficient becomes
 \beq
 \label{nchi}
N_c =-2M_p^2\dot H - (\bar\rho_{\calo} + \bar p_{\calo}).
\eeq
This is a general feature: the normalization of the $\pi$ Lagrangian will be given by the difference between the total `kinetic term', $(\bar \rho+\bar p)_{\rm tot} = -2M_p\dot H$ and (only) the contribution from $\cal L_{\calo}$ (the Lagrangian for $\calo$ independent of $\pi$). In this example $N_c$ is not equal to $\dot\bphi^2$  but rather
\beq \label{Ncchi} N_c = \dot\bphi^2 + 2f_2(t)\bar\calo_\chi,\eeq
where  $f_2(t) =\dot\bphi^2/2$ and $\bar\calo_\chi = \bar \chi^2/\Lambda_\chi^2$. But this is what we expect upon noticing that adding the term in (\ref{chiop}) to the usual Lagrangian, $-\frac{1}{2} (\partial\phi)^2 -V(\phi)$, renormalizes the kinetic part of the action by a factor
\beq
\label{nchi2}
-\frac{1}{2} (\partial \phi)^2 \to -\frac{1}{2} \left[1 +\frac{\bar\chi^2}{\Lambda^2_\chi}\right] (\partial \phi)^2,
\eeq
and therefore we obtain (using $\pi \sim \delta\phi/\dot\bphi$)
\beq
\frac{1}{2}\left(\dot\bphi^2 + 2 f_2(t)\bar\calo_\chi\right)\left(\frac{\delta\dot\phi^2}{\dot\bphi^2} + \ldots\right) \equiv c_s^2 N_c \left(\frac{\dot \pi^2}{2} + \ldots\right),
\eeq
as in Eq. (\ref{nchi2}). Note that from an effective field theory point of view, the consistency of this particular example with a slow rolling scalar field requires to have,  even in a situation where $\bar\chi^2 \gtrsim |\dot\bphi|$,  $\Lambda_\chi^2 \gg\bar\chi^2$, and therefore $c_s^2N_c$ remains essentially given by $\dot\bphi^2$. This is so as otherwise we should consider an infinite amount of terms. However this situation is not necessarily the case for all possible UV models, since we could instead of the term in (\ref{chiop}) write a generic expansion (see for example \cite{gel})
\beq 
(\chi^2/\Lambda_\chi^2) (\partial\phi)^2 \to M^4 F({\hat\chi}^2)P(X),
\eeq
with $\hat\chi = \chi/\Lambda_\chi$ and $X=-(\partial\phi)^2/M^4$ with $M$ some mass scale, such that $f_2(t)\bar\calo_\chi > \dot\bphi^2$. One of the most useful aspects of the EFT of inflation is that we do not need to worry about a specific realization of the background while studying its perturbations. As a result the scale  $N_c$ may be dominated by $f_2(t)\bar\calo_\chi$, rather than $\dot\bphi^2$.\\

Let us now return to the operator in Eq. (\ref{inter1n}). In principle it can also have a background value, i.e. $\sqrt{-g} f_1(t)\bar\calo_1(t)$. Since it only contributes a piece proportional to $\sqrt{-g}$ it can be absorbed into $\Lambda(t)$ in Eq. (\ref{tadpolec}), to ensure the background satisfies Einstein equations. Notice that the full $\calo_1$ is a scalar so that the coupling $f_1(t)\calo_1$ only develops $\pi$'s from $f_1(t) \to f_1(t+\pi)$. However, if we absorb $f_1(t)\bar\calo_1$ into $\Lambda(t)$ then its value gets fixed as in Eq. (\ref{lambt}), which we now have to expand in $t+\pi$. Hence somehow the pieces from $\bar\calo_1(t+\pi)$ must cancel out, and they do once we realize $\delta \calo_1(t)$ is not invariant under time reparameterizations and their background values are re-introduced from
\beq
\label{unitshift}
\delta \calo_1 \to \delta \calo_1 - \dot{\bar\calo}_1(t) \pi + \ldots.
\eeq
In other words, the fact that the $\calo$ operators have background expectation values means that if we split them into background plus fluctuations, the latter shift under a time diffeomorphism.\footnote{If instead we had chosen to work with $\tilde\pi$, such that by construction all the information about $\bar\calo$ is already incorporated in Eq. (\ref{orunit}), we would still have these background values appearing in the Lagrangian from the shift of $\delta\calo$ after re-inserting $\tilde\pi$, similarly to Eq. (\ref{unitshift}).} Of course if we do not split the operator in this manner, then since $\calo$ is a scalar, no $\pi$ field will be associated with it once we perform a time diffeomorphism.\\

Let us consider for instance a coupling $\phi^2\chi^2$ between the inflaton and a second scalar field (this will reappear later on), and allow for a non-zero expectation value $\bar\chi^2(t) \neq 0$. In our unitary gauge we have a term in the action $\bphi(t)^2 \langle\chi^2\rangle (t)$ (plus perturbations in the $\chi$'s), which we can think of as being included in $V(\bphi)$ (with time dependent coefficients). This corresponds to a $f(t)\calo$ type of coupling. As we mentioned above, we do not want to stream $\pi$ off the time dependence in $\bar\chi^2(t)$ (because the full operator is a scalar), but this will happen once we solve for $V(\bphi)$, i.e. $\Lambda(t)$ as in Eq. (\ref{lambt}). However, it is easy to see these extra terms cancel out against the ones induced from Eq. (\ref{unitshift}).\\

Let us finally briefly comment on the slow roll approximation, since in principle the coupling $\left(f_1(t)\bar\calo_1(t)\right)$ may break it. In general the slow roll condition can be satisfied provided
\beq \label{condV} \frac{d^2\left(f_1(t)\bar\calo_1(t)\right)}{dt^2} \lesssim\frac{\dot H}{H^2}.\eeq
This requires, in addition to $\epsilon=-\dot H/H^2\ll 1$ and $\eta \equiv \frac{\dot \epsilon}{\epsilon H}\ll 1$, that any explicit function of time $f(t)$ in the action, plus all background quantities associated to the ADOF, change very little in a Hubble time. Schematically we write:  $\epsilon_f \equiv \frac{\ddot f}{\dot f H}\ll 1$ and $\epsilon_{\calo} \equiv \frac{\dot{\bar\calo}}{H\bar\calo}\ll 1$. In practice we assume all the terms proportional to $\bar\calo_{1,2}(t)$, or in general stemming from $\calL_\calo$, are included in the background geometry $(H,\dot H)$ or into the coefficients $(c_s, N_c)$, and furthermore with their time dependence suppressed by slow roll parameters unless otherwise noted.

\subsection{Vectors}\label{vectors}

Moving into vector couplings, the one with the least number of metric fluctuations has the form
\beq
\label{linpim0}
\int d^4x \sqrt{-g}~\tilde f_1(t)\calo_\mu \delta g^{\mu0},
\eeq
where we have been careful in defining the vector with the index lowered as we stressed at the beginning of the section.
As we will see when we reinsert the $\pi$ field, something unusual about this operator is that it only entails terms linear in $\pi$, provided $\tilde f_1(t)$ is a constant.  At higher order the generalization is straightforward:
\beq
\int d^4x \sqrt{-g} \left(\tilde f_2(t)\calo_\alpha \delta g^{00} \delta g^{\alpha0} +\ldots \right).
\eeq

\subsection{Tensors}\label{tensors}

We can move on by considering generic tensors, with their indices contracted with  $g^{\mu0}$'s, as for instance
\beq
\int d^4x \sqrt{-g} ~\hat f(t)\calo_{\mu\ldots\nu} \delta g^{\mu0}\ldots \delta g^{\nu0}=\int d^4x \sqrt{-g}~\hat f(t)\calo^{0 \ldots 0}.
\eeq
Another type of terms, perhaps more interesting, are those coupled to the extrinsic curvature. For example,
\beq
\label{extcalo}
\int d^4x \sqrt{-g} ~s(t)\frac{1}{M_K} \tilde\calo_{\mu\nu} \delta K^{\nu\mu},
\eeq
where the factor of $M_K$, which we take to be much bigger than Hubble, accounts for the mass dimensions of $K^\mu_\nu$.
We can clearly continue adding factors of  $\delta g^{0\mu}$ and $\delta K^\mu_\nu$.

\section{The interaction terms for {\Large{$\pi$}}}\label{secpions}

As we already pointed out, in this paper we ignore the mixing with gravity and work in the decoupling limit (see appendix \ref{appdec}). Therefore, following our previous sketching of the procedure, to construct the interacting part of the action between the ADOF and the $\pi$'s we simply replace (see sec. \ref{eftsup})
\bea
g^{00} \to && -1 -2\dot\pi -\dot\pi^2 +\frac{1}{a^2}(\partial_i\pi)^2 \label{g00pi}\\
g^{0\mu} \to && -\delta^\mu_0 (1+\dot\pi) + \delta^\mu_i \frac{1}{a^2}\partial_i\pi\label{g0mpi}.
\eea
Also terms from the extrinsic curvature, that at linear order induces
\beq
\label{kij}
\delta K_{ij} \to a^2H\delta_{ij} \dot\pi - \partial_i\partial_j \pi + \ldots.
\eeq
\noindent As we anticipated, at quadratic order the Lagrangian for the $\pi$ field takes the form in Eq. (\ref{l2pi}). 

Next we include the interaction terms between the $\calo$'s and $\pi$ in the effective action. Let us start at quadratic level in the fluctuations. There are many operators that contribution at linear order in $\pi$. However, from Eqs. (\ref{g00pi},~\ref{g0mpi}, \ref{kij}) we note that all the terms at leading order in derivatives can be re-grouped basically as
\beq
\label{operlin}
\dot f_1(t) \calo\pi,~ f_2(t)\tilde\calo\dot\pi
,~\tilde f_1(t)\calo^i \partial_i \pi,~\ldots,
 \eeq
where the dots include higher derivative terms. Since the non-trivial features of the non-linear realization of time-diffeomorphisms comes from the connection between terms with different powers of $\pi$, at linear level we obtain basically all the terms allowed by rotational invariance.
The first term appears after expanding $f_1(t+\pi)\calo$ in powers of $\pi$ to first order, whereas the second term comes from $f_2(t) \tilde\calo \delta g^{00}$. Note that there is a contribution from Eq. (\ref{extcalotr}) to the second term, however, it is suppressed by a factor of $H/M_K \ll 1$. Terms like $\partial_i\pi\calo^i$, which follow from $\calo_\mu \delta g^{\mu0}$, may also generate contributions to $c_s$ as well as  $k$-dependent friction.  

For the purpose of understanding the generation of friction, we can concentrate on the linear order. We can therefore simply use integratation by parts and study an effective operator of the form \beq \label{gencalo} -\int d^4x \sqrt{-g}~\dot f(t)\calo(x)\pi(x),\eeq where $\dot f(t)$ provides an overall normalization scale which we assume remains constant protected by an approximate shift symmetry, but see secs. \ref{appsh} and \ref{ftcalo}. In most of the expressions below we assume $\dot f$ is absorbed into $\calo$ unless otherwise noted. Here $\calo$ accounts for a series of contributions, including $\partial_i{\calo}^i$ etc, so that we expect its Green's function to be quite generic.

We wish to understand now under which circumstances we recover an equation equivalent to (\ref{eom1}). The main difference, as we just mentioned, lies in the expansion of the universe. A crucial simplification will come from our assumption of a faster than a $H^{-1}$ time scale for dissipation, and therefore our analysis from sec.~\ref{warm} remains essentially unaltered.\footnote{This means that in practice we will work in the regime where terms like $H\delta\tilde\calo$, that appear after we integrate by parts and hit the $a$'s in the volume factors, are such $H \delta\tilde\calo \ll \delta\dot{\tilde\calo}$.}

\subsection{Modified dynamics \& local approximations}\label{localapp}

In sec. \ref{storyo} we started our discussion of the effect of terms such as $\calo\pi$ in the dynamics of $\pi$ in a flat background. The main difference now is the explicit time dependence introduced by the scale factor. For that reason, instead of working in frequency space, we find convenient to work in mixed Fourier space $(t,{\bf k})$, keeping time as usual.
Again we split the operator into pieces,
\beq
\label{calotxsr}
\calo (t,{\bf x})= \bar\calo(t)+ \delta\calo_S(t,{\bf x}) + \delta\calo_R (t,{\bf x}),
\eeq
with $ \bar\calo(t)$ the background expectation value, $ \delta\calo_S(t,{\bf x})$ the stochastic fluctuations, and $ \delta\calo_R (t,{\bf x})$ the those induced by $\pi$.
In what follows we omit the background piece $\bar\calo(t)$, which as explained in sec. \ref{scaladof}, we assume is absorbed in $H,\dot H, N_c$, and its time-dependence is  suppressed in the slow roll approximation $(\epsilon, \eta, \epsilon_f, \epsilon_{\calo}) \ll 1$. \\

Recall the first variation in Eq. (\ref{calotxsr}) represents the noise, whereas the second one is the response to the perturbation induced by the $\pi$ field, and is given as the integral of a Green's function as in Eqs.~(\ref{dret0},~\ref{dret}).  Varying the action we obtain the EOM
\beq
\label{gnoa0}
\ddot\pi_k(t) + 3H \dot\pi_k(t)+ \frac{c_s^2{\bf k}^2}{a^2} \pi_k -
\frac{1}{N_c}\int dt'a^3(t') G_{\rm ret}^\calo(t,t',{\bf k}) \pi_{k}(t') =- \frac{1}{N_c} \delta\calo_S(t,{\bf k}),
\eeq
with
\beq
\label{gnoa1}
G_{\rm ret}^\calo(t,t',{\bf k})  = i \int \frac{d^3 {\bf y}}{(2\pi)^3} e^{-i{\bf k}\cdot {\bf y}} [\delta\calo(t,{\bf y}),\delta\calo(t',{\bf 0})]\theta(t-t').
\eeq
(The overall normalization is given by $N_c$ as in Eq. (\ref{l2pi}).)\\

Our first approximation entails locality in space, and so we take the Green's function to be of the form
\beq
\label{gnoa2}
G_{\rm ret}^\calo(t,t',{\bf k}) = \frac{G_{\rm ret}^\calo (t,t')}{a^{3/2}(t)a^{3/2}(t')} + O(|{\bf k}|/M_\calo),
\eeq
with $M_{\calo}\gg k_\star$. (The factors of $a^{-3/2}$ account for the fact that we work in co-moving coordinates.) In other words, there is a `gap' in momentum space determined by the `mean free path' $l_\calo \sim 1/M_\calo \ll 1/k_\star$. We perform the same approximation for the correlation functions of the noise. For example for the two-point function we have
\beq
\label{twop1}
\langle \delta\calo_S(t,{\bf k})\delta\calo_S(t',{\bf q})\rangle \simeq \frac{\tilde{\nu}_{\calo}(t,t')}{a^{3/2}(t)a^{3/2}(t')} (2\pi)^3\delta^{(3)}({\bf q+\bf k}).
\eeq

To obtain a local approximation in time we assume that the characteristic time scale for the variation of the kernels, $\Gamma_\calo^{-1} \ll 1/H$, is much
 smaller than the one of the sources, i.e. the $\pi$ field. (Notice that, at least in principle, $\Gamma_{\calo}$ is not necessarily related to $l_{\calo}$.) Then by changing the integration variable $t'=t-\tau$ in the above EOM we can approximate
\beq
\label{pikap}
 \pi_k(t-\tau)\simeq \pi_k(t)-\dot\pi_k(t) \tau + \ldots,
\eeq

The first term would introduce a mass for $\pi$, as can be seen after using this approximation in Eq. (\ref{gnoa0}). However, as we mentioned in sec. \ref{warm}, in this paper we concentrate in models where $\zeta_k$  is not affected by the ADOF after horizon exit.  This requires that the equations for the $\zeta$ modes do not have a mass term.\footnote{This is guaranteed if we have a (softly broken) shift symmetry.} This imposes the condition that for a constant $\zeta$ the response from the $\calo$'s should vanish. More specifically we impose $\delta\calo^R_k \to 0$ as $k/a \to 0$, such that we do not generate a mass term for $\pi$. We enforce this at the level of the Green's function, imposing an emergent shift symmetry such that the effect of the first term vanishes, i.e.
\be
\int \frac{a^{3/2}(t-\tau)}{a^{3/2}(t)}G_{\rm ret}^\calo(t,t-\tau)d\tau = 0 .
\ee

On the other hand, the second term of Eq. (\ref{pikap}) produces our desired result, where the friction part is given by (see Eqs. (\ref{gnoa0}) and (\ref{gnoa2}))
\beq
\label{gamkap}
N_c \gamma \simeq- \int \frac{a^{3/2}(t-\tau)}{a^{3/2}(t)}\cdot\tau\cdot G_{\rm ret}^\calo(t,t-\tau)d\tau.
\eeq
In the flat space limit this corresponds to the condition
\beq
\label{gretto}
G_{\rm ret}^\calo(t,t') \simeq -\gamma N_c \partial_t\delta(t-t')+\ldots,
\eeq
or Eq. (\ref{expG}) in Fourier space, after re-introducing the factors of $N_c$.\\

 The noise part  will affect $\pi$ only through integrals of the Green's function whose variation time scale is assumed to be much longer than the characteristic scale corresponding to the noise. This means that $\pi$ will be sensitive only to the integral in cosmic time of $\nu_{\calo}(t,t')$, and therefore we can approximate
\beq
\label{twop2}
\tilde \nu_{\calo}(t,t') \simeq \nu_{\calo} \delta(t-t').
\eeq

If for example we would assume $\calo$ is in thermal equilibrium, at high temperature $T$ we could use the FD theorem which relates \beq \label{fdthg}\gamma \simeq \frac{\nu_{\calo}}{N_c T}.\eeq\\

Notice that the expansion of the universe, i.e. the factors of $e^{-\frac{3H}{2}\tau}$, helps to improve the locality of the expression in Eq. (\ref{gamkap}). That is to say, there is no significant influence between different Hubble times. In this paper we thus take $\gamma, \nu_{\calo}$ to be essentially constant up to slow roll effects, i.e. $\left(\frac{\dot\gamma}{\gamma H},\frac{\dot\nu_{\calo}}{\nu_{\calo} H}\right) \sim O(\epsilon)$. \\

At the end of the day the EOM becomes
\beq
\label{fineom}
\ddot\pi_k(t) + (3H+\gamma) \dot\pi_k(t)+ \frac{c_s^2{\bf k}^2}{a^2} \pi_k = -\frac{1}{N_c} \delta\calo_S(t,{\bf k}),
\eeq
plus the behavior of the noise dictated by Eqs. (\ref{twop1}, \ref{twop2}).

\subsection{The homogenous solution}\label{homog}

Here we show that the homogenous part of Eq. (\ref{fineom}) becomes negligible at horizon crossing for $\gamma \gg H$, which is the domain we are interested in this paper. To solve the equation we first make the change of variables $\pi = z^{\lambda/2} \varphi$, with $\lambda = 2 + \gamma/H \gg 1$. The equation for the perturbation reads
\beq
\label{phik}
\left(\frac{d^2}{dz^2}+ 1 - \frac{\lambda}{2z^2}(1+\lambda/2)\right)\varphi_k=0,
\eeq
with $z=-kc_s\eta$, and $\eta$ the conformal time. Naively, as in the case with $\lambda=2$, it appears as if the mode freezes out when $\lambda^2/z^2 \sim 1$, namely $z^2 \sim \gamma^2/H^2$, or $c_s k_\star \sim \gamma$. However, as we mentioned already, this expectation is incorrect. In fact, we can solve Eq. (\ref{phik}) exactly and the solution for $\pi$ looks like
\beq
\pi_k(z) = A^k_1 y_1(z) + A^k_2 y_2(z),
\eeq
where
\beq
\label{soleq}
y_1(z,\nu) = z^{\nu}J_\nu(z);~y_2(z,\nu) = z^\nu Y_\nu(z),
\eeq
$\nu=\frac{3}{2} + \frac{\gamma}{2H}$, and $J_\nu,Y_\nu$ are Bessel functions. By studying the asymptotic behavior we notice only $y_2$ tends to a finite value as $z\to 0$,
\beq
y_2(z \to 0 ,\nu) \to -\frac{2^\nu}{\pi} \Gamma[\nu] \simeq - 2^\nu \sqrt{\frac{2\nu}{\pi}} e^{\nu(\log\nu-1)}  ~~{\rm for}~\nu \gg 1,
\eeq
using Stirling's approximation. In order to estimate $\pi(z \to 0)$, and consequently its contribution to the two-point function, we need to specify the initial conditions to extract the value of $A_2^k$. The most conservative approach is to assume that at some given $z_0$ the mode is in the Bunch-Davies vacuum. (More precisely: $\langle \pi(z_0)\pi(z_0)\rangle \sim \langle \pi(z_0)\pi(z_0)\rangle_{\rm BD}$.) This requires\footnote{Note that the solutions in Eq. (\ref{soleq}) tend to $\cos z_0$ and $\sin z_0$, and therefore both are required to match into the Bunch-Davies vaccum, i.e. $e^{iz_0}$. This is slightly different than the analysis in sec. \ref{hompart}, however, notice that the factor of $z_0^{-\nu}$ resembles the exponential suppression $e^{-\frac{\gamma|t_0|}{2}}$ in Eq. (\ref{anormp}).}
\beq A^k_2 \sim z_0^{-\nu}~~{\rm for}~~ \nu \gg 1.\eeq 

The origin of the early time scale $z_0$ can be understood by taking our dissipative system to be characterized by a typically high energy scale, so that it decouples from fluctuations above this threshold. This implies that we can put $\pi$ in the Bunch-Davies vacuum above some this scale. Then as it approaches freeze out we have
\beq
\label{f2z0nu}
y_2(z \to 0,\nu) \simeq 2^\nu z_0^{-\nu}\sqrt{2\nu} e^{\nu(\log\nu-1)} \to \left(\frac{2\nu}{z_0}\right)^\nu\sqrt{2\nu} e^{-\nu},
\eeq
which is exponentially small for $z_0> 2\nu \gg 1$. Notice that even performing the matching at $z_0=2\nu \sim \lambda \gg 1$,  as $\eta \to 0$ the suppression is still exponential. In other words, the mode does not immediately freeze out, as naively suggested by Eq. (\ref{phik}), in fact it continues decreasing as it approaches $z \to \sqrt{\nu}$.\footnote{To show this more explicitly we can take the ratio $\frac{d\pi/dz}{\pi/z}$ which, using
\beq
\frac{d (z^\nu Y_\nu)}{dz} =  z^\nu Y_{\nu-1}, ~~ Y_\nu \sim -\frac{2^\nu\Gamma[\nu]}{\pi z^\nu}~~({\rm for}~z\sim 0,~\nu\gg 1),
\eeq
we see goes like
\beq
\frac{d\pi/dz}{\pi/z} \simeq \frac{z^2}{\nu}.
\eeq
This suggests the solution starts to deviate from the asymptotic value near $z\sim \sqrt{\nu}$, or $c_sk_\star \sim \sqrt{\gamma H}$, as we anticipated in sec. \ref{hompart}. }

Hence we conclude that the homogenous part effectively becomes unimportant for $\gamma \gg H$ which means that the two-point function can be easily dominated by the noise, as we will assume from now on.

\section{The power spectrum}\label{power}

The computation of the power spectrum follows the same step as in sec. \ref{noise}, except that we have to deal with a somewhat more elaborate Green's function. The particular solution of Eq. (\ref{fineom}) is given by
\beq \pi_k(\eta)=\frac{ k c_s}{N_cH^2}\int^{\eta}_{\eta_0} d\eta' G_{\gamma}(k c_s|\eta|,k c_s|\eta'|)\frac{\delta{ \calo}_S}{(k c_s{\eta'})^2 },\ee
where
\beq
\label{greengamma}
G_{\gamma}(z,z')=\frac{\pi}{2} z\left(\frac{z}{z'}\right)^{\nu-1} \left[Y_\nu(z) J_\nu (z')-J_\nu (z) Y_\nu(z')\right],
\eeq 
with $z=-k c_s\eta$ and  $z'=-k c_s\eta'$. Then with the use of Eq.~(\ref{twop1}, \ref{twop2}) we obtain
\beq \nonumber  P_{\pi}(k)\equiv \langle\pi_k\pi_k\rangle_\calo=\frac{\nu_{\calo}}{N_c^2(kc_s)^3}\int^{z_0}_z dz' (G_{\gamma}(z,z'))^2.
\eeq

For $kc_s\eta\to 0$ and $k c_s \eta_0 \to -\infty$ we find
\beq\label{pow1gamma}P_{\pi}(k)=\frac{\nu_{\calo}}{N_c^2(kc_s)^3}\frac{16^{\frac{\gamma}{H} } (\frac{\gamma}{H} +1)^3
 \Gamma \left(\frac{\gamma +H}{2H}\right)^4}{\pi  \Gamma ( \frac{2\gamma}{H} +4)}\quad \to\quad  \sqrt{\pi/4}\frac{ \nu_{\calo}^\star\sqrt{H_\star/\gamma_\star}}{N^2_c (k_\star c^\star_s)^3}\quad{\rm for}\quad \gamma \gg H,\eeq
or ($P_\zeta = H^2 P_\pi$)
\beq
\label{zeta2nu}
\Delta_{\zeta} \equiv k^3P_{\zeta}(k)\simeq \nu_{\calo}^\star \sqrt{\pi H_\star/\gamma_\star} \frac{H_\star^2}{2c_s^\star\left({c^\star_s}N_c\right)^2}.
\eeq
(Recall the $\star$ means that the quantity is evaluated at freeze out $c_s k/a(t_\star)\sim \sqrt{\gamma_\star H_\star}$.)

In Fig.~\ref{casom} the power spectrum is shown as a function of $\gamma$ and also $|k c_s \eta_0|$. Notice that the dependence on $\eta_0$ drops out once we take $|k c_s \eta_0|\gg 1$.\footnote{Note that even in the case where  $\gamma \ll H$, this contribution is still larger than the result for the Bunch-Davies vacuum, provided $\nu_{\calo}> \frac{3}{\pi} N_c{H}^2$. Hence, as long as $N_c<\epsilon H^2M_p^2$, it is enough to have $\nu_{\calo}\gtrsim\epsilon H^4M_p^2$.} 

\begin{figure}[htp]
\includegraphics[width=8.5cm]{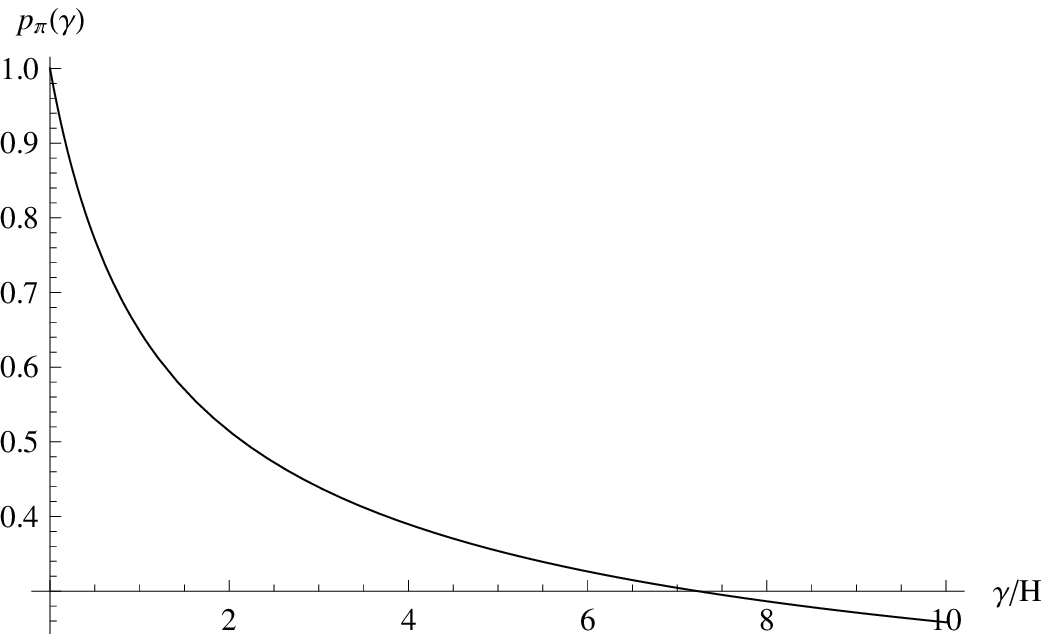}
\includegraphics[width=8.5cm]{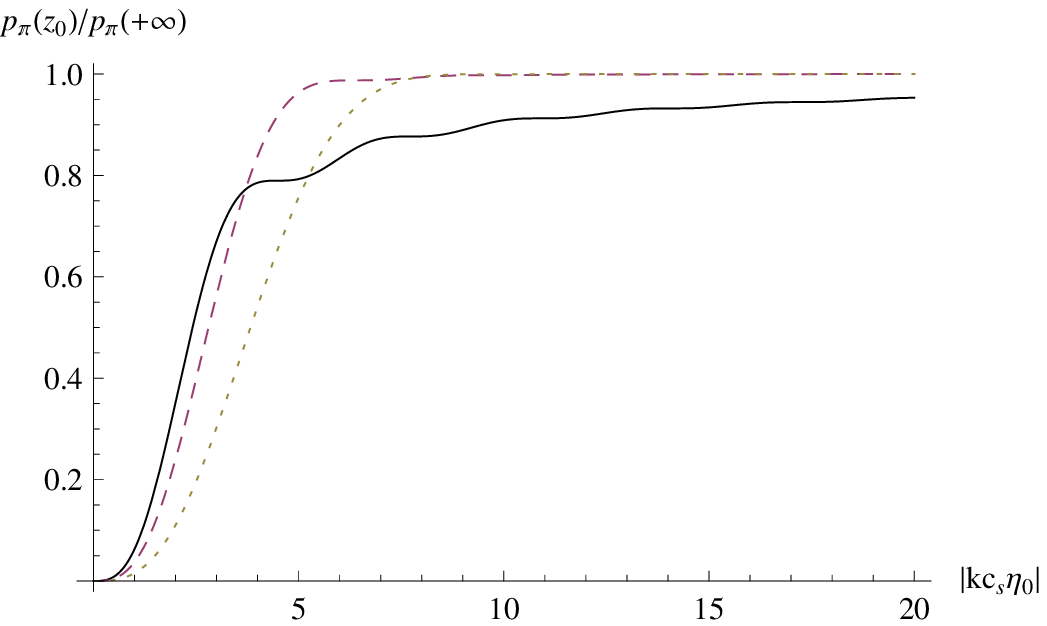}
\caption{Top:  Dependence on  $\gamma$ of power spectrum  given in Eq.~(\ref{pow1gamma}) normalized as  $p_{\pi}(\gamma)=6 N_c^2(kc_s)^3P_{\pi}(k)/(\pi \nu_{\calo})$, for  $z_0\to+\infty$; Bottom: The power spectrum $P_{\pi}(k)$ for  $|kc_s\eta|\to 0$ as a function of $z_0$, for $\gamma/H=0$ (solid), $\gamma/H=3$ (Dashed), and $\gamma/H=9$ (Dotted).}\label{casom}
\end{figure}

If we were to assume the ADOF are in thermal equilibrium at a (high) temperature $T$, using the FD theorem we would obtain
\beq
\label{tempz}
k^3\langle\zeta_k\zeta_k\rangle_T \simeq  \sqrt{\pi\gamma_\star H_\star} \frac{TH_\star^2}{2c_s^\star\left({c^\star_s}^2N_c\right)}.
\eeq

These are exactly (up to numerical factors) the results in Eq. (\ref{tpnuo}, \ref{tpwarm}). We note that an expression similar to Eq. (\ref{tempz}) was first introduced in \cite{berera1,berera2}, and plays a key role in warm inflation models \cite{warm}.\\

A few comments are in order. First of all, since we lump a series of operators into a single one, i.e. $\calo\pi$, the local approximations look significantly different depending on the type of terms we are dealing with.  However, once our assumptions are enforced the analysis is quite robust, and does not depend on the very details of the models. The only constraint we need to ensure is the scale invariance of the two-point function, which is guaranteed as long as $\gamma,\nu_{\calo}$ remain relatively constant during inflation (more below).\\

The distinction between models starts to play a role when we move to the non-linear level. For example, the friction term could derive exclusively from the operator $\calo_\mu \delta g^{\mu0}$, which does not lead to non-linear couplings between $\calo$ and $\pi$; or it may be produced by the scalar couplings $f(t)\calo$ or $\calo \delta g^{00}$, in which case we do generate quite distinct non-linear interactions. As we discussed before, $\calo \delta g^{00}$ appears to be the cleanest, for which the value of $f_{\rm NL}$ is tied up with the friction coefficient. This is not always the case for $f(t)\calo$, where the level of non-Gaussianities depends on the details of the model. However, as we argued before, in cases where there is only one preferred clock and the (non-linear) response $\delta\calo_R$ has an emergent shift symmetry and $f_{\rm NL}$ is equally enhanced. We make this analysis more precise in what follows.

\section{Non-Gaussianities}\label{nongaus}

The non-Gaussian features of the models we discussed in this paper are perhaps the most interesting results from the point of view of observational signatures. Indeed, while the two point function is essentially fixed by the symmetries of the quasi de Sitter space, and characterized just by two parameters, namely the overall amplitude and the tilt, higher order correlation functions enjoy a functional freedom that, if observed, it would allow us to decode a much larger amount of information about the physical mechanisms that produces them.

Studying the full spectrum of non-Gaussian signatures for the vast class of possible operators we introduced is difficult, mainly because of its generality, and thus lies beyond the scope of the present paper. Therefore we will concentrate on the leading (scalar) operators $\calo \delta g^{00}$ and $f(t)\calo$. Very interestingly, we will show that enforcement of the symmetries allows us to relate the linear and (some of) the non-linear contributions in the EOM, such that large dissipation will be linked to large non-Gaussianities. In the next sections we assume dissipation takes on the form $\gamma\dot\pi$. We will study the case of higher derivative dissipation in sec. \ref{localo2}.

\subsection{$\calo \delta g^{00}$}\label{Nolocalo2}

Here we analyze a situation in which the dominant interaction term has one time derivative of $\pi$, and it is produced by scalar operators of the form $-\calo\delta g^{00}$. This term has a piece which is linear in $\dot\pi$, more precisely $2\calo \dot\pi$. (In what follows we absorb the factor of 2 induced from Eq. (\ref{g00pi}) into the operator $\calo$, i.e. $\calo \to \calo/2$.) 

Notice that now we have a choice about the local approximation described in sec. \ref{localapp}, namely it may apply to $\langle\delta{\calo}_S\delta{\calo}_S\rangle$ and $G^{\calo}_{\rm ret}$, or  $\langle\delta\dot{{\calo}}_S\delta\dot{{\calo}}_S\rangle$ and $G^{\dot{\calo}}_{\rm ret}$. For the first case we will not obtain the usual dissipative term $\gamma\dot\pi$, but higher derivative terms (as in the ADL force). Here we analyze the second possibility and discuss the former in sec. \ref{localo2}.\\

Since we take the scale of time variation for $\delta\calo_S$ to be much shorter than $1/H$, the main contribution to the noise comes from $\delta\dot{\calo}_S \gg H\delta\calo_S$. We can then basically follow the same steps as in sec. \ref{localapp} to show that to linear order the equation for $\pi$ reduces to
Eq.~(\ref{fineom}) with ${\delta\calo_S}$ replaced by $\delta\dot{\calo}_S$. Then the power spectrum is  given in Eq.~(\ref{pow1gamma}), with  the
replacement $\nu_{\calo}\to\nu_{\dot{\calo}}$. More explicitly, to second order the equation for $\pi$ is given by
\beq
\ddot\pi + 3H\dot\pi  - \frac{c_s^2\nabla^2}{a^2}\pi = -N^{-1}_c\left(\delta\dot{\calo} (1+\dot{\pi})+\delta\calo\ddot\pi -\frac{1}{a^2}\partial_i(\delta\calo\partial_i\pi)\right),\label{pieqnlprev}
\eeq where $\delta\calo=\delta\calo_S+\delta\calo_R$. From Eq.~(\ref{g00pi}) we see that the force disturbing $\delta \calo$ is given by
$F=\dot{\pi}+\dot{\pi}^2/2-(\partial_i\pi)^2/2$. Using however the local approximation for the time derivative of the response part, $\delta\dot{\calo}_R$ as in  sec. \ref{localapp}, we obtain
\beq
\label{ocaldot}
\delta\dot\calo_R\simeq N_c\gamma\left(\dot{\pi} +\frac{\dot{\pi}^2}{2}-\frac{\partial_i\pi\partial_i\pi}{2 a^2}\right).
\eeq

As we already noted in sec. \ref{nonlineff}, working within this local approximation (and without making additional assumptions) we
are not able to compute all the contributions to non-Gaussianities, since in Eq.~(\ref{pieqnlprev}) there appear  not only
 $\delta\dot{\calo}$ but also ${\delta\calo}$. In general, in the perspective of EFT we do not foresee any fine cancellations and,
as we estimated in Eq. (\ref{ratiogk}), we expect  the contributions to the level of non-Gaussianities of the non-local terms
to be about the same as the local ones. Therefore, in what follows we concentrate on the terms involving only $\dot{\calo}$ and reduce Eq.~(\ref{pieqnlprev}) to
\beq
\ddot\pi + (3H+\gamma)\dot\pi  - \frac{c_s^2\nabla^2}{a^2}\pi =\frac{\gamma}{2}\frac{\partial_i\pi\partial_i\pi}{a^2}-\frac{3\gamma}{2}\dot{\pi}^2 -N^{-1}_c\delta\dot{\calo}_S (1+\dot{\pi}).\label{pieqnl}
\eeq
Let us start by focusing on the contribution of the first term on the RHS: $\gamma (\partial_i\pi)^2$. To analyze the non-Gaussianities we  decompose $\pi=\pi_1+\pi_2$, where the subscripts represent as usual the order of the solution for the fluctuations. Then
\be\label{pi1nl}\pi_1(k,\eta)=\frac{ k c_s}{N_cH^2}\int^{\eta}_{\eta_0} d\eta' g_{\gamma}(k c_s|\eta|,k c_s|\eta'|) \delta\dot{ \calo}_S,\ee where
$g_{\gamma}(k c_s|\eta|,k c_s|\eta'|)=G_{\gamma}(k c_s|\eta|,k c_s|\eta'|)/(k c_s{\eta'})^2$ with $G_{\gamma}$ defined in Eq. (\ref{greengamma}),
and (using Eq. (\ref{pi1nl}))
\begin{eqnarray}
\pi_2(k_3,0)&=&\frac{\gamma k_3 c_s^3}{2 N_c^2H^4} \int^0_{\eta_0} d\eta' {\eta'}^2  g_{\gamma}(0,k_3c_s|\eta'|)\int\frac{d^3{\bf q}}{(2\pi)^3} |{\bf q}|  |{\bf k}_3-{\bf q}| ({\bf q}\cdot({\bf k}_3-{\bf q}))\nonumber\\
&\times&\int^{\eta'}_{\eta_0} d\eta'' g_{\gamma}(qc_s|\eta'|,qc_s|\eta''|) \int^{\eta'}_{\eta_0} d\eta''' g_{\gamma}(|{\bf k}_3-{\bf q}| c_s|\eta'|,|{\bf k}_3-{\bf q}| c_s|\eta'''|) \nonumber\\&\times&\delta\dot{\calo}_S({\bf q},\eta'')\delta\dot{\calo}_S({\bf k}_3-{\bf q},\eta''').\label{pi2nl}
\end{eqnarray}
(Notice $\pi_2$ is sourced by terms quadratic in the perturbations: $\delta\dot{\calo}_S$, $\pi_1$.) \\

We want to compute the 3-point function for $\zeta\simeq -H\pi$ in the limit $\eta\to 0$:
\begin{eqnarray}\langle\zeta({\bf k}_1,0)\zeta({\bf k}_2,0)\zeta({\bf k}_3,0)\rangle&=&\langle\zeta_1({\bf k}_1,0)\zeta_1({\bf k}_2,0)\zeta_2({\bf k}_3,0)
\rangle+ \mbox{cyclic sum in $k_i$'s}\nonumber\\
&\equiv&(2\pi)^3 \delta^{(3)}({\bf k}_1+{\bf k}_2+{\bf k}_3) F(k_1,k_2,k_3). \end{eqnarray}
Then, after Eqs. (\ref{twop1}) and (\ref{twop2}) plus the condition that the noise is Gaussian (we will consider non-Gaussian noise later), i.e.
\begin{eqnarray}\label{os}&&\langle\delta\dot{ \calo}_S({\bf k}_2,\tilde{\eta'}) \delta\dot{\calo}_S({\bf k}_1,\tilde{\eta})  \delta\dot{\calo}_S({\bf q},\eta'')
\delta\dot{\calo}_S({\bf k}_3-{\bf q},\eta''')\rangle=(2\pi)^6\nu_{\dot{\calo}}^2\frac{\delta^{(3)}({\bf k}_1+{\bf k}_2+{\bf k}_3)}{a^4(\eta'')a^4(\eta''')}\\
&&\times \left\{\delta(\tilde{\eta'}-\eta'')\delta(\tilde{\eta}-\eta''')\delta^{(3)}({\bf k}_2+{\bf q})+
\delta(\tilde{\eta}-\eta'')\delta(\tilde{\eta'}-\eta''')\delta^{(3)}({\bf k}_1+{\bf q})\right\}\,\,\,(\mbox{for} \,\,\,k_3\neq 0)\nn,
\end{eqnarray}
defining $x_i=k_i/k$, with $k$ an arbitrary scale with units of momentum, we obtain (for $\eta_0\to -\infty$):
\begin{eqnarray}\label{fdeltagooshape}
&&F(x_1,x_2,x_3)= k^6 F(k_1,k_2,k_3)=-\frac{\gamma H^3\nu_{\dot{\calo}}^2}{2N_c^4c_s^8}x_1^2x_2^2x_3(x_3^2-x_2^2-x_1^2)\int_{0}^{+\infty} dy y^2  g_{\gamma}(0,x_3 y)\\
&&\times\int_{y}^{+\infty} dz\,{z}^4 \,g_{\gamma}(x_2 y,x_2 z)  g_{\gamma}(0,x_2 z)\int_{y}^{+\infty} dw\,{w}^4 \,g_{\gamma}(x_1 y,x_1 w) g_{\gamma}(0,x_1 w)+\mbox{cyclic sum in  $x_i$'s}\nn
\end{eqnarray} (we performed a change of variables: $y'=-kc_s\eta'$, $z=-kc_s\eta''$, $w=-kc_s\eta''$).\\

The associated parameter $f_{\rm NL}^{\rm eq}$ (defined for equilateral configurations) is given by
\be F(1,1,1)=\frac{18}{5}f_{\rm NL}^{\rm eq}\Delta^2_{\zeta},\ee
(with $\Delta_{\zeta}=k^3 P_{\zeta}$), and in terms of the Green's functions:
\begin{eqnarray}
f_{\rm NL}^{\rm eq}&=&\frac{5}{12}\left(\frac{\gamma}{Hc_s^2}\frac{2^{-8\gamma/H}\pi^2 {\Gamma\left(\frac{2\gamma}{H}+4\right)}^2}{\lp\frac{\gamma}{H}+1\rp^6{\Gamma\lp\frac{\gamma+H}{2 H}\rp}^8}\right)
 \int_{0}^{+\infty} dy {y}^2 g_{\gamma}(0, y)\int_{y}^{+\infty} dz z^4  g_{\gamma}( y,z)g_{\gamma}(0, z) \nonumber\\
&\times&\int_{y}^{+\infty} dw w^4  g_{\gamma}( y,w)g_{\gamma}(0, w),\label{fnlgoonoloc}
\end{eqnarray} where we have used the power spectrum given in Eq.~(\ref{pow1gamma}) (with $\nu_{\calo}\to\nu_{\dot{\calo}}$).
Performing a numerical integration we extract
\beq
\label{fnlNL}
f_{\rm NL}^{\rm eq}\simeq-\frac{\gamma}{4H c_s^2},\eeq
in the strong dissipative regime $\gamma\gg H$. Notice this can be large even when $c_s\simeq 1$. Using the analysis from Ref. \cite{Senatore:2009gt} we can also quote a bound at order of magnitude level:
\beq
\label{boundfnl}
\frac{\gamma}{H c_s^2} \lesssim 500.
\eeq

In Fig. \ref{shapegoonoloc} we plot the shape for two values of $\gamma$ ($\gamma=4H$ and $40H$). Notice there is a peak on equilateral configurations, and another peak around $x_2\simeq x_3\simeq 1/2$. The latter is one of the main features in the bispectrum, and it is due to the fact that the fluctuations are dominated by the noise.

For moderate values of $\gamma\gg H$, the shape resembles the orthogonal one described in \cite{Senatore:2009gt}, which is currently at $2\sigma$ level in the WMAP 7-yr data~\cite{Komatsu:2010fb}.\\

\begin{figure}[htp]
\includegraphics[width=8.5cm]{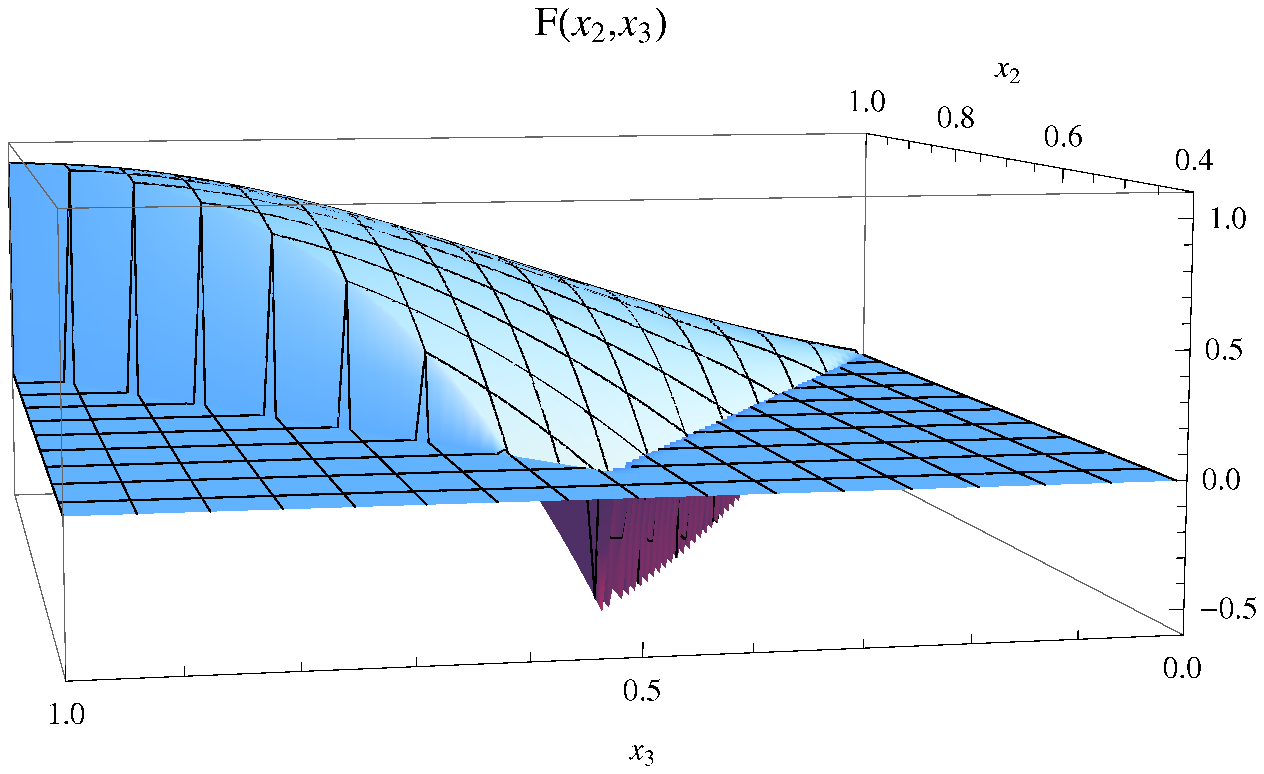}
\includegraphics[width=8.5cm]{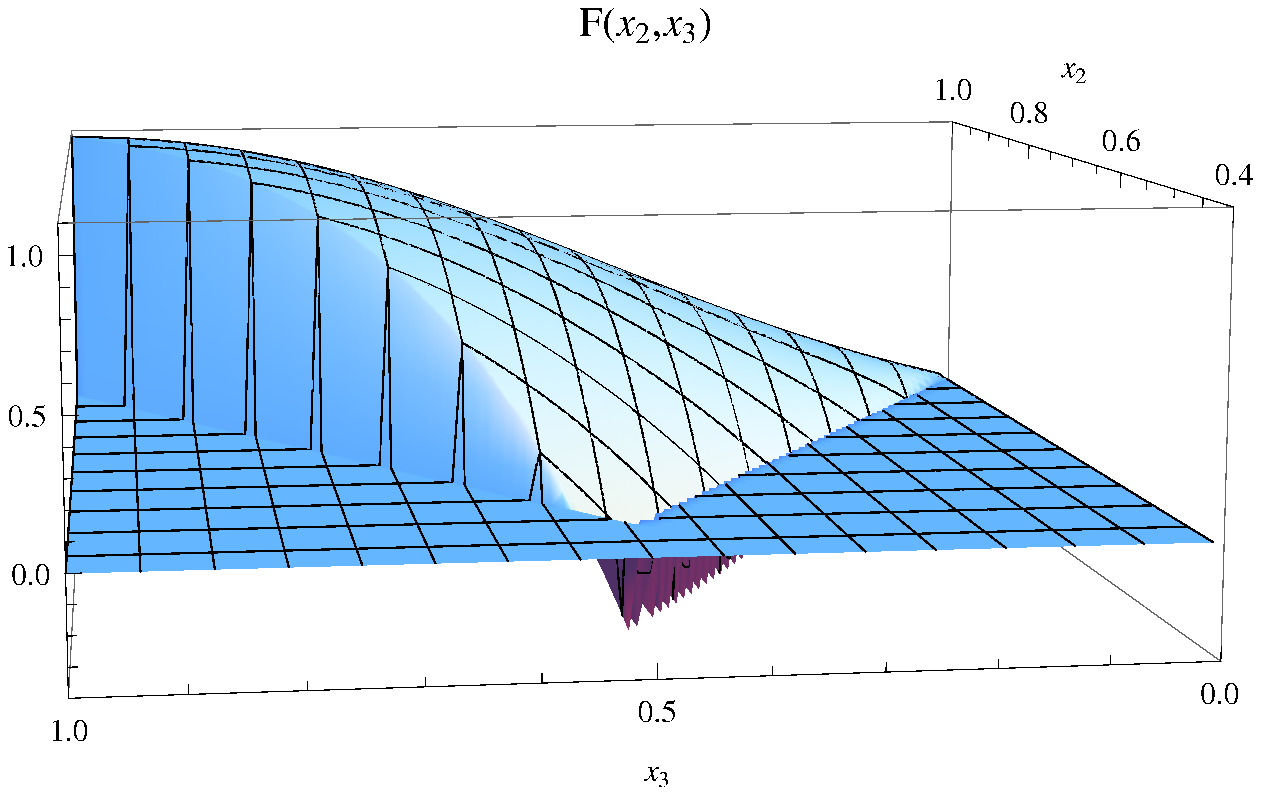}
\caption{The shape $F(x_2,x_3)=x_2^2x_3^2 \frac{F(1,x_2,x_3)}{F(1,1,1)}$ given by Eq. (\ref{fdeltagooshape}) for $\gamma= 4 H$ (top) and $\gamma=40 H$ (bottom). To avoid showing equivalent configurations twice, the function is set to zero outside the  region $1-x_2\leq x_3\leq x_2$.}\label{shapegoonoloc}
\end{figure}

Similarly, for the second source term on the RHS of Eq.~(\ref{pieqnl}) we obtain
\begin{eqnarray}
f_{\rm NL}^{\rm eq}&=&-\frac{5}{2}\frac{\gamma}{H}\left(\frac{2^{-8\gamma/H}\pi^2 {\Gamma\left(\frac{2\gamma}{H}+4\right)}^2}{\lp\frac{\gamma}{H}+1\rp^6{\Gamma\lp\frac{\gamma+H}{2 H}\rp}^8}\right)
 \int_{0}^{+\infty} dy  g_{\gamma}(0, y)\int_{y}^{+\infty} dz z^4  y\frac{d g_{\gamma}}{dy}( y,z)g_{\gamma}(0, z) \nonumber\\
&\times&\int_{y}^{+\infty} dw w^4  y\frac{dg_{\gamma}}{dy}(y,w)g_{\gamma}(0, w).\label{fnlgoonolocdot1}
\end{eqnarray} Despite appearances, the contribution from this term
to $f_{\rm NL}^{\rm eq}$ does not increase with $\gamma$ and becomes an order one sub-dominant effect. This follows from the properties of the Green's functions, and in particular because time derivatives scale as powers of $H$ (rather than $\sqrt{\gamma H}$), since this is the time dependence of the Green function at freeze out. This is indeed what we expected judging from our results in flat space of sec. \ref{noise} (``$\tau_{\rm eq}$'' $\sim 1/H$). See appendix \ref{greenf} for some collective details on these Green's functions.\\

For the last term in Eq.~(\ref{pieqnl}) we find
\begin{eqnarray}\label{fnlOdotpi}
f_{\rm NL}^{\rm eq}&=&\frac{5}{3}\left(\frac{2^{-8\gamma/H}\pi^2 {\Gamma\left(\frac{2\gamma}{H}+4\right)}^2}{\lp\frac{\gamma}{H}+1\rp^6{\Gamma\lp\frac{\gamma+H}{2 H}\rp}^8}\right)
 \int_{0}^{+\infty} dy  y^4\, (g_{\gamma}(0, y))^2\int_{y}^{+\infty} dz z^4  y\frac{d g_{\gamma}}{dy}(y,z)g_{\gamma}(0, z),\label{fnlgoonolocdot2}
\end{eqnarray} which is also a contribution of order one that does not increase with $\gamma$. The shape corresponding to this term
is shown in Fig.~\ref{ngg10} for $\gamma=10 H$.

As expected (because the contributions involve time derivatives of $\pi$, and also because $\delta{\cal O}$ is local) all of these shapes are suppressed as we approach the squeezed limit. That is when one of the momenta, say $k_1\equiv k_L$ (long mode), is much smaller than the other two (short modes), i.e. $k_L \ll k_S$ (with $k_S=|{\bf k}_2-{\bf k}_3|/2$). However an interesting question remains, and that is whether the suppression entails one (or more) power(s) of $k_L$ compared with the local shape, since this particular scaling could potentially distinguish signatures from single field models in measurements of the scale-dependent bias \cite{creminelli11}. We return to this point in the following section. (Let us point out that from Fig.~\ref{ngg10} the shape itself approaches zero in the squeezed limit $k_L/k_S\to 0$, thus  $F(x_1,x_2,x_3)$ cannot scale like $1/x_L^2\equiv (k_S/k_L)^2$, but rather as a softer power.)

\begin{figure}[htp]
\includegraphics[width=8.5cm]{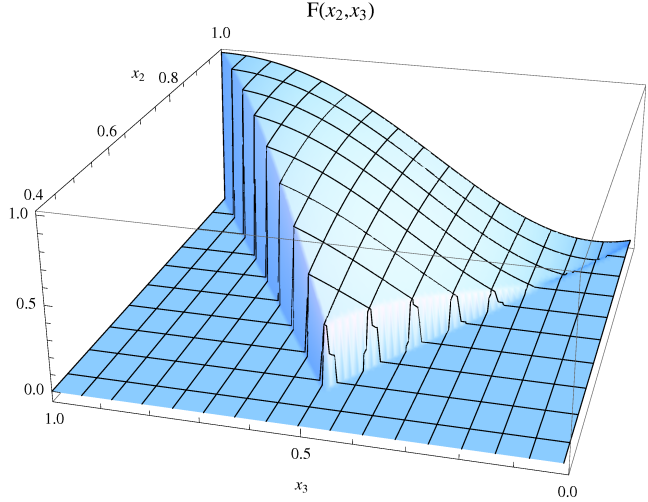}
\caption{The shape  $F(x_2,x_3)=x_2^2x_3^2 \frac{F(1,x_2,x_3)}{F(1,1,1)}$ for $\gamma= 10 H$ corresponding to the last term in Eq. (\ref{pieqnl}).}\label{ngg10}
\end{figure}


\subsection{$f(t)\calo$~I: linear response}\label{ftcalo}

Let us now consider the contribution from an interaction of the form $f(t)\calo$ as in Eq.~(\ref{inter1n}). We divide the possibilities in two: linear and non-linear response. We treat the latter in the next section.

To compute non-linear effects in linear response theory we use the force affecting the ADOF to second order in $\pi$. This is given by $F=\dot{f}\pi+\ddot{f}\pi^2/2+\ldots$, where the dots stand for terms with higher powers of $\pi$ proportional to higher temporal derivatives of $f(t)$. As we mentioned in sec \ref{appsh}, in general we neglect these terms in the slow roll approximation. Therefore in what follows we will treat $\dot f$ as essentially constant.\\

Working within the local approximation for $G_{\rm ret}^\calo$, the response part $\delta\calo_R$ can be written as
\be
\label{ddff}
\delta\calo_R\simeq \frac{\gamma N_c}{\dot{f}}\left(\dot{\pi}+\frac{\ddot{f}}{\dot{f}}\dot{\pi}\pi\right),
\ee
up to terms suppressed by higher powers of the slow roll parameter(s). (Recall we assume the presence of an emergent shift symmetry to neglect terms which do not involve derivatives.) Thus the equation for $\pi$ to second order reads
\be\label{pia}
\ddot{\pi}+(3H+\gamma)\dot{\pi}+\frac{k^2c_s^2}{a^2}\pi+2\gamma\frac{\ddot{f}}{\dot{f}}[\pi\dot{\pi}]_k=-N_c^{-1}\left(\dot f\delta \calo_S+\ddot{f}[\pi\delta \calo_S]_k\right),
\ee where  the noise part $\delta\calo_S$ satisfies Eqs.~(\ref{twop1}, \ref{twop2}). By comparing this equation
with Eq.~(\ref{fineom}), it follows that the power spectrum is the same as the one given in Eq.~(\ref{pow1gamma}), up to an overall factor of $\dot{f}^2$,
namely $\nu_{\calo} \to \nu_{f\calo}=\dot f^2\nu_{\calo}$. Therefore, to guarantee the scale invariance of the power spectrum we require $\dot \nu_{f\calo}/(\nu_{f\calo}H) \simeq O(\epsilon)$ (more on this below).\\

 As we stressed in sec. \ref{appsh}, the contribution to $f_{\rm NL}$ from the non-linear terms above will be suppressed by $\ddot f/(\dot f H) \simeq O(\epsilon)$. Notice that at this order we can no longer ignore other effects, such as the mixing with gravity. Moreover, we show in the next section that non-Gaussianities will be dominated by the non-linear response. Nevertheless, we include here the computation for the terms in Eq. (\ref{pia}) for two reasons. First of all we explicitly show that even though the non-linear terms are proportional to $\gamma$, it factors out in the final answer; but more importantly because it appears to give us a contribution in the squeezed limit which could be non-negligible in cases where $\dot f$ is not a constant.

Let us start analyzing the last source term in Eq.~(\ref{pia}): $\ddot{f}\pi\delta \calo_S$. Since this term does not involve enough derivatives of $\pi$, it
 is easy to see that the shape, given by $F(x_2,x_3)=x_2^2x_3^2 \frac{F(1,x_2,x_3)}{F(1,1,1)}$, will have its maximum contribution in the squeezed limit. In  such limit the parameter $f_{\rm NL}$ is given by
\begin{equation}\label{squ}
f_{\rm NL}^{\rm sq} =\lim_{x_3\to 0,\,\ x_2\to 1}\frac{5}{6} \frac{F(1,x_2,x_3)}{(P_{\zeta}(1)P_{\zeta}(x_2)+P_{\zeta}(1)P_{\zeta}(x_3)+P_{\zeta}(x_2)P_{\zeta}(x_3))}.
\end{equation}
Then following a procedure similar to that of the previous sections we find \beq f_{\rm NL}^{\rm sq}\simeq-\frac{5 }{6}\frac{\ddot{f}}{H\dot{f}} \label{fnlddf},\eeq
whatever the value for $\gamma$. This agrees with our previous estimation in Eq. (\ref{estfto}). Similarly, one can compute the contribution of the last term on the LHS of Eq. (\ref{pia}). Even though this term is proportional to $\gamma$, the parameter $f_{\rm NL}$ does not result in a significant increase compared to the one above. This follows from our heuristic arguments in sec. \ref{appsh}, but can also be seen directly from the properties of the Green's function, see appendix \ref{greenf}. Since the approximate shift symmetry requires $\ddot{f}/(H\dot{f})\simeq O(\epsilon)$, this leads to very small $f_{\rm NL}^{\rm sq}$ in Eq. (\ref{fnlddf}), as we mentioned before.\\

One may nonetheless worry about the result in Eq. (\ref{fnlddf})
had we assumed $\dot f, \nu_{\calo}$ were not approximately constant, while keeping constant the product $\nu_{f\calo}=\dot f^2\nu_{\calo}$, which is the combination that appears in the power spectrum. However, notice that the above computation is incomplete since at non-linear level there are contributions that arise from the fact that  $\pi$ also affects the probability density functional for the noise. (These are suppressed in the limit $\dot\nu_{\calo} \ll H\nu_{\calo}.$) If all the contributions are taken into account, at the end of the day we should have $f_{\rm NL}^{\rm sq}$ proportional to $(n_s-1)$ to be consistent with the fact that we did not include any type of perturbation outside of the horizon other than the clock \cite{consist,Cheung:2007sv}. 

To better understand the properties in the squeezed limit, let us compare the contributions from $\delta{\calo}\dot{\pi}$ and $\delta{\calo}\pi$ type of non-linearities, which appear commonly in our study. It is not difficult to see that if we fix the $\calo$ operators to be the same in both cases, the relevant difference between the two is that while the shape for the latter contains $g_{\gamma}(x_L y, z)$, the former involves $ y \partial_{y}g_{\gamma}(x_L y, z)$ (where $z$ and $y$ are integration variables). Using  \beq
\frac{y\partial_{y}g_{\gamma}(x_L y, z)}{  g_{\gamma}(x_L y, z)}\sim \frac{x_L^2y^2}{(1+\gamma/H)},
\eeq  for $y x_L \ll 1$, and taking into account that the integrals are dominated by values of $y$ and $z$ near $\sqrt{\gamma/H}$ (see appendix \ref{greenf}), we conclude that the squeezed limit for $\delta{\calo}\dot{\pi}$ is suppressed by a factor of $x_L^2 = (k_L/k_S)^2$ with respect to the contribution from $\delta{\calo}\pi$, in the region $x^2_L \ll H/\gamma$. This agrees with the findings in \cite{creminelli11}. We will discuss the squeezed limit and consistency conditions in more detail elsewhere.\\

We study next the non-linear response for a $f(t)\calo$ coupling, which turn out to provide the largest contributions to non-Gaussianities.

\subsection{$f(t)\calo$~II: non-linear response}\label{nonlrep}

So far we considered the response $\delta\calo_R$ to linear order in $\pi$. This entails the knowledge of only the two point function $G_{\rm ret}^\calo$ (see Eq. (\ref{gnoa1})). However, we can also induce non-linearities by going to higher $n$-point functions of the $\delta\calo$'s, for instance including their three point function
\beq
C^{\calo}_{\rm ret} (x,y,z) \equiv \langle [\delta\calo(z),[\delta\calo(y),\delta\calo(x)]]\rangle \theta(t_y-t_z)\theta(t_x-t_y),
\eeq
while keeping the leading order force $F(\pi) = \dot f \pi$ (in what follows we treat $\dot f$ as constant). Then we have
\beq
\label{calo12}
\dot f \delta\calo_R(x) = -\dot f^2 \int G^\calo_{\rm ret}\pi(y) dy + \dot f^3 \int dydz~C^{\calo}_{\rm ret}(x,y,z)\pi(y)\pi(z) + \ldots
\eeq
where the ellipses represent higher order corrections.

A priori, even assuming both are local in space and time, we would not expect any relationship between $G^{\calo}_{\rm ret}$ and $C^{\calo}_{\rm ret}$.
However, as we argue below, we will have a connection in cases where the response of the ADOF is governed by a single (preferred) clock.\\

As we emphasized throughout the paper to obtain a dissipative term from an operator $f(t)\calo$ we require (assuming a local Green's function)
\beq
\dot f  \delta^{(1)}\calo_R = N_c \gamma \dot\pi.
\eeq
Notice that this expression can be suggestively re-written as
\beq
\label{onecalR}
\dot f  \delta^{(1)}\calo_R = N_c \gamma~\delta^{(1)}\left\{n^\mu\partial_\mu(t+\pi)\right\},
\eeq
where (note $n^0= 1$ up to $O(\pi^2)$)
\beq
\label{nmu2} n^{\mu}=\frac{-g^{\mu\nu}\partial_{\nu}{\tilde t}}{\sqrt{-g^{\nu\rho}\partial_{\nu}{\tilde t}\partial_{\rho}{\tilde t}}},
~~\tilde t = t+\pi, \eeq   
such that Eq. (\ref{onecalR}) makes the symmetries manifest: $\calo$ is a scalar operator. This allows us now to extrapolate the response to all orders in $\pi$, that is\footnote{In principle this can be generalized to $F_\calo\left(\sqrt{-g^{\nu\rho}\partial_{\nu}{\tilde t}\partial_{\rho}{\tilde t}}\right)$. For simplicity we restrict to $F_\calo(x)=x$.}
\beq
\label{gretto2p}
\dot f  \delta^{(n)}\calo_R = N_c \gamma~\delta^{(n)}\left\{n^\mu\partial_\mu(t+\pi)\right\} = N_c\gamma~\delta^{(n)}\left(\sqrt{-g^{\nu\rho}\partial_{\nu}{\tilde t}\partial_{\rho}{\tilde t}}\right).
\eeq

It is then straightforward to show the above equation implies a relationship between linear and non-linear terms in the EOM. For example in the unitary gauge, i.e. $\tilde t=t$, 
\beq
\label{dpcalfn}
\dot f  \delta^{(n)}\calo_R = N_c \gamma~\delta^{(n)}\sqrt{-g^{00}},
\eeq
which allows us to read off the $\pi$ interactions using $g^{00}(\pi) = -1-2\dot\pi -\dot\pi^2 + (\partial_i\pi)^2$. Hence
\bea
-\dot f \delta^{(1)}\calo_R &=& -N_c \gamma \dot\pi,\\
-\dot f \delta^{(2)}\calo_R &=& \frac{1}{2} N_c \gamma (\partial_i\pi)^2 
\label{dpcalo2}.
\eea
We thus get a non-linear term that resembles the one we obtained for the case of an exact shift symmetry. (See the first term on the RHS of Eq. (\ref{pieqnl}), compare also with Eq. (\ref{ocaldot}).\footnote{Notice that for the type of non-linear response in Eq. (\ref{dpcalfn}) we did not obtain a $\dot\pi^2$ term. However, these can be generated if we allow for a generic function $F_\calo(\sqrt{-g^{00}})$.}) Then the analysis of the non-Gaussianities follows as in sec. \ref{Nolocalo2} (see also sec. \ref{shsym}), so that
\beq
f_{\rm NL} \simeq -\frac{\gamma}{c_s^2 H},
\eeq
as anticipated in sec. \ref{nlrep}. The shape will also resemble Fig.~\ref{shapegoonoloc}.\\

The previous argument is perhaps better illustrated if we particularize to the traditional model of scalar field inflation. The crucial point is that the response to the perturbation must be compatible with diffeomorphism invariance. Then in order to induce dissipation we need factors of $\dot\phi$ which in turn lead to large non-Gaussianities (for $\gamma \gg H$) as sketched in \ref{nlrep}. 

The reasoning goes as follows. Let us consider an interaction Hamiltonian that takes the form (with $\calo$ a scalar operator)
\beq
H_{\rm int} = \phi\; \calo \label{hint}.
\eeq
The EOM for the inflaton thus becomes
\beq
\label{eomphi}
D_\phi \phi = \calo,
\eeq
where $D_\phi \phi=0$ represents the EOM in the absence of $\calo$. There is certainly a dynamical system behind $\calo$, with its own $\calL_\calo$. Nevertheless, in the spirit of EFT we do not make any assumption other than the coupling in Eq. (\ref{hint}). We start now by solving for the background, i.e. $\bphi(t)$.  Assuming a local response we have
\beq
\label{gcalo}
\langle \bphi|\calo |\bphi\rangle = F_\calo(\bphi, \dot{\bphi}),
\eeq
where the brackets emphasize we are computing the response in the background given by $\bphi$. Plugging into Eq. (\ref{eomphi}) we get
\beq
\label{dphibpi}
D_\phi \bphi = F_\calo(\bphi, \dot{\bphi}).
\eeq
In order to have the type of (velocity dependent) dissipation we study in this paper we will impose
\beq
\label{fphip}
 F_\calo(\bphi, \dot{\bphi})\simeq |\dot\bphi|.
\eeq
We discuss a specific example of this behavior in sec. \ref{matching} (albeit with a more elaborate response function). The next step is to perturb $\bphi \to \bphi +\delta\phi$. The crucial piece of the argument is what replaces Eq. (\ref{gcalo}). There are several ways to attack this question. Perhaps the simplest is the following. Since $\calo$ is a scalar operator (by construction) the expression in Eq. (\ref{gcalo}) must be covariantized. (Notice Lorentz invariance is only broken by the presence of a preferred frame, i.e. $n^\mu$ in Eq. (\ref{nmu2}) with $\tilde t$ replaced by $\phi$.) Therefore, as we argued, we should rewrite Eq. (\ref{gcalo}) as
\beq
\label{gcalo2}
\langle \phi|\calo |\phi\rangle = F_\calo\left(\phi, n^\mu \partial_\mu \phi = -g^{\mu\nu}\partial_\nu\phi\partial_\mu\phi/\sqrt{-g^{\mu\nu}\partial_\nu\phi\partial_\mu\phi}\right),
\eeq
where we assumed the function $F_\calo$ is not essentially modified, which is a valid (local) approximation provided $\delta\phi$ is a {\it smooth} long wavelength perturbation.\footnote{In other words, we need the scale of variation of the extrinsic curvature of constant time surfaces (recall $K^{ij} \simeq \nabla^i_{(3)} n^j$) to be much larger
than the typical length scale $M^{-1}_\calo$, which controls the local approximation for the interaction between $\phi$ and $\calo$, such that we have a well defined derivative expansion. For instance if we take the perturbations at horizon crossing, namely $\delta\phi_{k_\star}$, and say $M^2_\calo \simeq g |\dot\bphi|$ with $g$ some coupling constant, then we require $k_\star^2 \ll M_\calo^2$, or $\gamma < g |\dot\bphi|/H$. (In all generality, even if suppressed, we should in principle include also corrections induced by $\delta K_{\mu\nu}$.) See sec. \ref{localtrap} for more details.}

Once again we go to the unitary gauge, and choose coordinates such that $\phi = \bphi$, i.e. $\delta\phi=0$. This means
\beq
\label{gcalo3}
\langle \phi| \calo |\phi \rangle = F_\calo\left(\bphi,\dot\bphi\sqrt{-g^{00}}\right).
\eeq
We then introduce the perturbation $\delta\phi$ (our $\pi$) using Eq. (\ref{dphibpi}) and the St\"uckelberg trick. On the LHS of we get the usual term, whereas on the RHS we use the requirement of Eq. (\ref{fphip}), then
\beq
D_\phi \delta\phi  \simeq \delta \left(|\dot\bphi| \sqrt{-g^{00}(\pi)}\right),
\eeq
similarly to what we obtained from Eq. (\ref{dpcalfn}),  with $\pi = \delta\phi/\dot\bphi$. Then, as advertised, we get a dissipative term $\gamma \dot\pi$ and also $\gamma (\partial_i\pi)^2$ with its subsequent enhancement of non-Gaussianities as shown in sec. \ref{Nolocalo2}.\\

There are some subtleties in the previous discussion that must be stressed out. Strictly speaking, the presence of dissipation requires
 only a linear coupling to $\pi$. In our discussion the non-linear term obtained from $n^\mu\partial_\mu \pi$ in Eq. (\ref{gretto2p}) was intrinsically related to $n^\mu$ being the time-like vector orthogonal to the uniform slices of the physical clock that controls the end of inflation. However, in general the presence of ADOF may also include time-like vector operators, let us collectively denote them as $u_\calo^\mu$, that might have non-zero expectation values in the background. Hence,
rather than $n^\mu \partial_\mu \pi$ we may have $u_\calo^\nu \partial_\nu \pi$  in Eq. (\ref{gretto2p}).\footnote{This may happen if the response of $\calo$ is controlled
by $u^\mu_\calo$.} In this case there will not be a guaranteed relationship between the linear term $\gamma\dot\pi$ (now derived from $\langle u_\calo^\mu\rangle = (1,0,0,0)$)
 and the non-linear counterparts. In this situation large dissipation may not necessarily lead to strong non-Gaussianity.\\

It is worth pointing out that had we chosen to work directly with the Goldstone boson $\tilde\pi$ that we introduced in sec.~\ref{newdof}, we would have had no vector-like operator taking a non-zero expectation value. This means that (up to normalization) the only vector taking a background expectation value is
 $n^\mu~\sim~g^{\mu\nu}\partial_\nu(t+\tilde\pi)$, which appears to induce the type of term in Eq. (\ref{dpcalo2}). However the two descriptions are equivalent, and in fact, upon noticing (schematically) $\tilde \pi \sim \pi + \delta O$, even in this case we cannot in general prevent an interplay between $\tilde \pi$ and fluctuations of vector operators $\delta u^\calo_\mu$ such that the term in Eq. (\ref{dpcalo2}) does not get generated at the non-linear level. Therefore, in order to ensure a connection between dissipation and non-Gaussianities, we will demand that there are no such cancellations. 
 
 There is no simple way to implement this condition, although in practice this amounts to requiring that the $\delta\calo$'s are sensitive mostly to the field $\pi$ whose fluctuations control the end of inflation. In this paper we referred to this condition as having a preferred clock.

Hence we conclude that the existence of such clock inevitably entails large non-Gaussianities in the strong dissipative regime, either from $\calo g^{00}$, as explicitly shown in sec. \ref{Nolocalo2}, or via non-linear response for $f(t)\calo$ as we just described. We will study relaxing this assumption in future work.

\subsection{Non-Gaussian noise}\label{ngaussnoise}

As a final source of non-Gaussianities, we have to consider the effect of the intrinsic non-Gaussianity of the noise fluctuations.
Let us consider, as in Eq. (\ref{ngnoise}), a non-Gaussian noise $\delta\calo_S$ that is local, that is
\beq
\langle\delta\calo_S(t_1, {\bf k}_1)\delta\calo_S(t_2, {\bf k}_2)\delta\calo_S(t_3, {\bf k}_3)\rangle\simeq (2\pi)^3\delta^{(3)}({ \bf k}_1+ {\bf k}_2+ {\bf k}_3)\nu_{{\calo}^3}\frac{\delta(t_1-t_2)}{a^{3}(t_1)}\frac{\delta(t_2-t_3)}{a^3(t_2)}.
\eeq
Then using
\beq \pi(t, {\bf k})=\frac{\dot{f} k c_s}{N_c H^2}\int^{\eta}_{\eta_0} d\eta' g_{\gamma}(kc_s |\eta|, k c_s|\eta'|)\delta\calo_S(\eta',{\bf k})
\eeq we find (for $\eta\to 0$ and  $\eta_0\to -\infty$)
\beq\label{NGNoise}
F(x_1,x_2,x_3)=-\frac{\dot{f}^3 H^5\nu_{{\calo}^3}}{N_c^3c_s^6}x_1x_2x_3\int_0^{+\infty} dz z^8 g_{\gamma}(0,x_1 z)g_{\gamma}(0,x_2 z)g_{\gamma}(0,x_3 z).\eeq

\begin{figure}[h!]
\includegraphics[width=8.5cm]{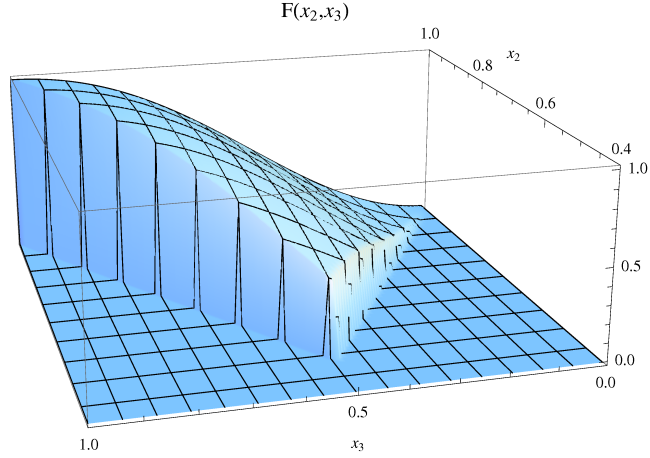}
\caption{The shape $F(x_2,x_3)=x_2^2x_3^2 \frac{F(1,x_2,x_3)}{F(1,1,1)}$ obtained from Eq. (\ref{NGNoise}) for $\gamma\simeq 10 H$. To avoid showing equivalent configurations twice, the function is set to zero outside the  region $1-x_2\leq x_3\leq x_2$.}\label{nonGausNoise}
\end{figure}

The shape is plotted in Fig.~\ref{nonGausNoise}. As expected there is a peak on equilateral configurations, and
\beq
f_{\rm NL}^{\rm eq}\simeq \frac{\gamma N_c\nu_{{\calo}^3} }{\dot{f} \nu_{\calo}^2},
\eeq
as we anticipated in Eq. (\ref{fnlnoise}).

\subsection{Higher-derivative dissipation}\label{localo2}

Let us consider now the same situation as in sec.~\ref{Nolocalo2}, where the dominant interaction has one derivative of $\pi$: $\delta g^{00}\calo$. However, unlike before, we will now assume a local approximation for the two-point functions of $\calo$ (or ${\rm Im}G_{\rm ret}^\calo \sim \omega$), such that we get
\beq\label{eq:help}\delta\calo_R\simeq-\frac{N_c}{a^3\Gamma_\calo}\partial_t(a^3\dot{\pi}),\ee
where $\Gamma_\calo$ (with units of mass) depends on the specific details of the model. Given that the interaction term goes like $\pi\delta\dot\calo$, the EOM
becomes\footnote{We also generate corrections to the speed of sound of order $H/\Gamma_\calo$ that we neglect here.}
\beq \ddot{\pi}+\frac{c_s^2k^2}{a^2}\pi+3H\dot\pi- \frac{1}{\Gamma_\calo a^3}\partial_t^2(a^3\dot\pi)=-\frac{\partial_t(a^3\delta\calo_S)}{a^3N_c},\eeq
Note this equation is now reminiscent of the ADL force. The presence of higher derivative terms forces us to restrict $\omega \simeq H \ll \Gamma_\calo$,
and treat the higher derivate term as a small perturbation. In this case freeze out occurs at $\omega \sim H$ and the leading effect is not on the two-point
function, that here is dominated by the vacuum solution, but on the three-point function. This can be estimated by comparing the non-linear term in the EOM to the linear term. The second order equation can be written as
\be\ddot{\pi}_2+\frac{{c}_s^2k^2}{a^2}\pi_2+H\lp3+\frac{ k^2c_s^2 \delta}{a^2H^2}\rp\dot{\pi}_2=J_k,\ee
where the term $J_k$ contains the quadratic sources. From the  coupling between $\pi$ and $\delta \calo$ we have
\be \label{source}J_k=-\frac{1}{N_c}\left\{\frac{\partial_t(a^3{\delta}_2\calo)}{a^3}+\frac{\partial_t(a^3\dot{\pi}_1{\delta}_1\calo)}{a^3}-\frac{\partial_i({\delta}_1\calo\partial_i\pi)}{a^2}\right\},\ee
where ${\delta}\calo=\delta\calo_S+\delta\calo_R$, and the subscripts represent to what order in $\pi$ the contributions are to be computed. For estimating purposes we consider the last one, which dominates for small speed of sound models. 
After using (\ref{eq:help}) the resulting level of non-Gaussianity is given by
\be
f_{\rm NL}\zeta\sim\left.\frac{\frac{1}{N_c}\partial_i({\delta}_1\calo\partial_i\pi)}{\ddot\pi}\right|_{\omega\sim H}\sim\frac{H^4 \pi^2}{c_s^2 \Gamma_\calo}\cdot \frac{1}{ H^2\pi}\sim\frac{H}{\Gamma_\calo}\cdot\frac{1}{c_s^2}\zeta\ll \frac{1}{c_s^2}\zeta,
\ee
which tells us that this can be at most a subleading signal.

 \section{Matching dissipative effects during inflation}\label{matching}

Even though from the EFT standpoint the conditions on $G^{\calo}_{\rm ret}$ and noise kernels may be taken as a given, it is still desirable to be able to identify those situations where we can be assured these conditions hold within certain degree of approximation. For instance we expect such description to be valid in cases where the memory effects decays sufficiently fast, as in Drude's model of Eq.~(\ref{drude}). This would be the case provided the dynamics of the $\calo$'s is such that the back reaction on the $\pi$ field becomes negligible after a (short) interaction time. In this section we study a specific realization in the spirit of the trapped inflation model of \cite{trapped}, and also briefly discuss the warm inflation paradigm \cite{berera1,berera2,warm}. See appendix \ref{mag00} for a class of models with $\gamma\dot\pi$ dissipation with a $\calo\dot\pi$ coupling.

\subsection{Local trapped inflation: $f(t)\calo \to \sum_i(\phi-\phi_i)^2\chi^2_i + \calL(\chi_i)$}\label{localtrap}

We are after an example in which we can apply the local approximation in time. In trapped inflation the energy of the inflaton is being transferred into the ADOF both because the particles are created and also due to the increase of their mass with time. In the original model this is important until the particles have diluted enough due to the expansion of the universe. The characteristic time scale for this to happen is given by $H^{-1}$, which does not allow us to use the local response functions that we used so far. However, this characteristic time scale can be significantly reduced if the produced particles decay into yet other degrees of freedom, with decay rate much faster than Hubble. With this in mind let us consider the trapped inflation action
\beq \label{seftrap} S_{\rm trap}=\int d^4x\sqrt{-g} \left\{\sum_i\left[-\frac{1}{2}\partial_{\mu}\chi_i\partial^{\mu}\chi_i-\frac{ g^2(\phi-\phi_i)^2}{2}\chi_i^2\right]-\frac{1}{2}\partial_{\mu}\phi\partial^{\mu}\phi-V(\phi)\right\},\eeq
where $\phi$ is the inflaton and $V(\phi)$ is its potential, and we add an interaction term, $S_{\rm int}(\chi_i,\varphi)$, that characterizes the coupling between $\chi_i$'s and some additional degrees of freedom $\varphi$ that leads to a decay rate for the $\chi_i$'s satisfying $\Gamma_{\chi_i} \gg H$. The only condition we need to impose is the requirement that $S_{\rm int}$ does not modify the leading order picture from $S_{\rm trap}$. In particular we need to ensure it does not generate a large mass for $\chi_i$ when $\phi \simeq \phi_i$. For example we can have
\beq
\label{decayr}
S_{\rm int} =  \sum_i \frac{1}{\Lambda_\varphi} \partial^2\chi_i \varphi^2,
\eeq
with $\varphi$ some scalar field which we assume has some small mass, $m_\varphi \lesssim \sqrt{g |\dot\phi|}$, and does not couple directly to $\phi$. In what follows we show that this model admits a local approximation. (To simplify notation in this section we drop the bar for the unperturbed values of $\phi$.)\\

As in the original trapped inflation scenario, particles associated to the field $\chi_i$ are created when $\phi$ approaches $\phi_i$.
The computation of the production of particles follows the same line as in \cite{trapped,beauty}. The time interval $\delta t_c=t-t_i$  during which modes
 of a given $\chi_i$ field do not behave adiabatically (so that the corresponding particles are produced) can be estimated as
\be \delta t_c\sim 1/\sqrt{g|\dot{\phi}|},\ee
which we want to be shorter than a Hubble time, $\delta t_c \ll H^{-1}$, hence \cite{trapped}
\be\label{c1}
H^2\ll{g|\dot{\phi}|}.
\ee
Note this condition implies that after particles are produced the mass  satisfies
\be M_{\chi_i}(t_i+\delta t_c)=g|\phi-\phi_i|\simeq g|\dot{\phi}|\delta t_c\gg H.
\ee
Moving to the decay rate, using Eq. (\ref{decayr}) we can estimate
\beq
\Gamma_{\chi_i} \simeq  \frac{M_{\chi_i}^3}{\Lambda_\varphi^2}. \label{gammab}
\eeq
To ensure the decay time scale is longer than the time it takes to create the particles  we need
\beq
\label{c2n}
 \frac{1}{\delta t_c}\simeq \sqrt{g|\dot\phi|}\gtrsim \Gamma_{\chi_i}\simeq\frac{(g|\dot{\phi}|)^{3/4}}{ \Lambda_\varphi^{1/2}} \quad\to\quad  \Lambda_\varphi \gtrsim \sqrt{g|\dot\phi|},
\eeq
where we have approximated $M_{\chi_i}\simeq g |\dot{\phi}|\tau_{\chi_i}\equiv M_{\chi_i}(\tau_{\chi_i})$, with $\tau_{\chi_i}$ the lifetime of the $\chi_i$-particles, while at the same time $\Gamma_{\chi_i} \gg H$, which is necessary for our local approximation.\footnote{Notice that $\tau_{\chi_i} > \delta t_c$ requires $\Gamma_{\chi_i} < M_{\chi_i}(\tau_{\chi_i})$, which also consistent with the particle interpretation.}  
After the particles are produced, the occupation number for each species is given by
\beq
\label{ocupnum}
|\beta_k|^2=e^{-\frac{\pi k^2}{a^2(t_i)\kappa^2(t_i)}},
\eeq where $\kappa(t_i)=\sqrt{g|\dot{\phi}(t_i)|}$, and the number density
\be\label{nchim} n_{\chi_i}(t_i+\delta t_c)=M_{\chi_i}\bar\chi^2_i\simeq \frac{\kappa^3(t_i)}{(2\pi)^3}.\ee
Then, taking into account the decay rate of the particles for later times, we have
\be n_{\chi_i}(t,t_i)\simeq \frac{\kappa^3(t_i)}{(2\pi)^3}\lp\frac{a(t_i)}{a(t)}\rp^3 e^{-\Gamma_{\chi}(t-t_i)} \; \Theta(t-t_i),\ee
where in order to obtain an estimation we have approximated $\Gamma_{\chi_i}$ by a constant $\Gamma_{\chi}$.
The EOM for the unperturbed inflaton becomes
\be\label{infeqLTI}\ddot{\phi}+3H\dot{\phi}+V'(\phi)+g\sum_in_{\chi_i}(t,t_i)=0.\ee

As the sum over the particle production events is difficult to deal with, we will replace it by an integral. Defining $\Delta=\phi_{i+1}-\phi_i$, the last term on the LHS of Eq. (\ref{infeqLTI}) can be thus approximated by
\beq
\sum_in_{\chi_i}(t,t_i)\simeq~\int^t dt'\frac{|\dot{\phi}(t')|}{\Delta}\frac{\kappa^3(t')}{(2\pi)^3}e^{-(3H+\Gamma_{\chi})(t-t')}.
\eeq

This is a good approximation only if the variation of the integrand is small between production events. This is quantified by the conditions
\be\label{c3c4} \frac{ (3H+\Gamma_{\chi})|\Delta|}{|\dot{\phi}|}\ll1,\,\,\,\frac{|\ddot{\phi}\Delta|}{\dot{\phi}^2}\ll1.\ee
Hence, for $\Gamma_{\chi} \gg H$ and $|\ddot{\phi}|\ll\Gamma_{\chi} |\dot{\phi}|$ we can approximate the integral as
\beq
\label{gnchi}
g \sum_i n_{\chi_i}(t,t_i)\simeq \frac{g|\dot{\phi}(t)|}{\Gamma_{\chi}\Delta }\frac{\kappa^3(t)}{(2\pi)^3} \simeq \frac{(g |\dot{\phi}|)^{5/2}}{\Gamma_{\chi}|\Delta|(2\pi)^3}.
\eeq
We discuss the consistency of our approximations in more detail in appendix \ref{consistency}.\\

We now define the (composite) operator $f(t){\calo}\equiv \sum_i f_i(t)\calo_i$, where\footnote{We chose to work with $f(t)\calo$, rather than each individual $f_i\calo_i$, in order make direct contact with our analysis of the EFT in previous sections.} 
\beq
f_i(t) = \frac{g^2}{2}(\phi(t)-\phi_i)^2,~~\calo_i =  \chi^2_i, \eeq
whose background expectation values are given by (see Eq. (\ref{nchim}))
\beq
\sum_i f_i(t)\bar\calo_i = \frac{g^2}{2} \sum_i (\phi(t)-\phi_i)^2\bar\chi_i^2 =  \frac{1}{2}\sum_i M_{\chi_i} n_{\chi_i}.
\eeq

Then, using Eq. (\ref{gnchi}), the expression in Eq. (\ref{infeqLTI}) becomes (where we add the gradient piece, that vanishes in the background, for later convenience) 
\beq
\label{inflocal}\ddot{\phi}+3H\dot{\phi}-\frac{c_s^2}{a^2}\partial_i^2\phi+V'(\phi)+\frac{(g |\dot{\phi}|)^{5/2}}{\Gamma_{\chi}|\Delta|(2\pi)^3}=0,
\eeq
which is local in time as we advertised. (The same applies for the EOM of the perturbations we study in the next section.)\\ 

To wrap up this section let us comment on the emergence of the shift symmetry, which plays an essential role in our analysis. Even though the EOM has a term $\sum_i g^2(\phi-\phi_i)\bar\chi_i^2$ which is apparently not invariant under $\phi \to \phi+c$, secretly it is, since we end up with Eq.~(\ref{inflocal}) which is manifestly invariant (ignoring the shift in $V(\phi)$ which we assume is small).  

Notice, first of all, the non-perturbative result
\beq
\bar\chi_i^2\simeq {n_{\chi_i}(\dot\phi(t_i))\over g|\phi-\phi_i|},
\eeq
which cancels the explicit factors of $\phi-\phi_i$. However this is not enough, since $n_{\chi_i}$ depends on $t_i$ (the time defined as $\phi(t_i)=\phi_i$), which in turn depends explicitly on $\phi$. Nevertheless, this dependence is ultimately removed by the presence of the sum, which is the key feature. At this point the EOM becomes invariant under $\phi \to \phi + c$. (This is the case because we can absorb the shift in $\phi$ into a redefinition of $\phi_i$, which is summed over a large number of periods.) We call this an emergent shift symmetry.

\subsubsection{Perturbations}

In order to obtain the equation for the perturbations at linear order we expand $\phi \to \phi+\delta\phi$ in Eq. (\ref{inflocal}). Taking into account the contribution of the noise we obtain
\be
\label{pertinflocal}
\ddot{\delta\phi}+\frac{c_s^2k^2}{a^2}\delta\phi+3H\dot{\delta\phi}+V''(\phi)\delta\phi+\frac{5}{2}\frac{g^{5/2} |\dot{\phi}|^{3/2}}{\Gamma_\chi|\Delta|(2\pi)^3}\dot{\delta\phi}=-g \Delta n_{\chi},\ee
with \beq \label{Dnoise} \Delta n_{\chi} = g \sum_i M_{\chi_i} \Delta \bar \chi_i^2, \eeq the variance of the number of particles produced.\footnote{Notice
that in generalizing the result derived for the background into the one for the perturbations, we are assuming that effects due to the extrinsic curvature
 of the surfaces of constant $\phi$ are suppressed by powers of $k_\star/\kappa\ll 1$  i.e. $\gamma \ll g|\dot\bphi|/H$, at the time of freeze-out with
$\kappa\sim (g|\dot\phi|)^{1/2}$.} Hence we arrive at an equation as in (\ref{eom1}), with an effective damping rate $\gamma$ given by
\be
\gamma=\frac{5}{2}\frac{g^{5/2} |\dot{\phi}|^{3/2}}{\Gamma_\chi|\Delta|(2\pi)^3}.
\ee
In this expression we ignored the spatial variation of the field in $\Delta n_\chi$ at linear order, provided the physical wave vector $k_{\rm ph}=k/a$ satisfies $k_\star\ll \kappa$, and also $k_{\rm ph}\Delta\ll|\dot{\phi}|$, in order to be able to replace sum into integrals.\\

Notice that while each individual $\dot f_i$ is not a constant the power spectrum still obeys scale invariance. This can be seen by redefining $\hat \calo_i \to n_{\chi_i}$, in which case the new $\hat f_i(t) \simeq g M_{\chi_i}(\phi)$ does obey $\ddot {\hat f}_i/(\dot{\hat f}_iH) \ll 1$ in
 the slow roll approximation. Indeed, the response at linear order in $\delta\phi$ becomes
\beq
g \sum_i \delta n_{\chi_i} \simeq \frac{5}{2}\frac{g^{5/2} |\dot{\phi}|^{3/2}}{\Gamma_\chi|\Delta|(2\pi)^3} \dot{\delta\phi} +\ldots \quad\to\quad \sum_i \dot f_i \delta\calo_i^R =\dot{\hat f}  \sum_i \delta n_{\chi_i} \simeq N_c \gamma \dot\pi +\ldots,
\eeq
where we used $\delta\phi/\sqrt{N_c} \simeq \pi$ and defined $\dot {\hat f} = g\dot\phi$, with the normalization $N_c \simeq \dot\phi^2$ as explained in sec. \ref{newdof}. Likewise, for the noise (see Eq. (\ref{Dnoise}))
\beq \label{noiseng} \frac{1}{N_c}\sum_i \dot f_i \delta\calo_i^S\quad \to\quad \frac{1}{N_c} \dot{\hat f} \Delta n_{\chi}.
\eeq

So far so good. What we need now is a local approximation for the two point function of $\Delta n_{\chi_i}$, the noise, which should also include the effects of the decay rate of the particles. The expression is given by~\cite{trapped}
\be\langle\Delta n_{\chi_i}(t,{\bf k})\Delta n_{\chi_j}(t',{\bf k}')\rangle\simeq(2\pi)^3\delta^{(3)}({\bf k}+{\bf k}')\frac{\delta_{ij}\kappa^3 e^{-2\Gamma_\chi|t-t'|}}{a^{3/2}(t) a^{3/2}(t')}\theta(t-t_i)\frac{a^3(t_i)}{a^3(t)}\theta(t'-t_j)\frac{a^3(t_j)}{a^3(t')}.\ee
This kernel sources $\delta\phi$ only through the integral in cosmic time of the Green's function. We can then approximate this equation by (see Eq. 3.20 in \cite{trapped})
\be  \langle\Delta n_\chi(t,{\bf k})\Delta n_\chi(t',{\bf k}')\rangle\simeq  (2\pi)^3\delta^{(3)}({\bf k}+{\bf k}')\frac{\delta(t-t')}{a^{3}(t)}\frac{\kappa^3 N_{\rm hits}}{2\Gamma_\chi}\equiv \nu_{\chi}(2\pi)^3\delta^{(3)}({\bf k}+{\bf k}')\frac{\delta(t-t')}{a^{3}(t)},
\ee
with
\beq
\label{nunhits}
\nu_{\chi} \equiv \frac{\kappa^3 N_{\rm hits}}{2\Gamma_\chi},
\eeq
and $N_{\rm hits}\sim\dot{\phi}/(H\Delta)$. The computation of the power spectrum then follows as in sec. \ref{power}.\\

Finally let us add a few words on our assumption that $\varphi$ is a light field. If they were too light one might wonder whether they could contribute
to the density fluctuations. Notice that  since we require $H \ll \Gamma_{\chi_i} < M_{\chi_i}(\tau_{\chi_i})$,
we do not necessarily need $m_\varphi$ to be as light as Hubble, in fact $m_\varphi \lesssim M_{\chi_i}(\tau_{\chi_i})/2$ would be just fine. In this case they quickly  redshift away. (Let us stress that even if they were effectively massless they could still be irrelevant for the late time curvature perturbations.)

\subsubsection{Non-Gaussianities}

To compute non-Gaussianities we follow \cite{trapped} and replace $\phi$ by $\phi+\delta\phi_1+\delta\phi_2$ in Eq. (\ref{inflocal}), and expand to linear and quadratic order in $\delta\phi_2$ and $\delta\phi_1$ respectively. To this purpose we need to compute particle creation in a time dependent background to second order in the perturbation, namely $\delta^{(2)} n_\chi$. In Eq. (\ref{pertinflocal}) we neglected the space variation of the field in the computation of $\delta n_\chi$, but at the non-linear level those gradient terms will provide us with the largest non-Gaussianities. (Recall the term proportional to $\gamma (\partial_i \pi)^2$ in Eq. (\ref{ocaldot}) is the one that induces the largest effects since $k_\star \sim \sqrt{\gamma H}$.)

Given that the $\calo_i \equiv \chi_i^2$ are scalars, as we argued in sec. \ref{nonlrep}, we are then able to extend the result in Eq. (\ref{gnchi}) to a spatially varying field, namely $\dot\phi \to n^\mu \partial_\mu \phi$ with $n_\mu \sim \partial_\mu \phi$, hence
\beq
\label{gdotp}
\frac{(g |\dot{\phi}|)^{5/2}}{\Gamma_{\chi}|\Delta|(2\pi)^3} \to \frac{g^{5/2}\left( \sqrt{-(\partial\phi)^2}\right)^{5/2}}{\Gamma_{\chi}|\Delta|(2\pi)^3}.
\eeq
(This is analogous to the covariant version of Schwinger pair production in an electromagnetic field with a purely electric background  $E \sim \dot\phi$, which turns out to be a scalar function of ${\cal F} = F_{\mu\nu}F^{\mu\nu}$ and ${\cal B} = F_{\mu\nu} F^\star_{\mu\nu}$.) 

One might worry about the validity of this procedure once we add inhomogeneities. In our example the particles are created in the state given by Eq. (\ref{ocupnum}), where (momentum) gradients
 of the unperturbed distribution are suppressed by factors of $1/(g|\dot\phi|)^{1/2} = 1/\kappa$. In fact, $\kappa$ is the scale that controls the validity of the local approximation. If we ignore these higher derivative (extrinsic curvature) terms, which are suppressed by $k_\star/\kappa$, we can then choose coordinates such that $\tilde t = t+\pi$. Hence for equal $\tilde t$ surfaces we get $n_\chi\left(\partial_{\tilde t}\phi(\tilde t)\right)$ given by Eq. (\ref{nchim}). Then we just replace $\partial_{\tilde t} \to n^\mu \partial_\mu$ for any coordinate system, with $n^\mu \sim g^{\mu\nu}\partial_\nu \tilde t$. To incorporate all type of corrections we also need to include terms involving the extrinsic curvature, $\delta K_{\mu\nu}$, and so on.\\
 
From the expression in (\ref{gdotp}) we already obtain the terms (with $\pi = \delta\phi/\dot\phi$)
\beq\label{pertlocal}\ddot{\pi}_2 - c_s^2 \partial_i^2\pi_2+(3H+\gamma)\dot{\pi}_2+ \gamma \left(\alpha \dot{\pi}_1^2 -\frac{1}{2}(\partial_i \pi_1)^2\right)+\ldots ={\rm Noise},\eeq
with $\alpha$ some numerical coefficients (which may vanish). In particular we get $\gamma(\partial_i \pi_1)^2$ at the non-linear level as advertised\footnote{Notice that for the model at hand we also need to include the perturbations in $\Gamma_\chi (\dot\phi)$. This will slightly modify the form of the function of $\sqrt{-\partial\phi^2}$ in Eq. (\ref{gdotp}), and ultimately the coefficient $\alpha$ in Eq. (\ref{pertlocal}).}, leading to $|f_{\rm NL}| \simeq \frac{\gamma}{c_s^2H} \gg  1$ for $\gamma \gg H$, similarly to the calculation in sec. \ref{Nolocalo2}.\\

There are yet other sources of non-linearities which can be induced and we have not incorporated (represented by the ellipses). Since $\ddot f_i \simeq \dot\phi^2$, one may worry about terms like  $\ddot f_i \pi \delta\bar\calo_i$, which do not respect the shift symmetry. (Moreover, they could induce non-negligible non-Gaussianities in the squeezed limit). However it turns out these terms actually cancel out, and the theory has an approximate shift symmetry, as we found already at linear order. To show this let us return to the interacting part of the effective action in Eq. (\ref{seftrap})
\beq
S^{\rm int}_{\rm trap}=\int d^4x\sqrt{-g} \sum_i\frac{ g^2(\phi-\phi_i)^2}{2}\chi_i^2 - V(\phi),\eeq
and perform the shift $\phi \to \phi+ c$. Assuming $V(\phi)$ obeys the slow roll conditions, the action will remain invariant provided we shift at the same time $\phi_j \to \tilde\phi_j \equiv \phi_j - c$. The new action then has the same form as the original one, with $\phi_j \to \tilde\phi_j$. Since, as we saw previously, the strength of the trapping mechanism depends on the velocity as it passes through the sweet spot, the physics stays essentially unchanged as long as we sum over a large number of periods. Hence the EOM will be (approximately) shift invariant. The crucial point is that the change $\phi \to \phi+c$ is compensated by a concurrent shift in $t_i$, i.e. $t_i \to t_i +\delta t_i$ with $\delta t_i \simeq c/\dot\phi$, which is ultimately summed over a large number of periods \cite{trapped}. Although this is not entirely obvious, the would-be breaking terms already canceled out at linear order. 

Something similar occurs in the model of \cite{trapped} before making the local approximation. In that case the contribution to the EOM from particle production reads \cite{trapped}
\beq
\hat m^2 \delta\phi + \int^t dt' \hat m^2 \left(\frac{5}{2} \dot{\delta\phi}(t')-3 H \delta\phi(t')\right) \frac{a^3(t')}{a^3(t)},
\eeq
which appears to have a `mass' term and violate the shift symmetry. However, that is not the case since the change in that term cancels against the shift in the second term in the integral, using $a(t) \sim e^{Ht}$. In the local approximations these pieces do not even show up, since they explicitly cancel each other out. To include these effects to all orders we need to be careful with the  sum over $f_i\calo_i$. At the end of the day the same cancellation occurs at second order and so on, such that we restore the shift symmetry as we argued before.

Finally let us stress that our results here apply for the case of a local response. In principle, as in the trapped inflation scenario of \cite{trapped}, large non-Gaussianities may also be produced via the non-locality in time of the response.

\subsection{Warm inflation}\label{secwarm}

Another example where we have dissipation/fluctuation is warm inflation \cite{berera1,berera2,warm}. In this case $\calo$ represents ADOF in thermal equilibrium at a given temperature $T$ mutually interacting with the inflaton $\phi$. Even though $\bar T^{\mu\nu}_\calo$ may be large enough to modify the dynamics of $\phi$, it remains sub-dominant with respect to the potential energy, $V \sim M_p^2H^2$, and one could hope inflation might not be drastically perturbed as long as $T \ll V^{1/4}$. The interesting aspect of this approach is the possibility of having $T \gg H$, which implies that thermal fluctuations dominate over vacuum effects. In warm inflation the evolution of the inflaton is governed by an equation of the type \cite{warm}
\beq
\label{warmd1}
\ddot \phi + (3 H+ \gamma) \dot\phi + V'(\phi) = K \xi,
\eeq
where $\xi$ represents the (Gaussian) noise with $\langle \xi\rangle=0$ and $K \simeq \sqrt{\gamma T}$. Provided $\gamma$ does not depend significantly on the temperature, the computation of the power spectrum follows the same steps as in sec. \ref{power}. The case $\partial_T \gamma \neq 0$ is more elaborate since one has to include perturbations in $T$ already at first order (see \cite{warmf33,berera11} for more details).

The dependence of $\gamma$ on the temperature varies with the different realizations of warm inflation. The most promising example is given by the so called two-stage decay model \cite{warm}. Similarly to the trapped inflationary case (although due to different mechanisms) the rolling inflaton produces some (ultimately heavy) particles ($m_\chi \gg T$) through a $g^2\phi^2\chi^2$ coupling, which subsequently decay into lighter degrees of freedom, $\varphi$'s. As we discussed before, this `catalyzing' mechanism may allow us to approximate the evolution equations by local dynamics. However, contrary to the model in \cite{trapped}, in warm inflation the decaying product is assumed to have thermalized. The friction coefficient can be then computed and scales like \cite{warm}
\beq
\label{chitcube}
\gamma(T) \simeq g^2 h^4 \left(\frac{m}{m_\chi}\right)^4 \frac{T^3}{m_\chi^2},
\eeq
where $h$ and $m=g\bphi$ enter in the coupling between $\chi$ and $\varphi$, e.g. $\calL_{\chi\varphi} = hm\chi\varphi^2$.\footnote{In \cite{reheating} it was argued that {\it assuming} warm inflation occurs and the potential obeys a series of slow roll conditions~\cite{warm}, the thermal hypothesis becomes plausible and $\rho_{\rm rad} \simeq (\gamma/H) \dot\phi^2$ has a stable equilibrium provided $\gamma(T) \simeq T^3$, as in Eq. (\ref{chitcube}).}
Clearly this effect is rather inefficient, since $m_\chi \gg T$. One possibility is to increase the number of $\chi$ particles such that more than one field is excited, or equivalently the number of decaying channels. Unfortunately, if we denote by $\cal N$ this `enhancement factor', the assumption that warm inflation occurs for sufficient e-foldings ($N_e \gtrsim 50-60$) with $\gamma(T) \gg H$ requires ${\cal N} \gtrsim 10^6$\cite{bastero,bueno}.\footnote{Even though one may be able to relax the condition on the curvature of the potential (the so called $\eta$-problem), some sort of tuning reappears in the form of very peculiar conditions on the matter content of the theory. Moreover, another important challenge for warm inflation model building is to keep under control the radiative and thermal corrections to the effective potential which could potentially ruin the slow roll conditions. (More so if one is going to assume $\gamma$ becomes sufficiently large \cite{linde}, in view of the previous requirement.) Supersymmetry may be invoked to tame the radiative corrections \cite{hall}. However, Supersymmetry is broken at finite temperature (and for non-zero vacuum energy), therefore some extra tuning may be required. A detailed account of the tuning of thermal inflation models lies beyond the scope our present paper.}\\

Non-Gaussianities in the warm inflationary scenario have been also computed in the literature. In most cases the contributions are small, i.e. slow roll
suppressed \cite{warmf1} (since they stem off the non-linearities induced by the potential). Following the reasoning of our previous section, to include
 other type of non-linearities in warm inflation one needs to generalize the $\gamma \dot\phi$ coupling to the case where $\phi \to \phi + \delta\phi$, which
we would now write as $\gamma u^\mu \partial_\mu \phi$ with $u^\mu$ some four vector. In principle there are different possibilities for $u^\mu$ depending
 on the model. For the two-stage case we can have either $n^\mu$ (as before) or $u^\mu_\chi$, the four velocity of the $\chi$ particles. The latter indeed would enter through the computation of the $\gamma$ factor in Eq. (\ref{chitcube}). However, in the limit $T \ll m_\chi$, the $\chi$ particles are non-relativistic (namely they are created essentially at rest) hence $u^\mu_\chi \simeq n^\mu$.\footnote{This is a consequence of the validity of a derivative expansion for heavy $m_\chi$'s, which also supports the local approximations. The replacement $u^\mu \to n^\mu$ agrees with the expressions in \cite{berera11}.} This suggests a coupling of the form $\gamma(T) n^\mu \partial_\mu \phi$ with $\gamma(T)$ given by Eq. (\ref{chitcube}), which following our previous arguments, would lead to non-linear effects of order $|f_{\rm NL}| \simeq \gamma(T)/H.$

\section{Conclusions}

In this paper we generalized the EFT for single field inflation \cite{eft1, Cheung:2007sv, Senatore:2009gt, Senatore:2009cf, Senatore:2010jy, Creminelli:2006xe, Bartolo:2010bj, Bartolo:2010di,dan2,strongc,dan2new} to include dissipative effects. Following \cite{dis1,dis2} we introduced ADOF described by composite operators in the effective action compatible with the symmetries of the long distance physics. Our general assumptions were: {\it i}) the existence of a preferred clock, {\it ii}) negligible (or vanishing) contribution from the ADOF to the curvature perturbations at late time, {\it iii})  the time scale for dissipation and fluctuation induced by the ADOF is much smaller than a Hubble time with negligible memory effects, and iv) the validity of an expansion in spatial derivatives. The last two allow us to use a local approximation for the dynamics of the perturbations. Under these conditions then we showed that in the strong dissipative regime the two point function is dominated by the noise, and that it can be significantly larger than the standard result for slow roll single field inflation. In particular
\beq
k^3\langle\zeta\zeta\rangle_{\calo} \simeq  {\nu_{\calo}}_\star \sqrt{\pi H_\star/\gamma_\star} \frac{H_\star^2}{2c_s^\star\left({c^\star_s}N_c\right)^2},
\eeq
which departs considerably from the Bunch-Davies result. The reason is twofold. First of all the size of the perturbations for the canonically normalized
 $\pi$ field ($\pi_c = \sqrt{N_c} \pi$) is larger than in the Bunch-Davies vacuum, but moreover because the normalization scale $N_c$ can be smaller than the
 value it takes in single field inflation without ADOF for fixed $\dot H$ and $c_s$, namely $c_s^2 N_c \leq  2M^2_p|\dot H|$.\footnote{Notice this implies
that the cutoff scale for higher dimensional operators in the EFT, $\Lambda_U$, (such as those obtained from $(1+g^{00})^n$ type of terms) may be lower than
 in the case of single field inflation without ADOF (for a given value of $\dot H, c_s$), where $\Lambda_U \sim M_p|\dot H| c_s^5$ for $c_s \ll 1$ \cite{eft1,dan2}.
However, keep in mind the our EFT description breaks down when $E \gtrsim \Gamma_\calo$, hence the `lowering' of $\Lambda_U$ is only meaningful in cases where
$\Lambda_U \lesssim \Gamma_\calo$.} This is even more transparent if we assume the FD theorem applies, or formally define $T_\calo$ as $\frac{\nu_{\calo}}{N_c \gamma}$, in which case
\beq
\label{pzetaw}
k^3\langle\zeta\zeta\rangle_T \simeq  \sqrt{\pi\gamma_\star H_\star} \frac{T_\calo H_\star^2}{2c_s^\star\left({c^\star_s}^2N_c\right)} \simeq \frac{c_s k_\star T_\calo H_\star^2}{\Lambda_c^4},
\eeq
with $\Lambda_c^4 \equiv c_s^3 N_c$, similarly to what happens in warm inflation \cite{berera1,berera2,warm}. Then, introducing $\Lambda_b^4\equiv 2M_p^2|\dot H| c_s \geq \Lambda_c^4$, we get\footnote{One might be tempted to identify  $c_s^3 N_c$ with the symmetry breaking scale $\Lambda_b^4$ \cite{dan2}, but this would not be correct. The reason is that in order to define the scale at which time translations are broken one has to properly account for the contributions to $\bar T^{\mu\nu}$
 stemming from all degrees of freedom, including the $\calo$'s. Following the procedure outlined in \cite{dan2}, one can easily show that $\Lambda^4_b$ remains at $2M^2_p|\dot H|c_s$. (This is actually not that surprising, since the normalization of the Goldstone boson $\tilde \pi$ did not change, as show in Eq. (\ref{orunit}).)} 
\beq
\frac{\langle\zeta\zeta\rangle_T}{\langle \zeta\zeta \rangle_{\rm BD}} \simeq \left(\frac{c_s k_\star T_\calo}{H_\star^2}\right)\left(\frac{\Lambda_b}{\Lambda_c}\right)^4 \gg 1,
\eeq
already for $T_\calo \geq H$ when $\gamma \gg H$. The factor of $c_s k_\star T_\calo$ in the numerator can be understood as follows. The energy density for $\pi$ is given by $(\tilde\partial \pi_c)^2 \simeq {\tilde k}_\star^2\pi_c^2$ (where $\tilde\partial$ represents the derivative with respect to $\tilde x^i = x^i/c_s$), which goes like ${\tilde k}^3_\star T_\calo$ (a.k.a. Rayleigh-Jeans law). Hence $\pi_c^2 \simeq  {\tilde k}_\star T_\calo = c_s k_\star T_\calo$, which we compare with the quantum noise in the Bunch-Davies vaccum given by $H_\star^2$.

As a result it is no longer the case that the power spectrum provides us the values of $H_\star$ and $\epsilon_\star$ as in the standard scenario,
 but rather with a set of new parameters: $\gamma, \nu_{\calo}, N_c$ (or $T_\calo = \nu_{\calo}/(N_c \gamma)$). Therefore, the tensor/scalar ratio would no longer give us a {\it clean} measurement of the energy scale of inflation. (Also because the ADOF might significantly contribute to gravitational wave production \cite{gwssz}.)\\

We were also able to identify some specific signatures for non-Gaussianities. In particular, we showed that the level of non-Gaussianities can potentially be much larger than in the case of single field inflation without ADOF by a factor of $\gamma/H \gg 1$. Our main observation was the following:
 in order to dissipate we require $\dot\phi$'s in the EOM, however, we need not only perturbe $\phi$ but also `the dot', since the equal time surfaces are fluctuating. Hence, the non-linear realization of time diffeomorphisms requires the following combination
 \beq F(\partial_t\phi) \to  F(n^\mu \partial_\mu \phi) \to F(\dot\phi \sqrt{-g^{00}}),\eeq
which not only induces $\gamma \dot\pi$ dissipation, but also (among others) a non-linear term: $-\gamma(\partial_i \pi)^2$. (Notice the relative sign is dictated by the non-linear realization of the symmetry.) Since horizon crossing happens at $c_s k_\star \simeq \sqrt{\gamma H}$, this type of non-linear coupling leads to \beq  \label{fnlconc} \frac{\gamma(\partial_i \pi)^2}{c_s^2\partial_i^2\pi} \sim f_{\rm NL}\zeta \to |f_{\rm NL}| \simeq \frac{\gamma}{c_s^2 H} .\eeq
The shape is plotted in Fig.~\ref{shapegoonoloc}, and peaks at the equilateral configuration. (This is not surprising given the fact that the
non-linearities involve derivatives of $\zeta$.) However, there is also a significant contribution at folded triangles  $x_1=1, x_2\simeq x_3\simeq 1/2$. Other non-linear terms may depend explicitly on the noise, such as $\delta\dot{\calo}_S \dot\pi$, and are plotted in Fig. \ref{ngg10}. Despite the fact that it scales with a single power of $\dot\pi$ one can show that its contribution in the limit $k_L \ll k_S$ is suppressed by $(k_L/k_S)^2$ with respect to the local shape, in agreement with the results in \cite{creminelli11}. We will analyze the squeezed limit and consistency conditions in the presence of dissipation in future work.\\

In this paper we also studied specific realizations of the type of operators introduced in the EFT and the matching procedure. In particular we analyzed a local version of trapped inflation where the produced particles decay after they are created, which leads to (approximately) localized response functions.  We showed how the term $\gamma (\partial_i \pi)^2$ gets generated, with the subsequent $\gamma/(c_s^2H)$ imprint on $|f_{\rm NL}|$. Crucial aspects of the model include: {\it i}) The ADOF responsible for dissipation do not contribute to the density perturbations at late time, {\it ii}) The emergence of a shift symmetry at the level of the perturbations, and {\it iii}) The response functions were predominately sensitive to the preferred clock $\phi$, whose fluctuations uniquely control the end of inflation, via $n_\mu \simeq \partial_\mu (t+\pi)$.

Intuitively, the necessary gradients of $\pi$ appear as a result of the fluctuations of the clock, the field $\phi$ itself, which sets the equal time surfaces where the unperturbed computation is assumed to hold to a good approximation. The derivative expansion remains valid as long as the typical length scale for the variation of the extrinsic curvature (of equal time surfaces) is larger than the typical wavelength of the produced particles, namely $1/\kappa \ll 1/k_\star$. (Our conclusions also apply to the two-stage model of warm inflation, provided one succeeds in producing sufficient e-foldings while having $\gamma \gg H$ in a consistent fashion.)

One might wonder about the possibility of having a `second clock' controlling the response functions for the ADOF. As long as we are only concerned about effects on the dynamics of the one clock driving inflation (assuming this second clock produces negligible direct contributions to $\zeta$), one can incorporate its presence in the $\calo$-system by replacing $n^\mu\partial_\mu \to u^\alpha_\calo \partial_\alpha$. We will study this in more detail in future work.\\

In a nutshell, departing from the vanilla single field scenario opens new possibilities which may well be realized in nature. Once again, the EFT machinery is a wonderful tool to reduce the plethora of conceivable realizations to a theory of low energy degrees of freedom coupled to a set of composite operators whose correlation function encode all the information about the dissipation/fluctuation properties of each specific model. In our case we reduced the number of additional parameters to three: $\gamma, \nu_{\calo}, N_c$. (Also $\Gamma_\calo$ and $M_\calo$, controlling the validity of local approximations.) By taking the ratio between $\zeta$-correlation functions, such as the two and three point functions, we manage to cancel out most of our ignorance on the underlying dissipative mechanism, thanks to the link between different $n$-point functions induced by the symmetries. In this fashion we were able to show (assuming the noise is Gaussian) that the bispectrum peaks at equilateral configurations and moreover $|f^{\rm eq}_{\rm NL}| \sim \frac{\gamma}{c_s^2H}$ for a vast class of models. 
 
 A detection of the generic type of signatures we discussed in this paper, as a result of incorporating dissipative effects during inflation, would increase our understanding of the dynamics of the early universe and also lead us towards a more precise description of the inflationary epoch.
 
\begin{center}
{\bf Acknowledgements}
\end{center}
It is our pleasure to thank Daniel Baumann, Esteban Calzetta, Paolo Creminelli, Raphael Flauger, Dan Green, Sean Hartnoll, Lam Hui, Fernando Lombardo, Diego Mazzitelli, Alberto Nicolis and Eva Silverstein for helpful discussions. This work was supported by: Universidad de Buenos Aires, CONICET and ANPCyT (DLN); NSF grant AST-0807444, DOE grant DE-FG02-90ER40542 and NASA grant NNX10AH14G (RAP); NSF grant PHY-0503584 (LS);  NSF grants PHY-0855425, AST-0506556 \& AST-0907969, and by the David \& Lucile Packard and the John D. \& Catherine T. MacArthur Foundations (MZ). DLN is grateful to the Institute for Advanced Study and Stanford Institute for Theoretical Physics for hospitality over the course of
this project. RAP, LS and MZ appreciate the hospitality of the CERN theory group during the final stages of this work.

\appendix

\section{Examples with Ohmic behavior}\label{app0}

Let us consider the well-known case in which the system (described by $\pi$) is linearly coupled to a ``bath" of harmonic oscillators, such that  \cite{weiss,calzetta}
\beq
\label{Ltot}
\calL_{\rm tot} =  \calL_\pi(\pi,\dot\pi) + \calL_{\rm int}(q,\pi) + \calL_q(q,\dot q),
\eeq
with
\beq
\calL_\pi = \frac{1}{2} \dot\pi^2 - \frac{1}{2}\omega_0^2 \pi^2,~ \calL_q = \sum_\alpha \frac{1}{2} \dot q_\alpha^2 - \frac{1}{2}\omega_\alpha^2 q_\alpha^2,~\calL_{\rm int} = -c_\alpha \sum_\alpha \pi q_\alpha.
\eeq
Then the EOM become
\bea
\ddot\pi +\omega_0^2 \pi + \sum_\alpha c_\alpha q_\alpha &=& 0 \label{eompia}\\
\ddot q_\alpha +\omega_\alpha^2 q_\alpha + c_\alpha\pi &=& 0 \label{eomqa}.
\eea
Plugging the solution to Eq. (\ref{eomqa}) back into Eq. (\ref{eompia}) we obtain an equation of the so called Langevin form:
\beq
\label{eom2n}
\ddot\pi +\omega_0^2 \pi +  \int dt' \tilde\gamma(t-t')\pi(t') = J(t),
\eeq
where
\beq
J (t)= -\sum_\alpha c_\alpha q^{\pi=0}_\alpha(t),
\eeq
with $ q^{\pi=0}_\alpha (t)$ representing the classical trajectories of the oscillators  in the absence of $\pi$ and
\beq
\label{gammattp}
\tilde\gamma(t-t') \equiv \sum_\alpha c_\alpha^2 G^\alpha_{\rm ret}(t-t').
\eeq
In the above expression we introduced
\beq
G_{\rm ret}^\alpha(t-t') = \theta(t-t') \frac{\sin(\omega_\alpha(t-t'))}{\omega_\alpha},
\eeq
the retarded Green's function for an harmonic oscillator of frequency $\omega_\alpha$.
If we now assume random initial conditions, taken independently for each oscillator in the environment, we can consider $J(t)$ as a stochastic variable.
Then, as $\pi$ is being affected by the `noise' induced by  $J(t)$,  its dynamics acquires a stochastic character.
In the jargon of particle physics, we have integrated out the bath of oscillators and obtained an effective EOM for the $\pi$ field, with a generalized friction term plus noise. (However, as we emphasized in this paper, this EOM does not derive from a Lagrangian of the form $\calL(\pi,\dot\pi)$.)\\

As it stands the EOM for $\pi$ is non-local (unless very particular properties for the $c_\alpha$ coefficients are assumed \cite{weiss, calzetta}). Nevertheless, as we argue below, there are situations where we get
\beq
\label{imgam2}
{\rm Im} \tilde\gamma(\omega) \simeq  \gamma \omega,
\eeq
with $\gamma$ a constant, such that the EOM takes the desired form
\beq
\label{eom1n}
\ddot\pi + \gamma\dot\pi+ \omega_0^2\pi = J
\eeq
with $\langle J\rangle =0$ (and after we redefine $\omega_0$ to include a renormalization piece).\\

A Lagrangian description of the sort of Eq. (\ref{Ltot}) arises in the so called Rubin's model, where a heavy mass $M$ is coupled to a half-infinite chain of harmonic oscillators of mass $m$ and spring constant $m\omega_R^2/4$, after diagonalization ($\omega_R$ is the highest attainable frequency) \cite{weiss}. In such case one has \cite{weiss}
\beq
\label{imrubin}
{\rm Im} \tilde\gamma_R \simeq \omega~\frac{m \omega_R}{M} \sqrt{1-\omega^2/\omega_R^2}~\theta(\omega_R-\omega),
\eeq
which is of the type in Eq. (\ref{imgam}) for $\omega \ll \omega_R$ with $\gamma \sim (m/M)\omega_R$.\footnote{Notice the asymptotic behavior is given by $\gamma_R(t) \sim \frac{m}{M} \sqrt{\frac{\omega_R}{t^3}}\sin\omega_R t$ for $t \gg \omega_R^{-1}$, which is milder than in Eq.~(\ref{drude}).}\\

The reason non-local effects make appearance in the EOM relies on the ability of the environment to back react in our system. Hence in order to have a local approximation, we should consider a situation where the energy of the system is damped into the environment and the latter has a negligible back reaction effect (or it becomes important on a time scale much longer than the length of the experiment).

As an example, and in the spirit of Rubin's model, let us take the case of a ring of mass $M$ attached to a half-infinite rope. If we denote by $y(x,t)$ the height of the rope as a function of $x>0$ and time, and place the ring at $\pi(t) \equiv y(0,t)$, the EOM read
\bea
\frac{\partial^2 y}{\partial x^2} &=& \frac{\partial^2 y}{\partial t^2} \\
M \ddot \pi &=& F \left(\frac{\partial y}{\partial x}\right)_{x=0}
\eea
where $F$ is the tension per unit of length on the rope (and we work in units where the speed of propagation is taken to be one). From the wave equation we know
\beq
\left(\frac{\partial y}{\partial x}\right)_{x=0}
= \pm \left(\frac{\partial y}{\partial t}\right)_{x=0},
\eeq
and choosing boundary conditions such that only outgoing waves are allowed (i.e. $y(x,t) = f(x-t)$) we obtain
\beq
\ddot \pi + \gamma \dot \pi =0,
\eeq
with $\gamma = F/M$. We can now simply add a spring of frequency $\omega_0$ attached to the ring to return to Eq. (\ref{eom1n}), in this case with $J=0$. (Notice that the rope represents the continuum limit of Rubin's model, where we take $m\to 0$ and $\omega_R\to\infty$ while keeping  $m\omega_R$ finite, so that Eq. (\ref{imrubin}) leads to a constant $\gamma_R$ for all times.)

In a more realistic setting we may imaging fixing the rope (now of length $L$) at an end, such that waves will bounce back on a time scale of order $t_B\sim L$, introducing non-local effects. In such scenario, and as long as we are interested in time scales $t \ll t_B$, our local approximation remains valid.

\section{The optical theorem}\label{appA}

Even though in this paper we deal with dissipation, our results are still consistent with unitarity. To make the connection more transparent let us consider the forward scattering amplitude for the $\pi$ particles, which we depict in Fig. \ref{fig1} as a `self-energy' diagram, with  $\pi$ propagators  represented by the wavy lines.

\begin{figure}[h!]
    \centering
    \includegraphics[width=6.5cm]{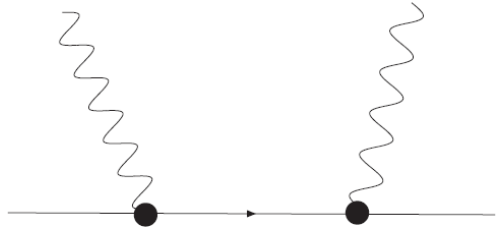}
\caption{The wavy lines correspond to $\pi$  and the dark interactions are insertions of $\calo\pi$ couplings. The line connecting the dots represents Feynman's time order product of Eq. (\ref{tfeynm}).}
\label{fig1}
\end{figure}

As it turns out, this amplitude is proportional to the time order product
\beq
\label{tfeynm}
\langle T (\delta\calo(x)\delta\calo(y))\rangle,
\eeq
also known as Feynman propagator, $iG_{\rm F}(x-y)$, which appears after we integrate out $\delta\calo$ in the usual path integral formalism with $\calL_{\rm int} = \delta\calo \pi.$ Unitarity, in the form of the optical theorem, tells us that the imaginary part of this amplitude must be related to the total power loss \cite{dis1,dis2}. Hence, using
\beq
{\rm Im}{\cal A}_{\pi\to\pi} \simeq {\rm Im}\, G_{\rm F}(\omega)
\eeq
and (for $\omega>0$)
\beq
\label{imgimf}
{\rm Im}\,G_{\rm F}(\omega) = {\rm Im}\,G_{\rm ret}^\calo(\omega),
\eeq
multiplying by a factor of $\omega$ (to go from rates to energies) we obtain
\beq
\frac{dE}{dt} \simeq \omega{\rm Im}\,{\cal A}_{\pi\to\pi} \simeq \omega{\rm Im}G_{\rm ret}^\calo(\omega).
\eeq
If we require this expression to match the condition \beq \frac{dE}{dt} = \gamma \dot\pi^2, \label{dedtap}\eeq
that follows from a dissipative term of the form $\gamma\dot\pi$, we conclude (for $\omega>0$)
\beq
{\rm Im}G_{\rm ret}^\calo(\omega) \simeq \gamma \omega
 \eeq
as expected. In general the expression in Eq. (\ref{dedtap}) may be more elaborate, but the procedure generalize to any function of $\omega$ \cite{dis1,dis2}.\\

The reader may be puzzled about the appearance of the Feynman propagator rather than the retarded propagator. However Feynman boundary conditions can be used to relate the imaginary part of self-energy diagrams with the total radiated power (or energy loss), similarly to what is done in the EFT for gravitational radiation of \cite{nrgr2,nrgr2p,nrgr2p2,radgo,eftrev}. The crucial point is that the boundary conditions are so chosen to ensure `in' and `out' vacuum states for the $\delta\calo$'s (e.g. no external gravitational radiation), but at the same time producing an imaginary part which precisely account for the total radiated power. Another way to see this is to notice that the total rate induced by the coupling $\delta\calo \pi$ is proportional to (ignoring factors of ${\bf k}$ for the sake of argument)
\be
\sum_N    \langle \pi,0|\pi\delta\calo(t)|0,N\rangle \langle \pi,0| \pi\delta\calo(0)|0,N\rangle^\star = \sum_N  \langle 0|\delta\calo(0)|N\rangle \langle N|\delta\calo(t)|0\rangle = \langle 0| \delta\calo(0)\delta\calo(t)|0\rangle,
\ee
which is nothing but the sum over the square of the emission amplitudes ${\cal A}_{\pi \to N}$, for all $N$ possible (intermediate) states. Then, using the relation (valid for $\omega>0$)
\beq
\int dt\, e^{i\omega t} \langle 0| \delta\calo(0)\delta\calo(t)|0\rangle = 2\,{\rm Im}\; i \int dt\, e^{i\omega t} \langle 0| T(\delta\calo(0)\delta\calo(t))|0\rangle,
\eeq
we reproduce our previous result (after multiplying by a factor of $\omega$). This is nothing but the optical theorem at work. What turns out to be a bit more subtle is how to obtain the correct retarded boundary conditions of Eq. (\ref{dret0}) from the path integral approach. However, this is possible in the so called `in-in' formalism \cite{calzetta, weiss, Zhang}. (See \cite{radre} for a discussion in the case of gravitational radiation reaction.)

\section{Retarded Green's function} \label{appB}

In this appendix we provide some basic features of retarded Green's functions, and in particular we discuss an example where ${\rm Im}G^\calo_{\rm ret} \simeq \omega$ as we used throughout the paper.

Let us start with some axiomatic properties for the retarded Green's functions and let us work at zero spatial momentum.  First of all, from causality we know $G_{\rm ret}^\calo(\omega)$ is an analytic function of $\omega$ for ${\rm Im}\, \omega>0$. Moreover, the imaginary part is odd in $\omega$:
\beq
\label{prop2}
{\rm Im}\,G_{\rm ret}^\calo(-\omega)=-
{\rm Im}\,G_{\rm ret}^\calo(\omega).
\eeq
Also, the real and imaginary parts are related via Kramers-Kronig relations,
\bea\label{kk}
\text{Re}\,G_{\rm ret}^\calo(\omega) & = & P \int_{-\infty}^\infty \frac{d\omega'}{\pi} \frac{\text{Im}\,G_{\rm ret}^\calo(\omega')}{\omega'-\omega}, \\
\text{Im}\,G_{\rm ret}^\calo(\omega) & = & - P \int_{-\infty}^\infty \frac{d\omega'}{\pi} \frac{\text{Re}\,G_{\rm ret}^\calo(\omega')}{\omega'-\omega},
\eea
where $P$ stands for the principal value. This implies
\beq
\label{prop1}
\int_{-\infty}^{+\infty} d\omega' \frac{{\rm Im}\, G_{\rm ret}^\calo(\omega')}{\omega'} < \infty,
\eeq
from which we obtain
\beq
{\rm Im}\,G_{\rm ret}^\calo(\omega \to 0) \to 0. \label{wtozero}
\eeq
Notice that Eq. (\ref{wtozero}) precludes the existence of a pole at $\omega=0$, however, it does not rule out the behavior of Eq. (\ref{tildG}) away from the origin, as we show in appendix \ref{mag00}.\\

Let us now study the example of electric conductivity. As it is well known  the relationship between the current and electric field in a conductor is given by 
\beq
{\bf j}_e = -i \omega \sigma(\omega) {\bf A}(\omega),
\eeq
with ${\bf E}(\omega) = -i\omega {\bf A}(\omega)$. Then using linear response theory one can show (Kubo formula)
\beq
{\rm Re}\, \sigma(\omega) = \frac{{\rm Im}\, G^j_{\rm ret}(\omega)}{\omega},
\eeq
where $G^j_{\rm ret}$ is the retarded Green's function for the electric current ${\bf j}_e$. In general this Green's function can be parameterized as \cite{forster}
\beq
G^j_{\rm ret}(\omega) = \frac{\alpha_j M_j(\omega)}{\omega + M_j(\omega)},
\eeq
with $M_j(\omega)$ a `memory function', and $\alpha_j$ a constant. Depending on the system, for some range of frequencies one can approximate $M_j(\omega) \sim i/\tau_j + O(\omega\tau_j)$ with $\tau_j$ the `memory time', such that
\beq
\label{Gjret}
G^j_{\rm ret}(\omega) \simeq -\frac{\sigma_j}{i\omega-\omega_D} \quad\Rightarrow\quad {\rm Im} G^j_{\rm ret}(\omega) \simeq \frac{\sigma_j\omega}{\omega^2 + \omega_D^2},
\eeq
with $\omega_D \simeq 1/\tau_j$, $\sigma_j = \alpha_j \omega_D$. For $\omega \ll \omega_D$ we obtain the behavior as in Drude's model in Eq. (\ref{drude}).

Notice that we also have Re $G^j_{\rm ret} \simeq \sigma\omega_D$ (which is obviously consistent with the dispersion relations of Eq. (\ref{kk})). This means, in principle, that we get a large correction in the real part of the Green's function. In cases where our ${\calo}$ operators have a Green's function of this form, this could potentially lead to a large mass for $\pi$. In our cases of interest we assumed there is a mechanism that forbids a large contribution such that $m_\pi$ remains $O(\epsilon)$, as in cases where there is an approximate shift symmetry.

\section{Decoupling limit}\label{appdec}

In single field inflation one can show that $\delta N$ and $\delta N_i$ are suppressed in the slow roll approximation, so that
 one may work with the theory of $\pi$ up to $O(\epsilon)$ effects, for models not {\it too close} to de Sitter~\cite{eft1}. Here we argue  that decoupling still occurs even
after we include ADOF, and make some general comments about the structure of the mixing terms. In this section we work in $M^2_p=1$ units
unless otherwise noted.\\

As we know, the full ${\bar T}^{\mu\nu}$ for the background takes the perfect fluid form
\beq
{\bar T}_{\mu\nu} = (\bar \rho_{tot}+\bar p_{tot}){\bar u}_\mu {\bar u}_\nu + \bar g_{\mu\nu} \bar p_{tot}
\eeq
(which follows from the isotropic and homogenous conditions). We will work in the Newtonian gauge where
\beq
g_{00} = -1-2\Phi,~ g_{0i}=0,~ g_{ij} = \bar g_{ij} (1-2\Psi).
\eeq

If we concentrate on first order scalar perturbations we have
\bea \label{stresslin}
\delta T_{ij} &=& -2\Psi \bar g_{ij} \bar p_{tot}  + \bar g_{ij} \delta p + \delta \tau_{ij}\\
\delta T_{00} &=& 2 \bar \rho_{tot} \Phi + \delta \rho \\
\delta T_{i0} &=&   - (\bar\rho_{tot}+\bar p_{tot}) \partial_i \delta u,
\eea
where we included a dissipative term $ \delta \tau_{ij}$, induced by the ADOF. (Keep in mind that the ADOF also enter in $\delta\rho, \delta p$.)
The EOM become \cite{weinberg}
\bea
\label{b6}
\partial_j\left(\delta p +2 \partial_0(\epsilon H^2\delta u) + 6\epsilon H^3\delta u + 2\epsilon H^2\Phi\right) +   \partial_i\delta\tau^i_j &=& 0\\
\delta\dot\rho + 3 H(\delta\rho+\delta p) + H\delta \tau^i_i + 2\epsilon H^2 \left(\frac{\nabla^2\delta u}{a^2} -3 \dot \Psi\right) &=& 0\label{b7}\\
\frac{\nabla^2\Psi}{a^2}-\frac{1}{2}\delta \rho + 3 \epsilon H^3 \delta u &=& 0\label{b8}\\
\frac{1}{a^2} \left( \partial_i\partial_j - \frac{1}{3} \delta_{ij} \nabla^2\right) (\Psi-\Phi)&=& \left(\delta\tau^i_j-\frac{1}{3}\delta_{ij} \delta\tau^l_l\right) \label{b9}
\eea
where we used $\bar \rho_{tot}+\bar p_{tot} = -2\dot H= 2H^2\epsilon$, ($\epsilon = -\dot H/H^2$). Let us assume for the moment that  $\delta \rho$, $\delta p$, $\delta u$ and  $\delta \tau_{ij}$
do not depend on the metric perturbations. Then, from these equations one can immediately note that the metric perturbations decouple
 in the $\epsilon \to 0$ limit. Moreover, the constraint equations (last two) give us the value of $\Phi$ and $\Psi$ in terms of $\delta \rho$, $\delta u$
and $\delta \tau_{ij}$. At the end of the day we end up with equations where the mixing with gravity is suppressed by $\epsilon$,
and moreover, is independent of the relationship between $\delta \rho$ and $\delta p$.\\

To analyze the relative importance of the contributions of these mixing terms, we can combine Eqs (\ref{b6} - \ref{b8}) to obtain ($k\neq 0$):
\bea
&&\delta\ddot \rho_k +3 H (\delta\dot\rho_k +\delta \dot p_k)-\frac{k^2}{a^2}\left(\delta p_k+ \frac{2}{3}\delta\tau_k\right)+ \frac{d}{dt}\left(H(\delta \tau^i_i)_k\right)-2\epsilon H^2 \delta\rho_k-3\epsilon H^2\delta p_k   \nn \\ && +2\epsilon H^2\delta\tau_k+10\epsilon H^2\frac{k^2}{a^2}\delta u-3\frac{a^2\epsilon H^2}{k^2}\left(\delta\ddot\rho_k + 2H \delta\dot\rho_k +4H^2\delta\rho_k\right)
+O(\epsilon^2)=0.
\eea

In this expression we replaced $\partial_i (\delta \tau^i_j)_k \to 2/3~\partial_j \delta \tau_k$, with $\delta\tau_k$ a scalar mode which
follows from the decomposition $(\delta\tau^{i}_{j})^{\rm TF}_k = (\hat k^i\hat k^j-\frac{1}{3}\delta_{ij}) \delta\tau_k + \ldots$ ($\hat k^2=1$),
and we used Eqs. (\ref{b8},~\ref{b9}) to solve for $\Phi$,
\beq
\Phi = \frac{a^2}{k^2} \left(\frac{1}{2}\delta\rho_k + \delta\tau_k\right) + O(\epsilon)\label{Phid}.
\eeq

Therefore the presence of gravity has two net effects. First of all, there is a ``mass'' term proportional to $\sqrt{\epsilon}H$; and secondly there are mixing factors, of order $\epsilon H^2/k_\star^2$
at horizon crossing (recall $k_\star=k/a(t_\star)$). Hence, as long as $\omega_\star \gg \epsilon^{1/2} H$ and $k_\star \gg \epsilon^{1/2} H$,
 we can ignore the mixing with gravity. For us, since $k_\star \sim \sqrt{\gamma H}/c_s > H$, this suppression is larger than the usual case upon noticing
\beq
{\rm mixing} \sim \frac{\epsilon_\star H_\star c_{s_\star}^2}{\gamma_\star} \ll 1 \label{mixingd}.
\eeq

So far we have assumed that terms involving the metric perturbations in  $\delta \rho$, $\delta p$, $\delta u$
and $\delta \tau_{ij}$ are negligible. To clarify this point let us take the example of single field inflation without ADOF \cite{eft1}. To linear order in the perturbations, the stress tensor obtained from the first contributions to the action defined in Eq. (\ref{act1}), i.e.\footnote{Note that the term $M^4_2(1+g^{00})^2$ allows for $c_s\neq 1$ (see Eq. (\ref{ncs})).}

\be \frac{1}{2} \int d^4 x \sqrt{-g}\{(\overline{p}-\overline{\rho}-(\overline{p}+\overline{\rho}) g^{00})+\,M_2^4 (1+ g^{00})^2\},  \ee
can be written as in Eq. (\ref{stresslin}), using
\begin{subequations}\label{var12}
\begin{align}
\delta\rho& =\dot{\overline{\rho}}\pi+(\overline{\rho}+\overline{p}+4 M_2^4)(\dot{\pi}-\Phi), \label{rho12} \\
\delta p& =\dot{\overline{p}}\pi+(\overline{\rho}+\overline{p})(\dot{\pi}-\Phi),\\
\delta u & =-\pi.
\end{align}
\end{subequations}
For these variables to be approximately independent of $\Phi$ we need  $\Phi \ll \dot\pi$.
Solving
for $\Phi$ using Eq. (\ref{Phid}) without the ADOF we obtain (in the slowly varying approximation and restoring $M_p$)
\beq
\Phi\sim \frac{a^2}{M_p^2k^2}\left(\bar\rho+\bar p + 4 M_2^4\right)\dot\pi . 
\eeq
Therefore the assumption  $\Phi \ll \dot\pi$  is self-consistent provided
\beq
\label{m2ks}
\frac{a^2}{M_p^2k^2} (\bar\rho+\bar p + 4 M_2^4) \ll 1 \quad \Rightarrow\quad k_\star \gg \frac{(\bar\rho+\bar p + 4 M_2^4)^{1/2}}{M_p}.
\eeq
Using $\bar\rho+\bar p \sim 2\epsilon M_p^2H^2$ we recover an expression similar to (\ref{mixingd}) (but without the factor  $c_{s_\star}^2$),
\beq
{\rm mixing} \sim \frac{\epsilon_\star H_\star }{\gamma_\star} \ll 1 \label{mixingd2}.
\eeq
Note that the above estimations do not apply  in the limit $\epsilon \to 0$, since the higher derivative contributions
can no longer be ignored. To analyze this case we add the term proportional to $\bar M_2^2$ in  Eq. (\ref{act1}), i.e.
\beq
-\frac{1}{2}\int d^4 x \sqrt{-g}\,\bar{M}_2^2 (\delta K_{\mu}^{\mu})^2.
\eeq
Setting $\overline{\rho}+\overline{p}\to 0$ and $c_s\to 0$ we get
\begin{subequations}
\begin{align}
\delta\rho &=4 M_2^4(\dot{\pi}-\Phi)-3\bar{M}_2^2 H\lb\frac{k^2}{a^2}\pi-3 (H\Phi+\dot{\Psi})\rb\\
&\simeq 4 M_2^4\dot{\pi}-3\bar{M}_2^2 H\frac{k^2}{a^2}\pi.
\end{align}
\end{subequations} (The second line is valid provided $\Phi \ll \dot{\pi}$ and $H\Phi+\dot{\Psi}\ll \frac{k^2}{a^2}\pi$.)
Using  again Eq. (\ref{Phid}) (without ADOF) we have
\beq
M_p^2\Phi\simeq \frac{a^2}{k^2}\delta\rho\simeq 4 M_2^4 \frac{a^2}{k^2}\dot{\pi}-3\bar{M}_2^2 H\pi .
\eeq
Therefore we see the assumption that we can neglect the dependence of $\delta\rho$ on the metric perturbations requires, not only $\bar{M}_2 \ll M_p$, but also~\cite{eft1}
\be\label{mixingM2}
 M_2^4 \frac{a^2}{k^2}\ll M_p^2 \quad \to\quad k_\star  \gg M_2^2/M_p.
 \eeq
In addition, this explains the apparent puzzle in the mixing expression of Eq. (\ref{mixingd}), that appears to exactly vanish in the de Sitter limit. (This we know is not the case, for example in the Ghost Condensate. See \cite{eft1} for more details.) Notice that the condition in (\ref{mixingM2}) does not make any reference to $c_s$, which is zero in the de Sitter limit. In general, for small $(\epsilon, c_s)$ the mixing with gravity will depend on which one is larger between $\bar\rho+\bar p$ and $M_2^4$, as in (\ref{m2ks}), and therefore in the near de Sitter limit the mixing may become enhanced with respect to $\epsilon$. 

The inclusion of ADOF does not modify the previous analysis considerably. Similar arguments can be made for vector and tensor modes.


\section{Comment on $\hat\calo\delta K^{\mu}_{\mu}$}\label{dodk}

Let us briefly analyze the importance of the terms with the extrinsic curvature. In particular we  compare  the operator  $s(t)\frac{\hat\calo}{M_K}\delta K^{\mu}_{\mu}$, given in Eq.~(\ref{extcalotr}), with $f(t)\calo_1$ of  Eq.~(\ref{inter1n}). For the sake of argument we take $\calo_1\sim\hat{\calo}$. By this we mean that for these operators all response and noise coefficients are of the same order. Then to linear order in $\pi$, from $f(t){\calo}_1$ we get $\dot{f}\pi\calo_1$, whereas from the one with the extrinsic curvature we have $\frac{s(t)}{M_{K}}\hat{\calo}\frac{\nabla^2}{a^2}\pi.$\\

Taking the ratio we see that if $k_\star$ satisfies \beq \label{ksmk} k_\star^2\ll~M_{K}\dot{f}(t)/s(t), \eeq we may be allowed to neglect operators with $\delta K$. As we have shown in this paper, the computation of the power spectrum and non-Gaussianities (due to the behavior of the retarded Green's function $G_{\gamma}$) is dominated by values for which, at freeze out, $k_\star\sim \sqrt{H\gamma}/c_s$. Therefore, to satisfy the condition in Eq. (\ref{ksmk}) we need  \beq \frac{s(t)\gamma H}{\dot f c_s^2} \ll~M_{K}. \eeq
Otherwise, the contributions from  $\calo\delta K^{\mu}_{\mu}$ could be important.


\section{Dissipative Green's functions in an expanding universe}\label{greenf}

In this appendix we discuss some properties of the Green's function relevant for the computation of non-Gaussianities in sec.~\ref{nongaus}, in particular $G_{\gamma}(x,y)=y^2 g_{\gamma}(x,y)$ ($x=-kc_s\eta$, $y=-kc_s\eta'$).
 In Fig.~\ref{GreGamma} we show some graphs of $G_{\gamma}(0,y)$ for three different values of $\gamma$. (For a fixed $x<(4+\gamma/H)^{1/2}$, \footnote{We keep the factor of 4 to recover the standard result in the case $\gamma\rightarrow 0$.} the behavior of $G_{\gamma}(x,y)$ is similar to the one shown for $x=0$.) We notice that $G_{\gamma}$ peaks at $y\simeq (4+\gamma/H)^{1/2}$, and as $\gamma$ increases its amplitude decreases, whereas its support increases.

\begin{figure}[h]
\includegraphics[width=8.5cm]{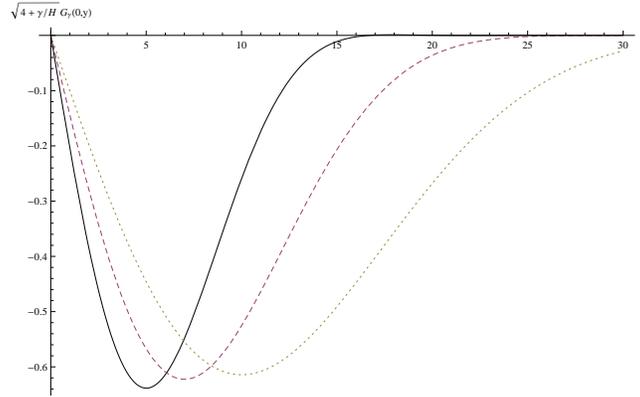}
\caption{The Green's function $G_{\gamma}(0,y)=y^2 g_{\gamma}(0,y)$  for three different values of $\gamma$: $4+\gamma/H=5^2$ (black), $4+\gamma/H=7^2$ (dashed) and $4+\gamma/H=10^2$ (dotted),  the first peak of  $G_{\gamma}$ is  around  $y=5,\,7,$ and $10$, respectively, supporting the fact that the location of the peak is at $\sqrt{4+\gamma/H}$.}\label{GreGamma}
\end{figure}

\begin{figure}[h]
\includegraphics[width=8.5cm]{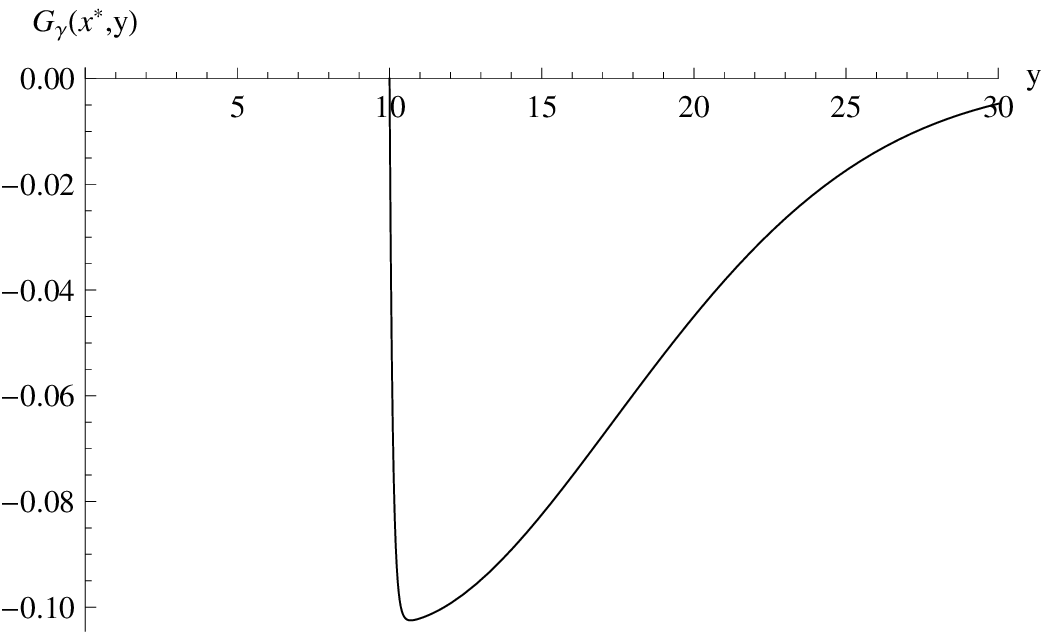}\includegraphics[width=8.5cm]{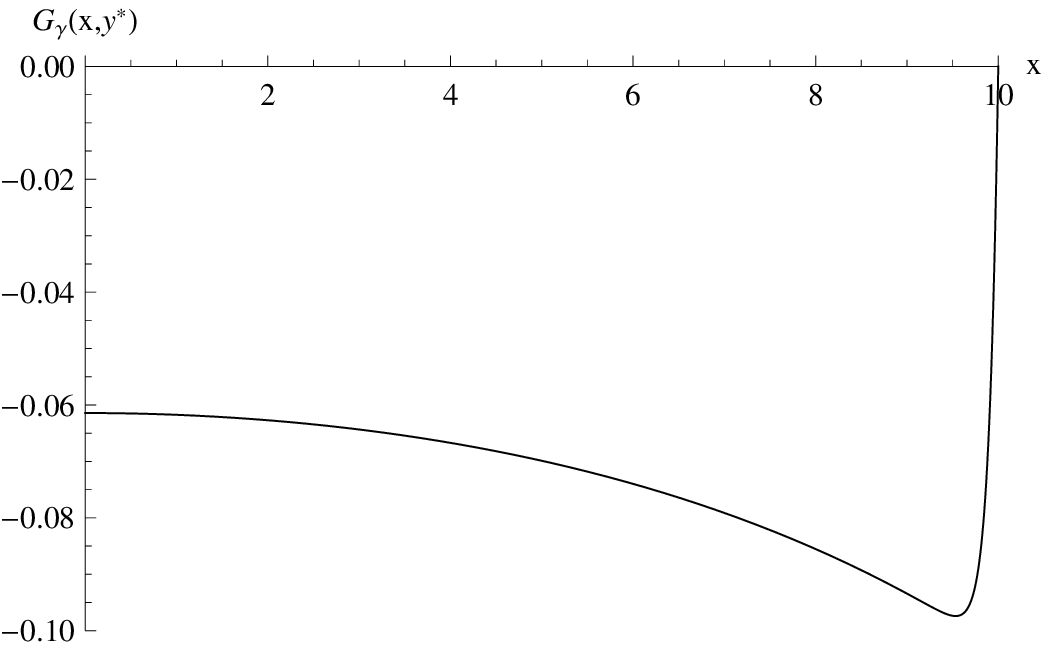}
\includegraphics[width=8.5cm]{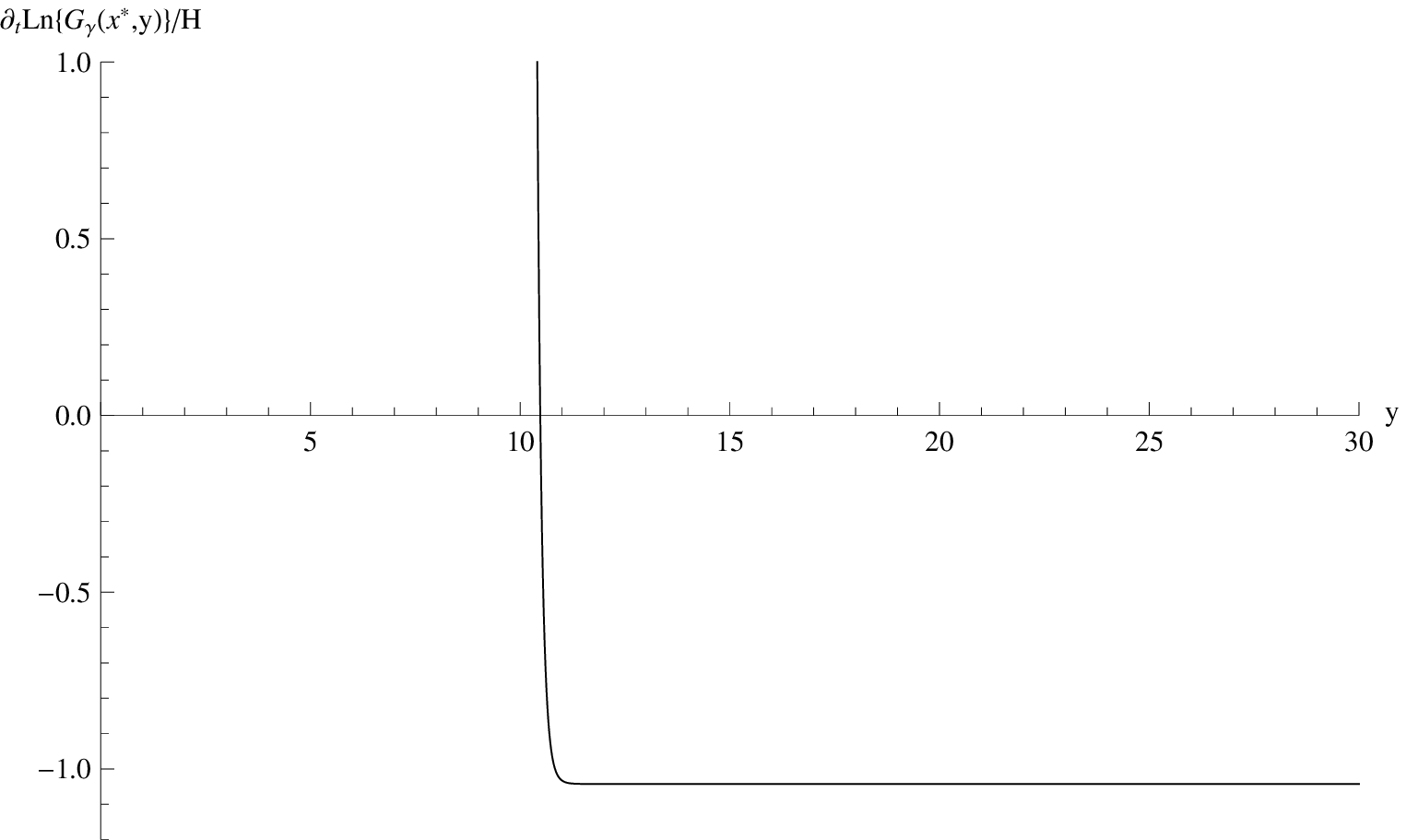}\includegraphics[width=8.5cm]{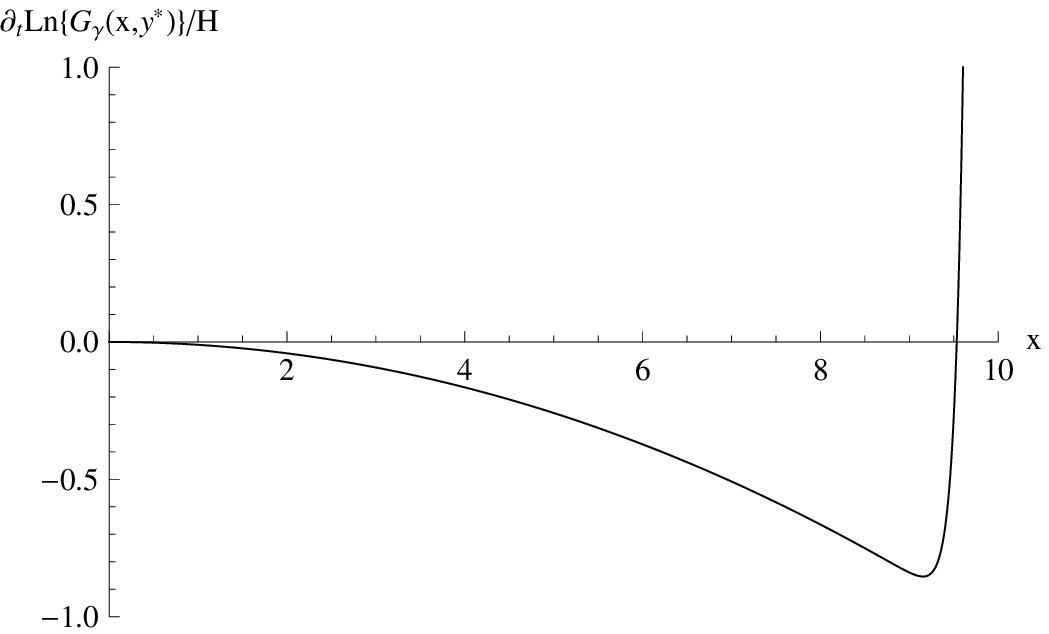}
\caption{The Green's function $G_{\gamma}(x,y)$ (on the top) and its temporal derivative, per Hubble time, $\partial_tG_{\gamma}(x,y)/H=-x\partial_xG_{\gamma}(x,y)$  (on the bottom); for  $x^\star=\sqrt{4+\gamma/H}$ as a function of $y>x^\star$ (on the left), and for  $y^\star=\sqrt{4+\gamma/H}$ as a function of $x<y^\star$ (on the right). All plots correspond to $x^\star=y^\star=10$.}\label{GreGamma2}
\end{figure}

For a particular value of $\gamma$, in Fig.~\ref{GreGamma2} we plot the Green's function $G_{\gamma}(x,y)$  and its temporal derivative per Hubble time, $\partial_tG_{\gamma}(x,y)/H=-x\partial_xG_{\gamma}(x,y)$, for  $x=\sqrt{4+\gamma/H}$ as a function of $y>x$, and for  $y=\sqrt{4+\gamma/H}$ as a function of $x<y$. To estimate the amplitude of integrals of  $g_{\gamma}$ it is useful to note  that
\be
\int_x^{+\infty} dy\, G_{\gamma}(x,y)=-1.
\ee
Then, for large values of $\gamma$ a rough estimation for $f_{\rm NL}^{\rm eq}$ can be obtained by counting each power of $y^2$ as $\gamma$, adding a factor of  $\gamma^{-1/2}$ for each $G_{\gamma}$ (i.e. $\gamma^{-3/2}$ for each $g_{\gamma}$) and a factor of $\gamma^{1/2}$ for each integral, with units made up by $\gamma/H$. Note that  having a time derivative per Hubble time of $G_{\gamma}$ cannot increase significantly $f_{\rm NL}^{\rm eq}$ (see Fig.~\ref{GreGamma2}). On the other hand, the contribution of an additional spatial derivative per Hubble length will increase $f_{\rm NL}^{\rm eq}$ by approximately a factor of $\gamma/c_s^2$ (since $k^2/(a^2H^2)=y^2/c_s^2$).


\section{Mixing}\label{mixing}

To study the effects of friction at leading order in $\pi$, in this paper we concentrated on the study of a generic
type of operator, i.e. $\calo\pi$, with $\calo$ including in principle a series of contributions (after integration by parts). However,
once we start adding higher order effects, we treated each of them separately. The reasons were both the fact that not all the terms generate the
same type of non-linearities, and also obviously for simplicity. In general however more than one operator will be present at the same time. Here we make a
few comments about the kind of effects that may appear as a result of including more than one
 type of ADOF concurrently.

\subsection{Scalars}\label{mixingscalars}

Let us start with scalar operators $\calo_A$, $A=1,\ldots N$. In general, in addition to the contributions of the response and noise given by the self-correlation functions, we will also have the ones given by the  mixed-correlations (e.g. $\langle\delta\calo^S_A\delta\calo^S_B\rangle$).  To be more precise, we are considering interaction terms of the form
\be S_{int}=-\int d^4x a^3(t) \calo_A F^A, \ee
where $F^B$ are taken as external forces that slightly disturb the dynamics of the ADOF associated to the $\calo$'s, and where repeated index are summed over. The linear response of $\calo_A$ due to the application of the external forces can therefore be written as
\be \delta\calo_A^R=-\int d^4x' a^3(t') G_{AB}^\calo(x,x') F^B(x'),\label{deltaO}\ee
and
\be G_{AB}^\calo(x,x')=i \theta(t-t')\langle[\delta\calo_A(x), \delta\calo_B(x')]\rangle.\ee
Additionally, we have the fluctuations of the ADOF, which are characterized by the two-point function of stochastic sources  $\delta\calo^S_A(x)$:
\be N_{AB}^\calo(x,x')= \langle\delta\calo^S_A(x)\delta\calo^S_B(x') \rangle.\ee

By concentrating in the case where a local approximation applies, as described in sec.~\ref{localapp}, we can write
\beq
\label{respoLoc}
 \delta\calo_A^R(t,{\bf k})\simeq \mu_{AB}(t) F^B(t,{\bf k})+\Gamma_{AB}(t)\frac{1}{a^{3/2}} \partial_{t}(a^{3/2}(t) F^B(t, {\bf k})),
\eeq
\beq \langle\delta{{\calo}}_A^S(t,{\bf k}_1)\delta{{\calo}}^S_B(t,{\bf k}_2)\rangle=(2\pi)^3\delta^{(3)}({\bf k}_1+{\bf k}_2)
{N}_{AB}^{\calo}(t,t'),\eeq
with
\be  {N}_{AB}^\calo(t,t')\simeq~ \nu_{AB}(t) \frac{ \delta(t-t')}{a^{3}(t)},\ee where $\mu_{AB}(t),\,\Gamma_{AB}(t)$ and $\nu_{AB}(t)$ are approximately time-independent.

\subsection{$ f(t)\calo_1$ and $\calo_2 g^{00}$}

For the sake of completeness, let us study the effects due to the mixing between $\calo_1$ and $\calo_2$, under the assumptions of
the local approximation described previously.
Note that if we ignore that mixing, the interaction $f_1\calo_1$ yields a term $\gamma\dot{\pi}$ in the EOM for $\pi$  (see sec. \ref{ftcalo}), while $\calo_2 g^{00}$
produces a higher derivative contribution  if we assume that both operators have a Green's function satisfying ${\rm Im}\,G(\omega)\propto\omega$ as $\omega\rightarrow 0$ (see sec. \ref{localo2}). Hence it is reasonable to consider the case where
 the term $f_1\calo_1$ dominates over $f_2\delta g^{00}\calo_2$ at linear order. This allows us to concentrate in a situation in which, at leading order
 in the corrections introduced by  $\calo_2$, we can neglect the contributions coming from the interaction
$\delta g^{00}\calo_2$ alone, and its effects enter only through the mixing with $\calo_1$. For example, this would be the case  if $\calo_1=\calo_2$ and
$f_2 H\ll\dot{f_1}$. So in practice we can work at leading order in $f_2$. Moreover, for simplicity, we  concentrate here only in the contributions obtained
 using the linear response approximation, although
as we have seen in sec.  \ref{localo2} the non-linear response contributions obtained from $ f_1\calo_1$ could produce the largest non-Gaussianities in this kind
of models\footnote{Note that from the analysis of sec. \ref{ftcalo}, if we have only the interaction $f_1\calo_1$ the level of non-Gaussianities obtained within the linear response approximation are negligible.}.

There are two types of mixed  response terms: one is given by how $f_1(t)$ (the force coupled to $\calo_1$) affects $\delta\calo_2$
 and the other by  how $f_2\delta g^{00}$ affects $\delta\calo_1$. Using the local approximation, these can be written as
\begin{subequations}
\begin{align}
\delta\calo_1^{R_{\rm mix}}&\simeq~ -f_2\tilde{\mu}_{12}\lp 2\dot{\pi}+\dot{\pi}^2-\frac{\partial_i\pi\partial_i\pi}{a^2}\rp\nonumber\\
&-f_2\Gamma_{12}\lp2\ddot{\pi}+2\dot{\pi}\ddot{\pi}-2\frac{\partial_i\pi\partial_i\dot{\pi}}{a^2}+2H\frac{\partial_i\pi\partial_i\pi}{a^2}\rp,\\
\delta\calo_2^{R_{\rm mix}}&\simeq~ \tilde\mu_{21}\lp\dot{f_1}\pi+\ddot{f_1}\frac{\pi^2}{2}\rp+\Gamma_{21}\lp\dot{f_1} \dot{\pi}+\ddot{f_1}\dot{\pi}\pi\rp,
\end{align}
\end{subequations} where $\tilde{\mu}_{AB}={\mu}_{AB}+3/2H\Gamma_{AB}$ and we assumed the existence of an emergent shift symmetry, as described in the main text. 

To linear order in $f_2$ the equation for $\pi_1$ becomes
\be
\ddot{\pi}_1+(3H+\tilde{\gamma})\dot{\pi}_1+\tilde{c}_s^2\frac{\nabla^2}{a^2}\pi_1=-\tilde{N}_c^{-1}
\left(\dot{f_1}\delta\calo_1^S-2f_2\frac{\partial_t(a^3\delta\calo_2^S)}{a^3}\right),
\ee where $\tilde{N}_c=N_c+2f_2\dot{f_1}(\Gamma_{21}-\Gamma_{12}))$,  $\tilde{c}_s^2={c}_s^2-2f_2\tilde{N}_c^{-1}\dot{f_1}c_s^2(\Gamma_{21}-\Gamma_{12})$,
and $\tilde{\gamma}= \tilde{N}_c^{-1}\left(\dot{f_1}^2\Gamma_{11}+2f_2\dot{f_1}(3H\Gamma_{12}-\tilde{\mu}_{12})\right)$. Note that  $\tilde{c}_s^2\tilde{N}_c=c_s^2N_c $
(recall we are working to linear order in $f_2$). To ease notation we omit tildes in what follows. The particular solution to the above equation is
\begin{eqnarray}
\pi_1(k,\eta)&=&\frac{\dot{f_1} k c_s}{N_cH^2}\int^{\eta}_{\eta_0} d\eta'g_{\gamma}(kc_s|\eta|,kc_s|\eta'|)\delta\calo_1^S({\bf k},\eta')\nonumber\\
&+&\frac{2 f_2 k c_s}{N_cH^2}\int^{\eta}_{\eta_0} d\eta'g_{\gamma}(kc_s|\eta|,kc_s|\eta'|)\frac{\partial_t \lp a^3(\eta')\delta\calo_2^S({\bf k},\eta')\rp}{a^3(\eta')},
\end{eqnarray}
and the resulting  power spectrum can be written as
\be
P_{\zeta}=P_{\zeta}^{f_2=0}\lp 1+6\frac{H f_2\nu_{12}} {\dot{f_1}\nu_{11}}\rp
\ee where $P_{\zeta}^{f_2=0}$ is the one obtained for $f_2=0$, which is given in Eq.~(\ref{pow1gamma}) where $\nu_\calo=\nu_{11}\dot{f_1}^2$. The source for $\pi$ reads
\begin{eqnarray}
J&=&N_c^{-1}\dot{f_1}(2f_2H\Gamma_{12}-f_2\tilde{\mu}_{12}) \frac{\partial_i\pi\partial_i{\pi}}{a^2}-2N_c^{-1} f_2\dot{f_1}(\Gamma_{12}-\Gamma_{21})\frac{\partial_i\pi\partial_i\dot{\pi}}{a^2}\nonumber\\
&+&2f_2N_c^{-1}\dot{f_1}\Gamma_{21} \dot{\pi}\frac{\nabla^2\pi}{a^2}+f_2N_c^{-1}\dot{f_1}(\tilde{\mu}_{12}-6H\Gamma_{21})\dot{\pi}^2+2f_2N_c^{-1}\dot{f_1}(\Gamma_{12}-2\Gamma_{21})\dot{\pi}\ddot{\pi}\nonumber\\
&-&\dot{f_1}\delta\calo_1^S-2f_2\frac{\partial_t\lp a^3\delta\calo_2^S\rp}{a^3}-2f_2\frac{\partial_t\lp a^3\dot{\pi}\delta\calo_2^S\rp}{a^3}+2f_2N_c^{-1}\frac{\partial_i\lp\delta\calo_2^S\partial_i\pi\rp}{a^2}.\label{mixsource}
\end{eqnarray}
Note that if $\calo_1=\calo_2$ the second term vanishes.\\

As we have learned from our computations in sec. \ref{nongaus}, the contribution of the non-linear terms with spatial derivatives become more important for generating non-Gaussianities than those with only time derivatives. Let us then analyze for simplicity only the first and the last term, which we expect to be among the leading ones.  Notice that the calculations for the first source term are exactly the same as the ones we have already performed in sec. \ref{Nolocalo2}, in that case for the first term on the RHS of Eq.~(\ref{pieqnl}). In fact, it can be written as
\be J_k^1=-s_1\gamma\frac{\partial_i\pi\partial_i{\pi}}{a^2},\ee
with
\be
s_1=-\frac{2f_2 H}{\dot{f_1}}\frac{\Gamma_{12}}{\Gamma_{11}} +\frac{f_2\tilde{\mu}_{12}} {\dot{f_1}\Gamma_{11}}, \ee
where we have $s_1\ll 1$ consistently with our linear expansion in $f_2$ after taking, for simplicity, $\gamma_{AB}\sim H \mu_{AB}$. From eq.~(\ref{fnlgoonoloc}) we obtain
\beq f_{\rm NL}^{\rm eq}\simeq\frac{s_1\gamma}{2H c_s^2},\eeq
which could be still large for $\gamma\gg H$, even when $c_s\lesssim 1$ and within the validity of our approximation, namely  $s_1 \ll 1$.\\

Similarly, for the last part of the source term we obtain
\begin{eqnarray}
f_{\rm NL}^{\rm eq}&=&-\frac{10}{3}\frac{s_2}{c_s^2}\frac{2^{-8\gamma/H}\pi^2 {\Gamma\left(\frac{2\gamma}{H}+4\right)}^2}{\lp\frac{\gamma}{H}+1\rp^6{\Gamma\lp\frac{\gamma+H}{2 H}\rp}^8}
 \int_{0}^{+\infty} dy' {y'}^6 (g_{\gamma}(0, y'))^2\nonumber\\&\times&\int_{y'}^{+\infty} dz z^4  g_{\gamma}( y',z)g_{\gamma}(0, z),\label{fnlmix2n}
\end{eqnarray}
where $s_2=\frac{f_2H}{\dot{f_1}}\frac{\nu_{12}}{\nu_{11}}\ll 1$, and again for simplicity we take all mass scales to be of the same order. After a numerical calculation, in the regime $\gamma/H\gg 1$, we can well approximate $f_{\rm NL}$ by
\beq
f_{\rm NL}^{\rm eq}\simeq- 3\frac{\gamma}{H} \frac{s_2}{ c_s^2},
\eeq
similar to our previous case. 

\subsection{Vectors \& Tensors}\label{vectens}

Finally, as we discussed in secs. \ref{vectors} and \ref{tensors},  there is a family of vector and tensor interactions terms that can be added. These can be written as
\begin{eqnarray}\label{family}
S^{\calo}&=&-\int d^4 x \sqrt{-g}\,\left(f_1(t)\calo^{0}_1+f_2(t)\calo^{00}_2+f_3(t)\calo^{000}_3+\ldots\right) \\
&-&\delta g^{00}\left(s_1(t)\calo^{0}_1+s_2(t)\calo^{00}_2+s_3(t)\calo^{000}_3+\ldots\right)+\ldots\nonumber
\end{eqnarray}

For the sake of simplicity (and notation), let us here ignore the factors $f_i(t)$'s and $s_i(t)$'s, whose time dependence is in general slow roll suppressed.\\

After introducing $\pi$, these terms become
\begin{subequations}\begin{align}
\calo^{0}_1&=\calo^{\mu}_1\partial_{\mu}(\tilde{t}+\pi)=\calo^{0}_1+\calo^{\mu}_1\partial_{\mu}\pi\\
\calo^{00}_2&=\calo^{\mu\nu}_2\partial_{\mu}(\tilde{t}+\pi)\partial_{\nu}(\tilde{t}+\pi)\\&=\calo^{00}_2+2\calo^{0\mu}_2\partial_{\mu}\pi+\calo^{\mu\nu}_2\partial_{\mu}\pi\partial_{\nu}\pi\nonumber\\
\calo^{000}_3&=\calo^{\mu\nu\rho}_3\partial_{\mu}(\tilde{t}+\pi)\partial_{\nu}(\tilde{t}+\pi)\partial_{\rho}(\tilde{t}+\pi)\\&=\calo^{000}_3+3\calo^{00\mu}_3\partial_{\mu}\pi+3\calo^{0\mu\nu}_3\partial_{\mu}\pi\partial_{\nu}\pi+\calo^{\mu\nu\rho}_3\partial_{\mu}\pi\partial_{\nu}\pi\partial_{\rho}\pi,\nonumber
\end{align}
\end{subequations}
etc. Adding all them up we get
\be\label{suma}
\calo^{0}_1+\calo^{00}_2+\calo^{00}_3+\ldots=\tilde{\calo}+\tilde{\calo}^{\mu}\partial_{\mu}\pi+\tilde{\calo}^{\mu\nu}\partial_{\mu}\pi\partial_{\nu}\pi+\tilde{\calo}^{\mu\nu\rho}\partial_{\mu}\pi\partial_{\nu}\pi\partial_{\rho}\pi \ldots,
\ee
with $\tilde \calo^{\mu\ldots\nu}$ made up from some combination of our original tensor operators. We can rearrange the second line of Eq.~(\ref{family}) in a similar fashion.\\

As an example, let us consider $\calo^0 \equiv \calo_\mu g^{\mu0}$, which produces the first $\pi$-dependent term in Eq.~(\ref{suma}). Notice that by simple inspection, at linear order in $\pi$, we can transform the analysis for this operator into the one for a scalar operator we studied throughout the paper. Indeed, after integrating by parts we obtain
\beq \label{oeff} \int d^4x \sqrt{-g} \calo^\mu \partial_\mu \pi \to \int d^4x \left(\partial_\mu \sqrt{-g} \calo^\mu\right)\pi \equiv \int d^4x \sqrt{-g} \tilde\calo \pi,
\eeq
with $\tilde\calo= \frac{1}{\sqrt{-g}}\partial_{\mu}(\sqrt{-g}{\calo}^{\mu})$, an effective scalar operator. By construction, this interaction obeys the shift symmetry. Moreover, assuming it satisfies the hypothesis of sec. \ref{localapp}, it may induce a large friction term for $\gamma \gg H$, and yet no $O(\pi^2)$ terms in the action. Naively this might suggest an absence of connection between friction and non-Gaussianities in this case, however as we emphasized throughout the paper, the non-linear response will induce non-linear couplings that indeed we expect to be large, leading to $f_{\rm NL} \simeq\gamma/c_s^2H$.


\section{$\calo\delta g^{00}$ model(s) with $\gamma\dot\pi$ dissipation}\label{mag00}

In this appendix we discuss an example of a $\tilde\calo g^{00}$ type of coupling with $\gamma\dot\pi$ dissipation. We argued in sec. \ref{storyo} this requires some peculiar analytic structure, in particular (see Eq. (\ref{tildG}))
\beq
{\rm Im} \tilde G_{\rm ret}^\calo (\omega) \sim 1/\omega,
\eeq
within a range of frequencies $\mu_\calo < \omega < \Gamma_\calo$ to be determined momentarily. Here we introduce one possible model where we couple the inflaton to a scalar field with strong dissipative dynamics induced by a second (hidden) sector that does not couple directly to $\pi$. We will explore this set up in more detail elsewhere. (We drop the tildes from now on, and due to the plethora of $\gamma$'s we use $\gamma_\pi$ to identify the dissipative coefficient for $\pi$.)\\

The basic idea is to use the EFT of inflation of \cite{eft1} but, in the spirit to the model in \cite{dan2}, let us couple $\pi$ to a scalar field $\sigma$ via a term $\sigma \delta g^{00}$ so that we generate a $\rho \sigma\dot\pi$ coupling with $\rho$ having units of mass \cite{dan2}. For its dynamics we write
\beq
\calL_\sigma = \frac{1}{2}(\partial_\mu \sigma)^2 - \frac{1}{2} \mu^2 \sigma^2 + \sigma J_\Psi,
\eeq
where $J_\Psi$ stands for the interaction between $\sigma$ and a dissipative sector described collectively by $\Psi$, which does not
couple directly to $\pi$.\footnote{In principle $\Psi$ could talk to the inflaton via gravitational interactions, but these are subleading.}
In essence this is a two-steps model where $\pi$ dissipate via a mixing term. This can also be described in the framework of \cite{multieft}, except that in addition we have a sector which couples to $\sigma$ and induces (strong) dissipation. (We could for example put the $\Psi$-system in thermal equilibrium at a given temperature $T_\Psi$.)\\

If we now compute the retarded Green's function for $\sigma$ we have (see Fig. \ref{odot})
\beq
\rho^2 G^\sigma_{\rm ret}(\omega,{\bf k}) = \frac{\rho^2}{\omega^2-{\bf k}^2 - \mu^2 + \Gamma_\Psi(\omega)},
\eeq
where $\Gamma_\Psi$ is the self-energy contribution from the $\langle[ J_\psi, J_\psi] \rangle$ correlator due to the $\sigma J_\Psi $ coupling (see below). We assume now that the imaginary part of this self-energy can be approximated as
\beq
\label{imsigm}
{\rm Im}\, \Gamma_\Psi(\omega) \simeq \gamma_\Psi \frac{\omega}{\omega+\Lambda_\psi},
\eeq
which is a more standard behavior (similar to Drude's model in Eq. (\ref{drude})) and behaves linearly in $\omega$ for $\omega < \Lambda_\psi$, see appendix \ref{appB}. Using analyticity of $\Gamma_\Psi$ in the upper half plane we also have (via Kramers-Kronig relations)
\beq
\label{resigm}
{\rm Re}\,\Gamma_\Psi(\omega) \simeq \gamma_\Psi \Lambda_\Psi \left(\frac{\Lambda_\Psi^2}{\omega^2+\Lambda_\Psi^2}\right) \simeq \gamma_\Psi \Lambda_\Psi,
\eeq
for $\omega \lesssim \Lambda_\Psi$.
Therefore
\beq
\label{imGs}
\rho^2 {\rm Im}\, G^\sigma_{\rm ret}(\omega,{\bf k}) \simeq  \frac{\gamma_\Psi \rho^2 \omega}{\left[\omega^2-{\bf k}^2 - \mu^2 + \gamma_\Psi \Lambda_\Psi \right]^2 + \gamma_\Psi^2\omega^2}.
\eeq

\begin{figure}[h!]
    \centering
    \includegraphics[width=6.5cm]{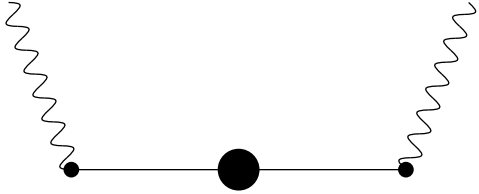}
\caption{The wavy lines correspond to $\pi$ and the dark interactions are insertions of $\sigma\dot\pi$ couplings. The larger dark circle represents the Green's function for the $\langle [J_\psi, J_\psi]\rangle$ correlator from to the $\sigma J_\Psi $ coupling.}
\label{odot}
\end{figure}

Our quest is to find a regime where we get $\gamma_\pi \dot\pi_k$ for the EOM, or in other words the scaling of Eq. (\ref{tildG}) in Eq. (\ref{imGs}). That is indeed the case when
\beq
\label{mutuned2}
\left[\omega^2-{\bf k}^2 - \mu^2 + \gamma_\Psi \Lambda_\Psi \right]^2 \ll \gamma_\Psi^2\omega^2.
\eeq

Naively this seems like a difficult task given the constraint $\omega\lesssim \Lambda_\Psi$, however we can always set $\mu$ to cancel out the large contribution, for example by tuning\footnote{It is not surprising we require some fine tuning while dealing with (unprotected) scalar fields. One possibility is to think of $\sigma$ as a pseudo-Goldstone boson, or introduce a supersymmetric version, with the scale of breaking near Hubble. (In that sense we resemble the EFT for supersymmetric multi-field inflation of \cite{multieft}, see also \cite{dan2new}.) We leave this open for future work.}
\beq
\label{mutuned}
\gamma_\Psi \Lambda_\Psi - \mu^2 \simeq H^2.
\eeq
Moreover we also assume $\omega_\star, k_\star \ll \gamma_\Psi$ (which we show is self-consistent), so that under our working hypotheses we get
\beq
\label{imGs2}
\rho^2 {\rm Im}\, G^\sigma_{\rm ret}(\omega) \simeq  \frac{\gamma_\Psi \rho^2}{\gamma_\Psi^2}\frac{1}{\omega},
\eeq
for $H \lesssim  \omega, |{\bf k}| \lesssim \gamma_\Psi (\lesssim \Lambda_\Psi)$, hence
\beq
\gamma_\pi \sim \frac{\rho^2}{\gamma_\Psi},
\eeq
which gives $\gamma_\pi \sim \gamma_\Psi$ for $\rho \sim \gamma_\Psi$. As we showed in this paper, with a $\gamma_\pi\dot\pi$ dissipation term we have $k_\star \simeq \sqrt{\gamma_\pi H}$, therefore the conditions which led to Eq. (\ref{imGs2}) are satisfied, provided $\gamma_\pi \gg H$.\\

To compute the power spectrum and non-Gaussianities we also need the noise part. This follows also from the (indirect) coupling between $\pi$ and $J_\Psi$ via $\sigma$. This can be shown by explicitly writing the EOM (ignoring the Hubble expansion for simplicity)
\bea
\label{ddpis}\ddot\pi_k + k_{\rm ph}^2\pi_k &=& -\rho\dot\sigma_k, \\
\label{ddsig}\ddot\sigma_k + (k_{\rm ph}^2+\mu^2)\sigma_k &=& J_\Psi + \rho\dot\pi_k.
\eea

As we argued above, in the regime we are interested in we can write (treating $J_\Psi$ as a particular operator~$\delta\calo$)
\beq
J_\Psi =  J^S_\Psi + \int G^\Psi_{\rm ret} \sigma \simeq J^S_\Psi +\gamma_\Psi\Lambda_\Psi\sigma - \gamma_\Psi\dot\sigma,
\eeq
where we used Eqs. (\ref{imsigm}) and (\ref{resigm}) and added as before a stochastic source term $J^S_\Psi$. Therefore, choosing $\mu$ according to Eq. (\ref{mutuned2}) we have
\beq
 \gamma_\Psi\dot\sigma_k \simeq \rho\dot \pi_k - J^S_\Psi,
\eeq
and plugging it back into Eq. (\ref{ddpis}) we finally obtain
\beq
\ddot\pi_k + k_{\rm ph}^2\pi_k +\gamma_\pi \dot\pi_k = J^S_\pi,
\eeq
with $J^S_\pi =\left(\rho/\gamma_\Psi\right) J^S_\psi$. Hence we are left with an equation as in (\ref{eom1}) and the computations throughout the paper follow. In particular we get non-Gaussianities of order $f_{\rm NL} \simeq \gamma_\pi/(c_s^2 H)$ as in sec.~\ref{Nolocalo2}.\\

Notice we can also construct a similar model for a ${\calo}_\mu g^{\mu0}$ type of operator replacing $\sigma$ by a gauge coupling $g A^\mu \partial_\mu \pi$. Then $J_\Psi$ would play as similar role as the usual electromagnetic current (see appendix \ref{appB}).

\section{Consistency of local trapped inflation}\label{consistency}

Here we summarize some basic constraints that guaranteed the validity of our approximations. For the background equations we need
\be\label{c5c6} |\ddot{\phi}|\ll 3H|\dot{\phi}|\ll V'(\phi),\ee
so that we have a constant velocity solution
 \be\label{E1} \dot{\phi}\simeq~ -\frac{(\Gamma_\chi|\Delta|(2\pi)^3 V')^{2/5}}{g}.\ee
In addition, we assume that the energy density is dominated by the potential energy,
\be\label{EnDens} 3M_p^2 H^2=\frac{\dot{\phi}^2}{2}+V(\phi)+\sum_i M_{\chi_i} n_{\chi_i}\simeq V(\phi).\ee The generalized slow roll parameter $\epsilon=-\dot{H}/H^2$ is given by
\be\epsilon=\frac{3(\dot{\phi}^2+\sum_i M_{\chi_i} n_{\chi_i})}{2 V(\phi)}.\ee
Then, $\epsilon\ll 1$ implies
\begin{subequations}\begin{align}
&\dot{\phi}^2\ll V(\phi),\label{c7}\\
&\sum_i M_{\chi_i} n_{\chi_i}\ll V(\phi),\label{c8}
\end{align}\end{subequations} which are the conditions required to make the approximation in Eq. (\ref{EnDens}).
To estimate the LHS of Eq. (\ref{c8}) we replace the sum by an integral as above and find
\be
\sum_i M_{\chi_i} n_{\chi_i}\simeq~\frac{g^{5/2}|\dot{\phi}|^{7/2}}
{\Gamma_{\chi}^2|\Delta|(2\pi)^3}\ll V(\phi).
\ee
Gathering all pieces together we collect Eqs. (\ref{c1}), (\ref{c2n}), (\ref{c3c4}), (\ref{c5c6}), (\ref{c7}), (\ref{c8}),
as well as the conditions $k_{\rm ph} \ll \kappa$ and $k_{\rm ph}\Delta \ll |\dot \phi|$, plus $\Gamma_\chi\gg H$.
In addition we also have the number of e-foldings, where we get
\be \label{efolds}N_e=\int \frac{H}{\dot{\phi}}d\phi=\frac{5 g m \phi ^2}{16 2^{7/10} \sqrt{3} \pi ^{6/5} \left(m^2 \Gamma_\chi  \Delta  \phi \right)^{2/5}},\ee  using $\phi\equiv\phi_I \gg\phi_E$, with $\phi_I$ ($\phi_E$), the inflaton field at the beginning (end) of inflation.

\begin{figure}[h!]
\includegraphics[width=8.5cm]{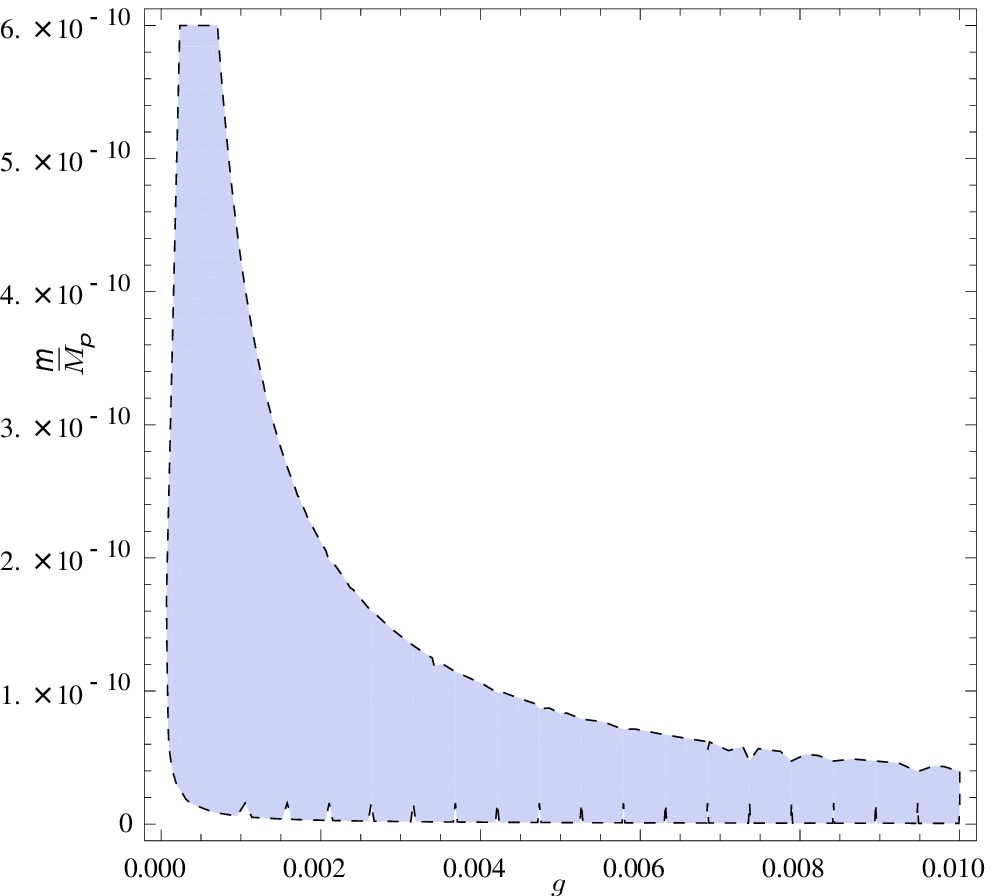}
\includegraphics[width=8.5cm]{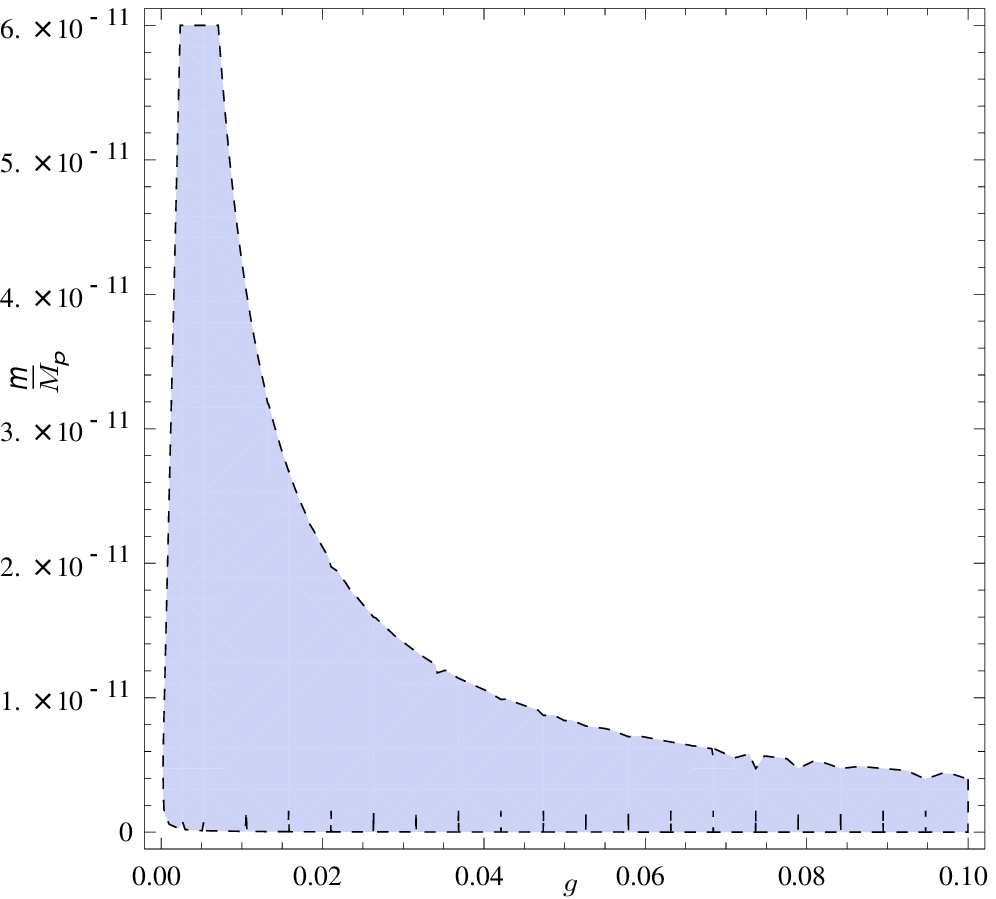}
\caption{The region of the free parameters  ($m$ and $g$) allowed for consistency   (with $``\ll"$ replaced by $``<"$).  The horizontal axis represents $g$  and the vertical axis   $m/M_p$. Top:  $\Gamma_\chi=100 H$ (case I); Bottom: $\Gamma_\chi=1000 H$ (case II).}
\label{RR}
\end{figure}

Next we analyze these constraints for the paradigmatic example
\be V(\phi)=\frac{m^2}{2}\phi^2.\ee

We have five parameters in the model: $m,\,\Delta,\,\Gamma_\chi,\,g$ and the initial value of the inflaton field, $\phi_I$. To reduce the parameter space we impose  $N_e=60$ and
\beq
P_{\zeta}\simeq\lp\frac{H}{\dot{\phi}}\rp^2P_{\delta\phi}\simeq \frac{10^{-9}}{k^3},
\eeq from where we obtain the condition (see Eq. (\ref{zeta2nu}), using $N_c \simeq \dot\phi^2$)
\beq\label{E2}
\lp\frac{H}{\dot{\phi}}\rp^2 \frac{g^2\nu_{\chi}}{4}\lp\frac{\pi H}{\gamma}\rp^{1/2}\simeq 10^{-9}.
\eeq
Then,
\begin{equation}\phi \simeq \frac{16 2^{3/16} 3^{15/16} \pi ^{3/4} \sqrt[4]{\Gamma_\chi  |\Delta| }}{\sqrt[8]{g^5 m}} M_p,~~
|\Delta|\simeq\frac{1179648\times 10^{34} \sqrt[4]{6} \pi ^{11} g^{13/2} m^{9/2} M_p}{\Gamma_\chi }
\end{equation}
For simplicity, we analyze two cases: $\Gamma_\chi=100 H$ (case I) and  $\Gamma_\chi=1000 H$ (case II). In Fig.~\ref{RR}  we plot the 2-dimensional region for the parameters $m$ and $g$ for which all conditions are satisfied (with $``\ll"$ replaced by $``<"$). The horizontal axis represents $g$  and the vertical axis  $m/M_p$: case I corresponds to the graph on the left and case II to the right. As an example, taking $\Gamma_\chi=100 H$, $m=10^{-11}M_p$ and $g= 10^{-2}$, we obtain $H\simeq 1.3\times 10^{-11} M_p$, $\gamma/H\simeq  135$, $\phi\simeq 3.3 M_p$,  $\delta t_c \simeq 1.98 \times 10^{-4} H^{-1}$, $|\Delta|\simeq 1.29\times 10^{-8}M_p$ and $\epsilon\simeq 1/32$. Also, for $\Gamma_\chi=1000 H$, $m=5\times 10^{-13}M_p$ and $g= 10^{-1} $, we get $H\simeq 3.34\times 10^{-13} M_p$, $\gamma/H\simeq 539$, $\phi\simeq 1.63 M_p$,  $\delta t_c \simeq 1.4\times 10^{-5} H^{-1}$, $|\Delta|\simeq 2.27\times 10^{-7} M_p$ and $\epsilon\simeq 1/32$. In all cases $\gamma/H \ll g|\dot\phi|/H^2$, which guarantees the validity of the local approximation $k_\star\lesssim \kappa$. As a final check let us estimate the
size of the curvature perturbations one obtains from the fluctuations in the additional scalar degrees of freedom we added into the theory. Following \cite{trapped} we can estimate this contribution by computing
\beq
M_p^2\frac{\partial_i^2}{a^2} \delta g_\chi \sim M_\chi \Delta n_\chi,
\eeq
where we take $M_\chi(\delta t_c) \simeq g |\dot\phi| \delta t_c \simeq \sqrt{g|\dot\phi|}$. In the above expression $\delta g_\chi$ is not the  curvature perturbation $\zeta_\chi$, but nonetheless it gives us an idea of the size of the curvature it induces. For the sake of comparison, let us evaluate this expression at $k/a \simeq H$, where we get
\beq
\langle\delta g_\chi\delta g_\chi\rangle \simeq \frac{(g|\dot\phi|)^{5/2}N_{\rm hits}}{HM_p^4}.
\eeq
On the other hand, from the analysis in sec. \ref{power} we have (see Eq. (\ref{nunhits}))
\beq
\langle\zeta_\phi\zeta_\phi\rangle \simeq \sqrt{H/\gamma} \frac{H^2}{\dot\phi^2} \frac{g^2(g|\dot\phi|)^{3/2}N_{\rm hits}}{\Gamma_\chi}.
\eeq
Hence, using $\dot\phi^2 \sim \epsilon_\phi V$ ($\epsilon_\phi < \epsilon$), then \beq
 \frac{\langle\delta g_\chi\delta g_\chi\rangle}{\langle\zeta_\phi\zeta_\phi\rangle} \simeq \frac{\Gamma_\chi k_\star}{M_\chi^2} \epsilon_\phi^2 \ll 1,\eeq  since $\Gamma_\chi\lesssim M_\chi\,, k_\star\lesssim M_\star$ and $\epsilon_\phi\lesssim 1$. Therefore curvature perturbations become subleading.

\end{document}